\documentclass[english,nofootinbib]{revtex4-2}
\usepackage[T1]{fontenc}
\usepackage[latin9]{inputenc}
\usepackage{color}
\usepackage{array}
\usepackage{booktabs}
\usepackage{textcomp}
\usepackage{mathtools}
\usepackage{amsmath}
\usepackage{stackrel}
\usepackage{graphicx}
\usepackage{esint}

\makeatletter

\providecommand{\tabularnewline}{\\}

\usepackage{babel}

\makeatother

\usepackage{babel}
\begin{document}
\title{Analysis for satellite-based high-dimensional extended B92 and high-dimensional
BB84 quantum key distribution}
\author{Arindam Dutta$^{1,2}$}
\email{arindamsalt@gmail.com}

\email{https://orcid.org/0000-0003-3909-7519}

\author{Muskan$^{1}$}
\email{muskan.1@iitj.ac.in}

\email{https://orcid.org/0009-0009-1630-8898}

\author{Subhashish Banerjee$^{1}$}
\email{subhashish@iitj.ac.in}

\email{https://orcid.org/0000-0002-7739-4680}

\author{Anirban Pathak$^{2\,}$}
\email{anirban.pathak@gmail.com}

\email{https://orcid.org/0000-0003-4195-2588}

\affiliation{$^{1}$Department of Physics, Indian Institute of Technology Jodhpur,
Jodhpur 342030, Rajasthan, India~\\
$^{2}$Department of Physics and Materials Science \& Engineering,
Jaypee Institute of Information Technology, A 10, Sector 62, Noida,
UP-201309, India}
\begin{abstract}
A systematic analysis of the advantages and challenges associated
with the satellite-based implementation of the high dimensional extended
B92 (HD-Ext-B92) and high-dimensional BB84 (HD-BB84) protocol is analyzed.
The method used earlier for obtaining the key rate for the HD-Ext-B92
is modified here and subsequently the variations of the key rate,
probability distribution of key rate (PDR), and quantum bit error
rate (QBER) with respect to dimension and noise parameter of a depolarizing
channel is studied using the modified key rate equation. Further,
the variations of average key rate (per pulse) with zenith angle and
link length in different weather conditions in day and night considering
extremely low noise for dimension ${\rm d}=32$ are investigated using
elliptic beam approximation. The effectiveness of the HD-(extended)
protocols used here in creating satellite-based quantum key distribution
links (both up-link and down-link) is established by appropriately
modeling the atmosphere and analyzing the variation of average key
rates with the probability distribution of the transmittance (PDT).
The analysis performed here has revealed that in higher dimensions,
HD-BB84 outperforms HD-Ext-B92 in terms of both key rate and noise
tolerance. However, HD-BB84 experiences a more pronounced saturation
of QBER in high dimensions. 
\end{abstract}
\maketitle

\section{introduction}

In an information-centric society, safeguarding communications and
data emerges as a fundamental necessity. This necessity spans various
applications, including but not limited to financial transactions,
upholding individual privacy, and preserving the integrity of vital
components within the Internet of Things. Cutting-edge classical cryptosystems
like Rivest-Shamir-Adleman (RSA) algorithm provide security that hinges
on the computational complexity of a problem and associated assumptions
about the computational power of the adversaries \cite{RSA83}. However,
these assumptions can be compromised once large-scale quantum computers
come into play \cite{XMZ+20}. The remedy for this challenge is provided
by a relatively recent cryptographic concept known as quantum key
distribution (QKD) \cite{BB84,E91,B92,BBM92,SAR+04}. Its security remains
unaffected by algorithmic or computational progressions \cite{CJC+20}.
QKD enables the creation of symmetric keys between remote entities
or parties, ensuring a level of confidentiality that is inherently
constrained by the fundamental laws of physics \cite{P13,DP+22}.
The polarization of light (photons) is a degree of freedom that is
often utilized to realize different schemes for QKD and other schemes
for secure quantum communication \cite{PRM+14,VZJ+17,ZZW18,HS20,BO22,TKO+17,DP23,DP+23,VRG+22,PWM+21,PCL+23}.
Drifting qubits encoded in photons have the potential to be distributed
over a distance of at most a few hundred kilometers through optical
fibers \cite{IMT+13,BLH+15,WJS+20,SPC+09}. To extend these distances
further, the utilization of quantum repeaters has been proposed \cite{BDC+98,SSR+11},
but quantum repeaters are not yet commercially available. Further,
maintaining light polarization possesses practical difficulties in
long-distance QKD protocols. Improving the key rate
and extending the range of QKD present a crucial challenge. This challenge,
which may seem insurmountable without the implementation of quantum
repeaters \cite{SSR+11}, revolves around surpassing the fundamental
rate--distance limit inherent to QKD. This limit dictates the maximum
secret key rate achievable between two parties over a given distance
using QKD and is defined by the secret-key capacity of the quantum
channel \cite{PLOB17} linking these parties. In addition to the traditional
QKD protocols, an alternative approach involves generating pairs of
phase-randomized optical fields at separate distant locations, which
are subsequently combined at a central measurement station. These
fields, characterized by identical random phases, are termed as \textquoteleft twins\textquoteright{}
and can be utilized to distill a quantum key. The key rate of this
twin-field QKD follows a similar distance-dependent trend as a quantum
repeater, scaling proportionally to the square root of the channel
transmittance \cite{LYD+18,XWL+23}. However, compensating for channel
fluctuations and locking laser fluctuations necessitates the use of
phase tracking and phase locking techniques in experimental setups,
significantly increasing complexity and impeding free-space implementation.
Further advancements include the development of an asynchronous measurement-device-independent
quantum key distribution protocol capable of surpassing secret key
capacity even in the absence of phase tracking and phase locking \cite{XLW+22,ZZW+22,XBL+23}.
While these schemes have been experimentally demonstrated to be effective
under conditions of significant quantum channel attenuation \cite{CZL+21,ZLX+23,ZHL+23,LZJ+23},
they hold promise for future applications in space-ground quantum
key distribution.

In the case of QKD protocols based on
optical fiber, the polarization state is susceptible to alterations
caused by random fluctuations of birefringence in the optical fiber
\cite{VR99,GK2000}. Further, the diminishing signal strength and
the interference from ambient noise experienced during QKD transmissions
via optical fibers hinder the attainment of substantial key rates
beyond networks of metropolitan proportions \cite{TTS+11,BAL17,ASW+23}.
An alternative solution involves the proper utilization of optical
satellite links, which can potentially overcome the limitations on
transmission distances encountered by ground-based photonic communication
schemes \cite{UJK+09,SWU13,SB19}. When dealing with open space conditions,
while polarization exhibits greater resilience against atmospheric
turbulence, in the reference frame of the satellite, variation of
polarization is observed due to the motion of the satellite. This
introduces a negative impact \cite{KSD+05,ZDD17,ZJF+21}, and it becomes
crucial to address these polarization fluctuations issues in both
free-space and fiber-based QKD systems \cite{P21,P+21}. Traditional
approaches for addressing this challenge encompass the utilization
of active polarization tracking devices \cite{XWF+09,TTS+11,DCC+17,LGL+18,NBB+22,LKB+22}.
An alternative approach through a proof-of-principle experiment was
proposed in 2023 \cite{CGC+23}. Here, quantum state tomography was
used to determine the optimal measurement bases for a single party.
Moreover, embedding quantum technology within space platforms offers
an avenue for conducting fundamental experiments in physics \cite{RJC+12,JPR+18}\textcolor{red}{{}
}and pioneering innovative concepts like quantum clock synchronization
\cite{GLM01,GLM+01,HLK09} and quantum metrology \cite{ABS+14}. Although
this endeavor presents significant technological challenges, a variety
of experimental investigations \cite{UTM+07,VJT+08,FUH+09,YRL+12,WYL+13,NMR+13,CLY+13,PKB+17}
alongside theoretical inquiries \cite{SWU13,BTD09} have showcased
the feasibility of this approach. These studies have demonstrated
the viability of this approach using state-of-the-art technology already
in use on the ground and approved for space operations \cite{TCT+22,VVM19}.
In fact, over the last decade, numerous experiments in free-space
conditions have been conducted to assess the practicality of QKD setups
on mobile platforms, encompassing diverse vehicles like hot-air balloons
\cite{WYL+13}, trucks \cite{BHG+15}, aircraft \cite{NMR+13,PKB+17},
and drones \cite{LTG+20}. Consequently, in what has been characterized
as the \emph{quantum space race} \cite{JH13}, multiple international
research groups in countries like Canada, Japan, Singapore, Europe,
and China have been actively participating and trying to establish
stable space-based communication channels \cite{BAL17,KHN+18}. Notably,
these efforts have seen the successful launch of satellites with payloads
capable of being used in quantum communication \cite{YCLL+17,LCL+17,TCF+17,LLR+17,YLL+20,YCL+17,RXY+17}.
Furthermore, the development of a quantum internet
is poised to be pivotal in introducing fundamentally novel internet
technology, facilitating quantum communication between any two locations
on Earth. This quantum internet, in conjunction with the existing
\textquotedblleft classical\textquotedblright{} internet infrastructure,
will interlink quantum information processors, enabling the realization
of capabilities that are demonstrably unattainable solely through
classical information methods \cite{WEH18}. Through this quantum
technology, the security objectives, specifically confidentiality,
integrity, authenticity, and non-repudiation---will underpin diverse
e-commerce transactions and the communication of sensitive information
\cite{YFL+23,CLW+24}. The advancement of quantum technology has been greatly influenced
by the healthy competition among the researchers \cite{TCT+16,TCTC+16,SVF+16},
driving notable progress in quantum nonlinear optics, entangled photon
generation techniques, and single photon detection in recent years.
Given these remarkable technological strides, it is imperative to
reevaluate the enhanced performance aspects of QKD through typical
free-space connections. This reevaluation particularly focuses on
the considerable rise in secure key generation rates compared to earlier
experiments \cite{PYB+05,MLK+06,UTM+07,ECL+08,SUF+09} (see Figure
1 of Ref. \cite{ELH+21}). While this analysis (refer to Figure 1
of Ref. \cite{ELH+21}) does not incorporate field tests utilizing
prepare-and-measure schemes, it is worth noting that both terrestrial
\cite{MWF+07,DBD+24} and satellite-based \cite{LCL+17,LCH+18} studies
have effectively demonstrated decoy-state key exchange across free-space
links at high rates. Furthermore, entanglement-based QKD protocols
eliminate the necessity to place trust in the source of the satellite
in a dual down-link scenario. Motivated by these facts, in the present
work we wish to investigate the effectiveness of two specific protocols
for QKD for the long-distance free-space secure quantum communication
to be implemented with the assistance of a satellite. Before we specifically
mention the protocols selected here for the investigation, we would
like to briefly mention the logical evolution of the relevant protocols
that led to the protocols of our interest. 

The first protocol for QKD was proposed by Bennett and Brassard in
1984 (BB84 protocol) \cite{BB84}. From the introduction of the BB84
protocol, there has been a continuous progression in both theoretical
and practical aspects of QKD \cite{SPC+09,SPR17,PAB+20}. Nonetheless,
owing to the formidable challenges posed by the generation, maintenance,
and manipulation of quantum resources using current technologies,
there has been a concerted effort to formulate QKD protocols with
more straightforward conceptual frameworks such that the protocols
would require fewer quantum resources. For instance, the BB84 protocol
itself involves four quantum states and two measurement bases. In
1992, Bennett introduced a notably simpler QKD protocol named B92,
which relies solely on two non-orthogonal states and two measurement
bases \cite{B92}. However, B92 exhibits a heightened susceptibility
to noise in contrast to alternative protocols like BB84, as indicated
in the original paper \cite{B92}. Subsequently, in 2009, Lucamarini
et al. \cite{LGT09} introduced an extended version of B92 (Ext-B92),
incorporating two extra non-informative states to more effectively
constrain Eve's information gain. BB84, B92, and Ext-B92 protocols
and most of the other existing protocols for QKD utilize qubits, which
are two-dimensional systems, as the means of communication between
Alice and Bob. Nevertheless, there have been limited investigations
concerning the susceptibility of qudit-based schemes (i.e., schemes
utilizing key encoding on ${\rm d}$-level systems) to eavesdropping
in the case of high dimensional systems. Initiatives are currently
underway to establish and investigate qudit systems within laboratory
settings \cite{ZYW+22,MML+23}. Quantum systems with dimensions higher
than two have demonstrated numerous benefits and intriguing characteristics
compared to protocols based on qubits (as briefly discussed in \cite{CLB+19}).
There have been several studies related to their continuous variable
counterparts (\cite{PAT+19} and references therein). Further, certain
protocols have exhibited the ability to tolerate high levels of channel
noise as the system dimension expands, as evidenced by various studies
\cite{PT200,SS10,VKM+18,STP+16}. Motivated by these facts, in this
article, we assess the performance of the key rate under different
scenarios for the HD-Ext-B92 and HD-BB84 protocols. We calculate the
key rate of the HD-Ext-B92 scheme without the inclusion of extra independent
variables, in contrast to the method outlined in Ref. \cite{IK21},
which is explained further in Appendix A. We utilize the channel transmission
$\eta$ to evaluate our results, focusing on light propagation through
atmospheric links using the elliptic-beam approximation originally
presented by Vasylyev et al. \cite{VSV16,VSV+17}. Additionally, we
incorporate the generalized approach and varying weather conditions
introduced in \cite{LKB19}. Specifically, we investigate the applications
of these models in quantum communication using \emph{Low Earth Orbit}
(LEO) satellites. Here, it may be noted that the methodology proposed
in \cite{VSV16,VSV+17,LKB19} has a notable impact on the transmittance
value, which is influenced by beam parameters and the diameter of
the receiving aperture. 

Before delving into our main text, it is important to state that a
satellite-based link is of two distinct types: the up-link and the
down-link. These links should not be considered symmetrical due to
the crucial distinction in the order of signal beam traversal through
the atmosphere and space. In the up-link scenario, the signal beam
first encounters the atmosphere, where it is subject to the effects
of turbulence and scattering particles. It then proceeds into the
expanse of space over long distances, where beam broadening becomes
the dominant factor affecting its characteristics. Conversely, in
the down-link scenario, the beam travels through space first and then
through the atmosphere. In this scenario, the primary factor influencing
the signal beam's journey through extended space is the pointing error.
This contrast in the order of traversal results in unique requirements
for the receiving equipment on the ground and in space \cite{BSH+13,LKB19}.

The remainder of this paper is structured as follows.
Section \ref{sec:II} provides a detailed exploration of the HD-Ext-B92
and HD-BB84 protocols, alongside an extensive analysis of how atmospheric
conditions affect satellite communication links and the elliptical
approximation of beam deformation at the receiver. We also investigate
the key rate and QBER under varying noise parameters to determine
the noise tolerance of these higher-dimensional protocols. Section
\ref{sec:III} presents a thorough evaluation of the performance of
these high-dimensional protocols, supported by illustrative results
from simulations. Finally, we summarize our paper with the findings
being consolidated and deliberated upon in Section \ref{sec:IV}.
Appendix A contains detailed calculations for deriving the key rate
of the HD-Ext-B92 protocol, while Appendix B covers the first and
second moments of the beam parameters for both the up-link and down-link,
crucial for our simulation results.

\section{Preliminaries: High-dimensional B92 and BB84 protocols and Elliptic
beam approximation \label{sec:II}}

Numerous researchers have extensively investigated the unconditional
security of QKD-based protocols, and their research, (see for examples,
\cite{TKI03,M13}) has consistently revealed increasingly robust results.
For instance, in \cite{M13}, a noise tolerance of 6.5\% was reported
for the B92 protocol. Depending on the user's selected key encoding
states, the noise tolerance for this B92 protocol can extend up to
11\% in the asymptotic scenario, as demonstrated in \cite{LGT09}.
This level of noise tolerance is comparable to that of BB84. In scenarios
with a finite key length, as indicated in \cite{AK20}, the protocol
still maintains a minimum noise tolerance of 7\%. In this context,
we summarize the key-rate analysis for HD-Ext-B92 and HD-BB84 protocols.
We modify the calculation for HD-Ext-B92 using a theorem to eliminate
any additional free parameters (as detailed in Appendix A). Additionally,
we briefly delve into the methodology of elliptical beam approximation,
designed to encompass satellite-based connections while accounting
for signal losses in various real-world scenarios, including diverse
weather conditions. This methodology is particularly tailored for
application in LEO satellite contexts. 

\subsection{High-dimensional extended B92 protocol and high-dimensional BB84
protocol \label{subsec:HD-Ext-B92=000026HD-BB84}}

Before going into the intricacies of higher-dimensional
protocols, let's briefly discuss the higher-dimensional quantum states
utilized in performing HD-QKD schemes \cite{HZL+20,CLB+19}. Traditional
two-level quantum systems, represented by discrete variable states
\cite{BPM+97,FDI+04,OMM+09}, and continuous variable states \cite{FSB+98,YBF07,LBT+11}
within a single degree of freedom (such as polarization), have historically
been employed for communicating information as qubits. However, there
is a growing interest in exploring quantum information within larger
Hilbert spaces, achieved either by increasing the number of qubits
or by utilizing d-level quantum systems, known as qudits. The decision
to expand into higher dimensions depends on the specific objectives
of the task at hand. The overarching aim is to enhance the available
dimensions to transmit more than one bit per photon from one party
(Alice) to another (Bob). Various photonic degrees of freedom, such
as orbital angular momentum \cite{DLB+11}, temporal mode \cite{MGT+17},
frequency mode \cite{KRR+17}, and spatial mode \cite{HXL+20,VSP+20},
inherent to single photons, are natural candidates for realizing high-dimensional
systems. However, implementing HD-QKD protocols ideally necessitates
a reliable single-photon on demand source. While significant experimental
efforts have been directed towards constructing such sources (see
\cite{LP21,TS21} and references therein), weak coherent pulses (WCPs)
generated by attenuating laser outputs are commonly used as an approximate
single-photon source in many commercial products. The quantum state
of a WCP resulting from laser attenuation can be characterized as
follows:

\[
|\alpha\rangle=|\sqrt{\mu}{\rm exp}\left(i\theta\right)\rangle=\stackrel[k=0]{\infty}{\sum}\left(\frac{e^{-\mu}\mu^{k}}{k!}\right)^{\frac{1}{2}}{\rm exp}\left(ik\theta\right)|k\rangle,
\]
here, the symbol $|k\rangle$ denotes a Fock state ($k$ photon state)
and the mean photon number is denoted as $\mu=|\alpha|^{2}\ll1$.
Essentially, Alice generates a quantum state that can be conceptualized
as a superposition of Fock states, characterized by a Poissonian photon
number distribution expressed as $p(k,\mu)=\frac{e^{-\mu}\mu^{k}}{k!}$.
When utilizing the WCP source for signal state generation, users encounter
a probability of multi-photon pulses within the signal state. In this
scenario, Eve may execute a photon number splitting (PNS) attack.
Eve initiates the attack by employing a quantum non-demolition measurement
(QND) to determine the photon number, subsequently obstructing the
single-photon pulses and retaining one photon from the multi-photon
pulses. To counter the threat of PNS attacks, the decoy state method
is implemented \cite{H03,LMC05}. Notably, different intensities are
employed for generating the signal particles and the decoy state,
resulting in distinct photon number distributions. The security procedure
involves intentional and random replacement of signal pulses with
multi-photon pulses (decoy pulses) by legitimate users. Subsequently,
they assess the loss of the decoy pulses. If the loss of decoy pulses
is anomalously lower than that of signal pulses, the entire protocol
is aborted. Conversely, if the decoy pulse loss aligns with certain
expectations, the protocol continues. The estimation of signal multi-photon
pulse loss is then conducted based on the decoy pulse loss, assuming
similar values for the two losses. Within HD-QKD protocols, the decoy
state serves as a crucial tool for scrutinizing potential eavesdropping
activities and ensuring channel security.

\subsubsection{HD-Ext-B92}

Here, we summarize the HD-Ext-B92 protocol and recap some important
steps involved in the \emph{parameter estimation} process proposed
in Ref. \cite{IK21}. In fact, in this section, after briefly discussing
the HD-Ext-B92 protocol we modify the derivation of the asymptotic
key rate given in \cite{IK21} (see
Appendix A). It is apt to note that \emph{negotiation
efficiency}{} stands as a crucial parameter in determining
the security and accuracy of the final cryptographic key \cite{ZHG18,ZWF+23,CDG+21,FLL+21,BZJ+21,ZM20}.
This parameter encompasses several procedural steps, including quantum
state preparation-measurement, data reconciliation, QBER estimation,
parameter estimation, error correction, and privacy amplification\footnote{For simplicity, while evaluating the performance
of the satellite-based HD-Ext-B92 and HD-BB84 protocols, we are intentionally
excluding the incorporation of error correction and privacy amplification
measures.}. Before explaining the protocol, we would like
to introduce the notations used and the methodology for achieving
key rate. $|m\rangle$ and $|n\rangle$ are the fixed d-dimensional
states and defined from d-dimensional computational basis states $\{Z\in|1\rangle,\cdots,|m\rangle,|n\rangle,\cdots,|{\rm d}\rangle\}$,
and $|\psi\rangle=\frac{1}{\sqrt{2}}\left(|m\rangle+|n\rangle\right)$
is a fixed state which is chosen from d-dimensional diagonal basis
($X$-basis) states. As previously elucidated, a
photon can manifest as a high-dimensional system through the utilization
of different photonic degrees of freedom. It is pertinent to mention
that $|m\rangle$ and $|n\rangle$ denote specific higher-dimensional
systems that signify distinct states within the photonic degrees of
freedom. These states can be precisely measured by Bob using d-dimensional
computational basis states. In the context of the HD-Ext-B92 protocol,
Alice is only required to transmit three high-dimensional states.
Simultaneously, Bob's task entails conducting either a computational
basis measurement or a partial basis measurement in an alternative
basis (POVM $X$). It is noteworthy that this partial measurement
need only discriminate a specific superposition state as defined in
the protocol and is not obliged to discern all the possible states
(d states).

\emph{State preparation and transmission: }Alice randomly chooses
key-round and test-round. The key-round is employed for generating
raw key bit and test-round is employed to estimate error for this
protocol that will help to improve the negotiation
efficiency. Alice generates a sequence using states $|m\rangle$ and
$|\psi\rangle$ to represent classical bit values 0 and 1 during the
key-round, respectively. To assess channel noise and security, she
randomly includes decoy states in the sequence\footnote{The decoy state is not required to be a higher-dimensional
state; it can be a two-dimensional quantum state, qubit.}. Subsequently, Alice transmits the enlarged sequence
to Bob while maintaining confidentiality of the basis information.
During test-round, she uniformly prepares states $|m\rangle$, $|n\rangle$,
or $|\psi\rangle$ with a random selection, inserts decoy states,
and transmits the sequence to Bob. The basis information remains confidential
until Bob performs measurements on the sequence.

\emph{Estimation of channel noise and loss:}
Alice communicates the position and basis information of the decoy
state to Bob through a public classical announcement. Bob performs
measurements on the decoy state and publicly discloses the obtained
results. The comparison of these results allows for the calculation
of channel noise. If the noise falls within the predetermined threshold,
the protocol advances to the next step; otherwise, the protocol is
aborted. The measurement of decoy states also serves to assess channel
loss. In both the key round and the test round, the loss of signal
particles corresponds to the loss occurring for the decoy state. This
ensures the security of the channel against PNS attack.

\emph{Measurement and classical announcement:} Following
the security check conducted using the decoy state, Bob will proceed
to measure each state within the received sequence. This measurement
involves the elimination of decoy states, accomplished either using
the $Z$ basis or by a POVM bases defined by $\left\{ |\psi\rangle\langle\psi|,I-|\psi\rangle\langle\psi|\right\} $,
and referred to as POVM $X$. Here, the symbol $I$ represents the
d-dimensional identity operator. Bob sets the bit value as $1$ when
he observes $I-|m\rangle\langle m|$ by using measurement basis $Z$,
i.e., any measurement outcome in $Z$ basis other than $|m\rangle\langle m|$\footnote{This process resembles the B92 protocol \cite{B92}.
Alice utilizes the $|0\rangle$ and $|+\rangle$ states to encode
0 and 1, respectively. Bob deciphers 0 and 1 based on his measurement
outcomes, which correspond to the $|-\rangle$ and $|1\rangle$ states,
respectively.}; and he sets bit value $0$ when his measurement outcome using POVM
$X$ is other than $|\psi\rangle\langle\psi|$. All other results
are not taken into account as conclusive measurements. Alice and Bob
discard the iteration for inconclusive outcomes in key-round, and
determine the channel error rate in test-round by announcing their
basis choices and measurement results using an authenticated classical
channel. To enhance negotiation efficiency, we conduct
parameter estimation and QBER analysis, taking into account a depolarizing
channel with the consideration of the noise parameter $q$. In instances
where a round is inconclusive or does not qualify as a key round,
the obtained results are employed for parameter estimation. Finally,
they run the error correction and privacy amplification protocols
to get the final secure key. Here are some crucial
formulations pertaining to parameter estimation, which contribute
to deriving the key rate equation of the HD-Ext-B92 protocol. Depolarizing
noise is a very general noise of the Pauli class of noise channels
and can be obtained by twirling them in two dimensional quantum state.
Any quantum channel can be twirled by the depolarizing channel. Our
aspiration to expand randomized benchmarking to d-dimensions led
to the selection of the depolarizing channel in the protocol.

\[
{\mathcal{D}_{{\rm q}}(\rho)=\left(1-\frac{{\rm d}}{{\rm d}-1}{\rm q}\right)\rho+\frac{{\rm q}}{{\rm d}-1}I.}
\]
The observable statistics can be expressed within
the context of a depolarizing channel scenario.

\[
{\begin{array}{lccclclcl}
 &  & p_{mm} & = & p_{nn} & = & p_{\psi\psi} & = & 1-{\rm q},\\
 &  & p_{mc} & = & p_{nc} & = & p_{\psi c} & = & \frac{{\rm q}}{{\rm d}-1},\\
p_{m\psi} & = & p_{n\psi} & = & p_{\psi m} & = & p_{\psi n} & = & \frac{1}{2}\left(1-\frac{{\rm q}\,{\rm d}}{{\rm d}-1}\right)+\frac{{\rm q}}{{\rm d}-1}.
\end{array}}
\]
Supposing $p_{ij}$ represents the joint probability
associated with Alice's and Bob's raw bits being $i$ and $j$, considering
the scenario without eliminating that specific iteration. The values
of observable probabilities under the simulated channel for parameter
estimation are,

\[
{\begin{array}{lcl}
p_{00} & = & \frac{1}{2M}\left(1-p_{m\psi}\right),\\
p_{01} & = & \frac{1}{2M}\left(1-p_{mm}\right),\\
p_{10} & = & \frac{1}{2M}\left(1-p_{\psi\psi}\right),\\
p_{11} & = & \frac{1}{2M}\left(1-p_{\psi i}\right).
\end{array}}
\]

In \cite{IK21}, authors proposed a collective attack by Eve in which
she can independently and identically attack each round of the protocol.
Eve also can delay measurement on her register (quantum memory) after
completion of the protocol. The Devetak Winter key rate equation \cite{DW05,RGK_05}
is used to compute the key rate in the asymptotic limit\footnote{For instance, we are interested in seeing the performance of satellite-based
communication in the infinitely generated raw key scenarios.}:

\begin{equation}
R\left(a,b,E\right)=\underset{N\longrightarrow\infty}{{\rm lim}}\frac{l}{N}={\rm inf}\left[S\left(a|E\right)-H\left(a|b\right)\right],\label{eq:Key-rate equantion}
\end{equation}
this analysis helps us to obtain the minimum value of the key rate
by subtracting conditional Shannon entropy $H\left(a|b\right)$ from
conditional von Neumann $S\left(a|E\right)$. Here, $S\left(a|E\right)$
is defined as the entropy or the uncertainty present in Alice's classical
register $a$ given Eve's quantum memory $E$ and $H\left(a|b\right)$
denotes the entropy present in Alice's register $a$ given Bob's classical
register $b$. Here, $l$ is determined as a number of secret key
bits over the transmission of $N$ number of raw key. In Eq. (\ref{eq:Key-rate equantion}),
$R$ elucidates the infimum value of key rate under all collective
attacks performed by Eve. We apply \emph{Theorem
}\cite{K16,IK21} introduced by Krawec and analyze
the parameter estimation to derive $S(a|E)$ and $H(a|b)$, consequently
enhancing negotiation efficiency, thereby contributing to the overall
improvement of the security and correctness of the final key. This
formulation also facilitates the determination of QBER for the HD-Ext-B92
protocol. Using these findings, we can establish the minimum value
for the key rate by employing Equations (\ref{eq:Value of S(A|E)})
and (\ref{eq:Value of H(a|b)}) from Appendix A in Equation (\ref{eq:Key-rate equantion}).

\subsubsection{HD-BB84}

In a two-level system, BB84 \cite{BB84} protocol is well studied
both in theoretical and experimental domains. Essentially two-dimensional
quantum states (qubits) are used to realize this scheme for QKD which
uses two mutually unbiased bases randomly. In a more general scenario,
higher dimensional quantum systems (qudits) can
be used to realize the same task (i.e., QKD), and such a modified
version of BB84 protocol is referred to as \emph{qudit}- (i.e., a
quantum state in ${\rm d}$-dimensional Hilbert space) based BB84
protocol or HD-BB84 protocol. Here, we briefly discuss the HD-BB84
protocol \cite{CBK+02} and the necessary formulae to compute the
secret key rate.

In this protocol, Bob generates a sequence of qudits,
where each qudit represents a higher-dimensional quantum state. These
states are prepared based on a randomly selected basis, chosen from
two mutually complementary bases: $Z=\left\{ |0\rangle,|1\rangle,\cdots,|D-1\rangle\right\} $
and $X=\left\{ |x_{0}\rangle,|x_{1}\rangle,\cdots,|x_{D-1}\rangle\right\} $.
The qudit sequence is generated using a WCP source, with the possibility
of a PNS attack. To counter the threat of a PNS attack, Bob strategically
introduces decoy states at random positions within the qudit sequence,
expanding it. Upon receiving the enlarged sequence, Bob communicates
the positions and basis information of the decoy states to Alice.
Subsequently, Alice performs measurements on the decoy states and
publicly announces the measurement outcomes. Both parties then compute
the losses and errors incurred during the communication channel. If
the error rate falls within the acceptable threshold, the protocol
advances to the next stage. After discarding the decoy particles from
the enlarged sequence, Alice proceeds with a measurement operation
on the qudits. The measurement is performed by randomly selecting
one of the two d-dimensional bases, namely $Z$ and $X$. This process
ensures the security and reliability of the quantum communication
protocol, especially in the presence of potential PNS attacks. Subsequently,
they announce their bases choice in a public authenticated classical
channel \cite{CBK+02}) and obtain correlated ${\rm d}$-ary random
variables when they use the same bases. With $\frac{1}{2}$ probability,
Alice and Bob use different bases and yield uncorrelated results which
are considered as discarded data after key-sifting sub-protocol. It
is crucial to emphasize that the loss should be consistent for both
decoy states and signal states, thereby ensuring the security of the
channel. This method ensures that any effort made by an eavesdropper,
Eve (who is unaware of the chosen basis), to obtain information about
Bob's state will result in an error in transmission, which can subsequently
be detected by the legitimate parties.

To ensure a smooth comprehension of readers we would like to provide
a concise overview of key points discussed in Ref. \cite{BCC+10}.
in Ref. \cite{BCC+10}, authors have modified the Maassen and Uffink
bound \cite{K87,MU88} to establish a new bound on the uncertainties
associated with the measurement results, contingent on the amount
of entanglement between the measured particle $(A)$, and the quantum
memory ($B)$. This relationship can be expressed mathematically as,

\begin{equation}
S\left(Z|B\right)+S\left(X|B\right)\ge\log_{2}\frac{1}{C}+S\left(A|B\right),\label{eq:First_Bound}
\end{equation}
where, $Z$ and $X$ are two possible observable like measurement
bases and $A$ refers to the qudit measured by Alice which is sent
by Bob and $B$ refers to the qudit which represents the quantum memory
of Bob. $S$ represents von Neumann entropy and $S\left(A|B\right)$
quantifies the amount of entanglement between $A$ and $B$. $C:={\rm max_{i,j}|\langle\phi_{i}|\psi_{j}\rangle|^{2}}$,
where $|\phi_{{\rm i}}\rangle$ and $|\psi_{{\rm j}}\rangle$ are
the eigenvectors of $Z$ and $X$, respectively. Using a result established
by Devetak and Winter \cite{DW05}, the minimum limit on the quantity
of key that Alice and Bob can extract from each state can be expressed
as $S\left(Z|E\right)-S\left(Z|B\right)$\footnote{Here, Z and X can be employed in a similar manner or with a similar
effect.}. This limit is applied when the eavesdropper is trying to obtain
the key from the composite quantum system\footnote{Eve performs an entanglement operation using her ancillary state $E$
with Alice's state ($A$) and Bob's quantum memory ($B$).} $\rho_{ABE}$, where $A$ is Alice's particle, $B$ is Bob's quantum
memory, and $E$ is Eve's ancillary state. Equation (\ref{eq:First_Bound})
may be reformulated as $S\left(Z|E\right)+S\left(X|B\right)\ge{\rm \log_{2}}\frac{1}{C}$
(see Supplementary Information of \cite{BCC+10}), and
the key rate equation may be written as,

\begin{equation}
\begin{array}{lcl}
r\left(A,B,E\right) & \ge & S\left(Z|E\right)-S\left(Z|B\right)\\
 & \ge & \left[{\rm \log_{2}}\frac{1}{C}-S\left(X|B\right)\right]-S\left(Z|B\right)
\end{array}.\label{eq:KeyRate_HD_BB84_Before_Fano}
\end{equation}
Both parties involved in the HD-BB84 protocol utilize complete bases
elements ($Z$ and $X$ bases) within a d-dimensional Hilbert space.
In this context, the lower limit for parameter estimation can be directly
achieved from Fano's inequality. It is possible to consider an arbitrary
quantum channel with a noise probability denoted as $q$. Fano's inequality
states that $S\left(Z|B\right)\le h\left(q\right)+q\log_{2}\left({\rm d}-1\right)$.
By applying this relation in Eq. (\ref{eq:KeyRate_HD_BB84_Before_Fano})
and considering the condition that ${\rm \log_{2}}\frac{1}{C}$ cannot
exceed ${\rm \log_{2}}{\rm d}$, we can derive

\begin{equation}
\begin{array}{lcl}
r & \ge & \log_{2}{\rm d}-2\left(h\left(q\right)+q\log_{2}\left({\rm d}-1\right)\right)\end{array}.\label{eq:Key_rate_HD_BB84}
\end{equation}
For the binary encoding and decoding scheme the conditional entropy
of Alice's measurement outcome given Bob's measurement result is equal
to $h\left(\varepsilon\right)$, where $\varepsilon$ is quantum bit
error rate (QBER) \cite{CRE_04,BMA+09}. Here, $q$ is the depolarizing
channel parameter i.e., the probability that outcome of the $Z$ by
Alice and Bob is not equal and $h$ is binary entropy. Through
an examination of the aforementioned formulas, we ascertain the lower
bound of the key rate for the HD-BB84 protocol. This analysis involves
the application of Fano's inequality, which establishes bound on parameter
estimation. Additionally, the determination of the QBER is crucial
for enhancing negotiation efficiency, thereby playing a pivotal role
in augmenting the overall security and correctness of the final key
in the HD-BB84 protocol.

Before delving
into the formal analysis of the aforementioned formalism, it is crucial
to elucidate the relationship among key rate, QBER, noise and negotiation
efficiency. The noise introduced in the quantum channel ($q$), results
in QBER ($\varepsilon$) in the raw key sequence after executing the
quantum protocol. The conditional entropy of Alice's measurement outcome
given Bob's measurement outcome, as well as the conditional von Neumann
entropy of Alice's quantum state given Bob's quantum state, can be
expressed as functions of $\varepsilon$. Additionally, it is evident
that the conditional entropy is directly influenced by the noise in
the quantum channel. We may now explicitly analyze the previously
mentioned formulae, considering negotiation efficiency and QBER within
the key rate equation. First, we need to elaborate on all the elements
in Eq. (\ref{eq:Key-rate equantion}). From Appendix A, $S(a|E)$
represents the conditional von Neumann entropy, defined as follows:

\[
S(a|E)\ge\underset{c\ne m,c\ne n}{\sum}\left(\frac{K_{c}^{0}+K_{c}^{1}}{M}\right)S_{c}+\left(\frac{K_{m}^{0}+K_{m}^{1}}{M}\right)S_{m}+\left(\frac{K_{n}^{0}+K_{n}^{1}}{M}\right)S_{n},
\]
where,

\[
\begin{array}{lclcclcl}
K_{c}^{0} & := & \langle E_{c}^{m}|E_{c}^{m}\rangle, &  &  & K_{c}^{1} & := & \frac{1}{2}\,\langle F_{c}|F_{c}\rangle,\forall\,c\ne m,n\\
\\
K_{m}^{0} & := & \frac{1}{4}\,\langle E_{m}^{m}|E_{m}^{m}\rangle, &  &  & K_{m}^{0} & := & \frac{1}{8}\,\langle F_{m}|F_{m}\rangle,\\
\\
K_{n}^{0} & := & \frac{3}{4}\,\langle E_{n}^{m}|E_{n}^{m}\rangle, &  &  & K_{n}^{1} & := & \frac{3}{8}\,\langle F_{n}|F_{n}\rangle.
\end{array}
\]
and
\[
\begin{array}{lcl}
S_{c} & = & h\left(\frac{K_{c}^{0}}{K_{c}^{0}\,+\,K_{c}^{1}}\right)-h\left(\frac{1}{2}+\frac{\sqrt{\left(K_{c}^{0}\,-\,K_{c}^{1}\right)^{2}+4\,{\rm Re^{2}}\langle E_{c}^{m}|\frac{1}{\sqrt{2}}F_{c}\rangle}}{2\,\left(K_{c}^{0}\,+\,K_{c}^{1}\right)}\right),\\
S_{m} & = & h\left(\frac{K_{m}^{0}}{K_{m}^{0}\,+\,K_{m}^{1}}\right)-h\left(\frac{1}{2}+\frac{\sqrt{\left(K_{m}^{0}\,-\,K_{m}^{1}\right)^{2}+4\,{\rm Re^{2}}\langle\frac{1}{2}E_{m}^{m}|\frac{1}{2\sqrt{2}}F_{m}\rangle}}{2\,\left(K_{m}^{0}\,+\,K_{m}^{1}\right)}\right),\\
S_{n} & = & h\left(\frac{K_{n}^{0}}{K_{n}^{0}\,+\,K_{n}^{1}}\right)-h\left(\frac{1}{2}+\frac{\sqrt{\left(K_{n}^{0}\,-\,K_{n}^{1}\right)^{2}+4\,{\rm Re^{2}}\frac{3}{4\sqrt{2}}\langle E_{n}^{m}|F_{n}\rangle}}{2\,\left(K_{n}^{0}\,+\,K_{n}^{1}\right)}\right).
\end{array}
\]
Further, $H\left(a|b\right)$ represents the conditional entropy of
Alice's measurement outcome given Bob's measurement outcome, defined
as follows:

\[
\begin{array}{lcl}
H\left(a|b\right) & = & H\left(p_{00},\,p_{01},\,p_{10},\,p_{11}\right)-h\left(p_{00}+p_{10}\right).\end{array}
\]
We have previously defined the values $p_{00},p_{01},p_{10}$ and
$p_{11}$ (also detailed in Appendix A) through the analysis of parameter
estimation for the HD-Ext-B92 protocol. These terms depend on $q$.
Moreover, for binary encoding and decoding, $H(a|b)\equiv h(\varepsilon)$.
Therefore, it can be concluded that the conditional entropy and key
rate depend on the value of $\varepsilon$ and, consequently, on $q$
as $\varepsilon$ depends on $q$. Now, if we incorporate negotiation
efficiency\footnote{Negotiation efficiency is defined as the effectiveness
with which the steps involved in establishing a secure key are executed.
These steps generally encompass sifting, error correction and privacy
amplification. High negotiation efficiency indicates that these processes
are conducted in a way that optimizes the conversion of raw key bits
into secure key bits \cite{ZHG18,ZWF+23,BZJ+21}.} ($\xi$) into the key rate for the HD-Ext-B92 protocol
(as shown in Eq. (\ref{eq:Key-rate equantion})), the secure key rate
equation can be written as,

\[
R_{\xi}\equiv\xi\,R\left(a,b,E\right)=\xi\left({\rm inf}\left[S\left(a|E\right)-H\left(a|b\right)\right]\right).
\]
Here, $H(a|b)$ depends on noise and the QBER value. To reduce QBER,
legitimate users perform error correction and privacy amplification,
which increase the mutual information between Alice and Bob. This
leads to a a decrease in $H(a|b)$ and maximizes the proportion of
raw key bits successfully converted into secure key bits. Therefore,
analyzing QBER of a quantum protocol enhances the \emph{correctness}
and \emph{security} of the raw
key and optimizes the conversion of raw key into secure key more efficiently,
thereby improving the negotiation efficiency. The QBER analysis during
the test rounds is a crucial aspect of quantum communication protocols,
as it helps to improve negotiation efficiency. This logic applies
similarly to the HD-BB84 protocol. From Eqs. (\ref{eq:KeyRate_HD_BB84_Before_Fano})
and (\ref{eq:Key_rate_HD_BB84}), it is clear that the conditional
von Neumann entropy $S\left(Z|B\right)$ depends on $q$ (as Fano's
inequality states that $S\left(Z|B\right)\le h\left(q\right)+q\log_{2}\left({\rm d}-1\right)$),
which affects the QBER value as well. Considering negotiation efficiency
in the key rate for the HD-BB84 protocol (Eq. (\ref{eq:Key_rate_HD_BB84})),
the secure key rate equation can be expressed as follows:

\[
r_{\xi}\equiv\xi\,r\ge\xi\left(\log_{2}{\rm d}-2\left(h\left(q\right)+q\log_{2}\left({\rm d}-1\right)\right)\right).
\]
Here, the legitimate parties perform parameter estimation and error
correction on the raw key sequence, which leads to a reduction in
QBER and a decrease in $S\left(Z|B\right)$. As a result, the key
rate increases. Consequently, this process increases the proportion
of raw key bits that are successfully converted into secure key bits.
More specifically, analyzing the QBER during the test rounds of a
quantum communication protocol improves the correctness and security
of the raw key, making the conversion process to secure key more efficient.
This optimization of the secure key conversion improves the negotiation
efficiency, $\xi$. From this analysis, we can conclude that QBER
analysis is crucial for improving the negotiation efficiency of a
quantum protocol.

The preceding discussion, along with an analysis
of the formulations, elucidates that Eq. (\ref{eq:Key-rate equantion}),
which incorporates Eqs. (\ref{eq:Value of S(A|E)}) and (\ref{eq:Value of H(a|b)})
from Appendix A, and Eq. (\ref{eq:Key_rate_HD_BB84}), representing
the key rate formula for HD-Ext-B92 and HD-BB84 protocols, inherently
encompasses the essential steps of parameter \emph{negotiation
efficiency}. By employing these key rate equations,
coupled with the transmittance in satellite-based communication (cf.
Eqs. (\ref{eq:Transmittance}) and (\ref{eq:PDT Equation})), we can
derive the average key rate for LEO satellite-based quantum communication.
The computation of the average key rate using Eq. (\ref{eq:Average key-rate})
facilitates the determination of the probability distribution of key
rates and the variation of key rates concerning various parameters
in satellite quantum communication, as detailed in Section \ref{sec:III}.
Before delving further this section, we conduct an in-depth analysis
of these key rate equations for both HD-QKD protocols in the subsequent
paragraph. Before proceeding,
it is crucial to emphasize the definition of noise tolerance and its
relationship with QBER. Noise tolerance in a quantum communication
protocol refers to the ability of the protocol to function correctly
and securely despite the presence of noise. Specifically, the value
of noise tolerance for a protocol is determined by the point at which
the secure key rate approaches zero. Noise can originate from various
sources, including environmental disturbances, imperfections in quantum
devices, and potential eavesdropping activities. Noise tolerance and
QBER are interdependent factors in quantum communication protocols.
Effective noise management ensures that QBER remains below the critical
threshold, thereby enabling secure and efficient quantum communication.
Error correction techniques are employed to correct errors in the
raw key, and their efficiency depends on the QBER. Higher QBER necessitates
more robust error correction, which can diminish the efficiency of
the key generation process. Privacy amplification is used to reduce
the information an eavesdropper (Eve) might have obtained; the amount
of privacy amplification required increases with higher QBER\footnote{A detailed analysis of privacy amplification and
error correction is beyond the scope of the current work.}, further reducing the final key length. Maintaining
noise within the noise tolerance limit helps to keep QBER below a
secure threshold value, which is specific to the protocol. In such
cases, extensive privacy amplification and error correction may not
be strictly necessary.

\begin{figure}[h]
\begin{centering}
\includegraphics[scale=0.5]{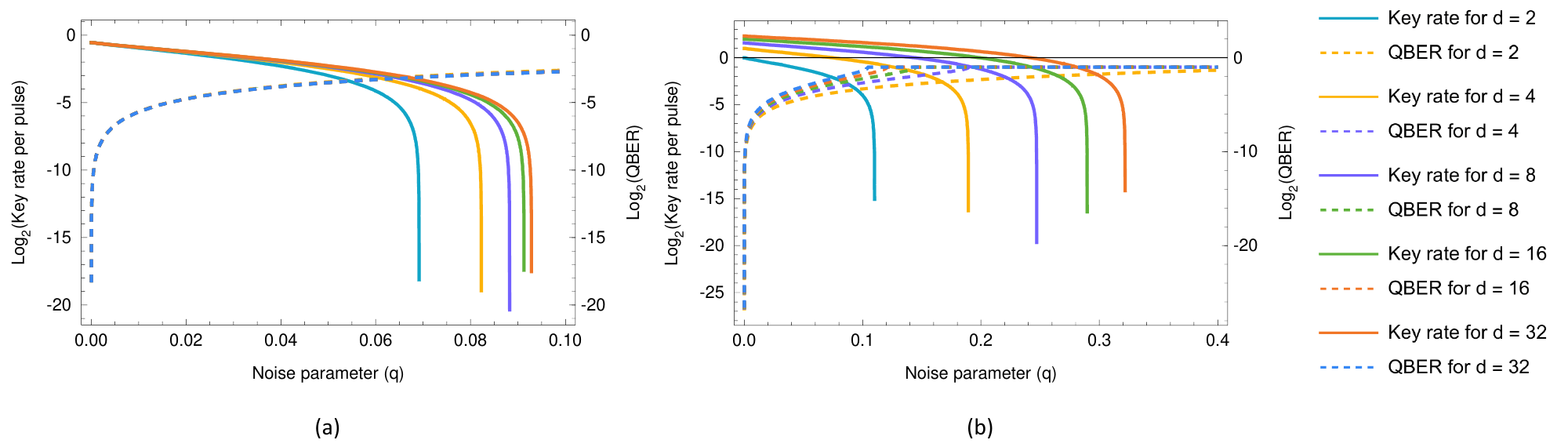}
\par\end{centering}
\caption{\label{fig:Keyrate_Noise_QBER}(Color online) Plot of variation of
key-rate and QBER with channel noise parameter (both plots share a
common legend): (a) key-rate and QBER analysis of HD-Ext-B92 protocol
with noise parameter (q) for different dimensions in Hilbert space
(the plot lines representing QBER for all dimensions are superimposed),
(b) key-rate and QBER analysis of HD-BB84 protocol with noise parameter
(q) for different dimensions in Hilbert space.}
\end{figure}

To analyze the behavior of the key rate per pulse and the QBER concerning
the noise parameter in both the above-discussed HD protocols, we utilize
the key rate equations (refer to Eqs. (\ref{eq:Key-rate equantion}),
(\ref{eq:Key_rate_HD_BB84}), and Appendix A) and the binary QBER
function $\left(h\left(\varepsilon\right)\right)$. Now, we analyze
the result illustrated in Figure \ref{fig:Keyrate_Noise_QBER} for
HD-Ext-B92 and HD-BB84 schemes. We can observe in the HD-Ext-B92 protocol
that events with mismatched bases are not disregarded, which occurs
when Alice and Bob employ different measurement bases. These events
can significantly enhance key generation rates \cite{IK21,BHP93,WMU08,MW08,TCK+14},
and therefore noise tolerance is also increased for this scheme which
is evident from the graph. We plot the variation of the key rate of
HD-Ext-B92 protocol with noise parameter (${\rm q}$) in a depolarizing
channel in the ${\rm d}$-dimensional Hilbert space; we also depict
the variation of QBER with the same noise parameter $({\rm q})$.
It may be observed that as the value of ${\rm d}$ increases, the
tolerance for noise also increases, showing a rise from $7\%$ to
$10\%$. It is apt to note that, the maximum tolerable noise is dependent
on the choice of both the depolarizing channel and of Eve's ancilla
state, since, these two factors significantly impact parameter estimation
and consequently affect the key rate. Nevertheless, our analysis is
confined to a specific choice of these two factors, which have been
outlined in Appendix A. In Figure \ref{fig:Keyrate_Noise_QBER} (a),
it becomes evident that the QBER remains constant across different
${\rm d}$ values. The plots representing the distinct ${\rm d}$
values (i.e., ${\rm d}=2,4,8,16,32$) overlap in QBER analysis, indicating
consistent outcomes for higher-dimensional cases of the HD-Ext-B92
protocol. Additionally, the graph (represented by a dotted line) demonstrates
that the QBER reaches a saturation point for a particular ${\rm q}$
value (for HD-Ext-B92). We have computed the initial point where the
QBER begins to rise for various ${\rm q}$ values and observed a physically
reasonable variation. For instance, when ${\rm q}$ is $\sim0.005$,
the QBER is approximately $0.015$. As the ${\rm q}$ value increases
to $\sim0.1$, the QBER saturates at approximately at $0.088$. Now,
we analyze the plots for HD-BB84 in Figure \ref{fig:Keyrate_Noise_QBER}
(b) and undertake a comprehensive comparison with HD-Ext-B92. A numerical
assessment reveals that the key rate increases as the value of ${\rm d}$
rises for HD-BB84. Conversely, in HD-Ext-B92, the minimum key rate
remains fairly consistent for all ${\rm d}$ values which is around
$0.7$. Further, the noise tolerance is increased significantly with
a greater value of ${\rm d}$ in HD-BB84. For instance, the tolerable
noise is $\sim11\%$ for ${\rm d}=2$ (for qubit) and with the increased
value of ${\rm d}=32$, this limit increases to $\sim32\%$. More simply,
the HD-Ext-B92 protocol operates securely and correctly with a maximum
noise tolerance ranging from $7\%$ to $10\%$, whereas the HD-BB84
protocol can tolerate up to $11\%$ noise for ${\rm d}=2$ and up
to 32\% noise for ${\rm d}=32$, as can be inferred from Eq. (\ref{eq:Key_rate_HD_BB84}). This
outcome demonstrates the advantage of opting for the HD-BB84 protocol
over HD-Ext-B92 when considering aspects like key rate and noise tolerance.
HD-BB84 surpasses the HD-Ext-B92 protocol. It is worth mentioning
that in the original scheme of HD-Ext-B92 \cite{IK21}, authors do
not employ two complete bases as HD-BB84 does. In their approach,
they utilize a simplified version in which Alice's requirement is
reduced to transmitting just three states, and Bob only needs to carry
out partial measurements within the second basis \cite{IK21}. Additionally,
it is important to highlight that they did not select an optimal basis
configuration. Alternate choices for the encoding state might yield
greater key rates for the HD-Ext-B92 protocol, as shown in cases involving
qubits \cite{LGT09,K16}. If we examine the QBER aspect within the
context of HD-BB84, it becomes apparent that the variation of QBER
with the noise parameter ($q$) rapidly converges to a saturation
value ($\sim0.25$) as the dimension of qudit increases. In contrast
to the HD-Ext-B92 protocol, the susceptibility of QBER to noise is
notably more vulnerable in the HD-BB84 protocol. Moreover, as depicted
in Figure \ref{fig:Keyrate_Noise_QBER}, when considering ${\rm d}=32$,
the saturation point of noise tolerance is attained in the HD-Ext-B92
protocol. In contrast, in the HD-BB84 protocol, the rate at which
noise tolerance increases becomes progressively lower as ${\rm d}$
increases. It is noteworthy that at ${\rm d}=32$, the QBER has not
yet reached its saturation point (for HD-BB84); this point will be
reached at higher values of ${\rm d}$. From Figure \ref{fig:Keyrate_Noise_QBER} (b), we observe that when
the key rate reaches zero, it indicates the noise tolerance values,
${\rm q}$, with QBER lines reaching their maximum value for the respective
dimensions. The conditional entropy of Alice's measurement outcome
given Bob's measurement outcome (and the conditional von Neumann entropy
of Alice's quantum state given Bob's quantum state) is directly influenced
by the noise in the quantum channel. These conditional entropies can
be expressed as function of QBER. Mathematically, the relationship
between noise tolerance and QBER in a quantum communication protocol
is illustrated by the effect of QBER on the secure key rate. A protocol
has a capability to handle noise up to a specific QBER threshold,
allowing error correction and privacy amplification to still generate
a secure key. The binary entropy function $h({\rm QBER})$ and the
specific security function $f({\rm QBER})$ measure the information
loss due to errors and potential eavesdropping, respectively.

\subsection{Satellite-based optical links: model used for the elliptic beam approximation\label{subsec:Elliptic_Beam_Model}}

In this article, we aim to analyze the performance of key rates in
various situations of HD-Ext-B92 and HD-BB84 protocols. The channel
transmission $\eta$ for the light propagation through atmospheric
links using elliptic-beam approximation as introduced by Vasylyev
et al. \cite{VSV16,VSV+17} will be employed to perform the analysis.
Further, in what follows, we impose the generalized approach\footnote{Using non-uniform link between a satellite and the ground station,
referred to in Eq. (\ref{eq:Down-Link and Up-Link condition}).} and different weather conditions as introduced in \cite{LKB19}.
This method yields an impact on the value of transmittance as the
transmittance is determined by beam parameters along with the diameter
of the receiving aperture. To provide readers with a clearer understanding
of both the elliptic beam approximation and its modified version in
a more comprehensive manner, in this section, we offer a succinct
explanation of the underlying theory.

\begin{figure}[h]
\centering{}\includegraphics[scale=0.6]{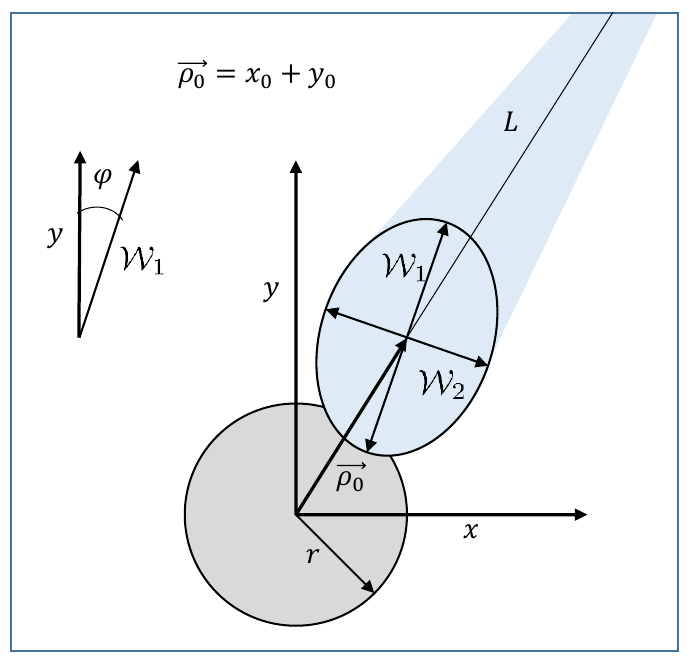}\caption{\label{fig:Elliptic_beam_impinge_circular_aperture}(Color online)
Diagram illustrating the received beam and the receiving aperture.
${\rm L}$ is the total link-length in the propagation direction,
$r$ represents the radius of the receiving aperture, $\rho_{0}=(x_{0},y_{0})$
signifies the position of the beam centroid, $\mathcal{W}_{1}$ and
$\mathcal{W}_{2}$ are the principal semi axes of elliptic beam profile,
and $\varphi$ is the orientation angle of the elliptic beam.}
\end{figure}

Temporal and spatial fluctuations in temperature and pressure within
turbulent atmospheric flows result in random variations of the air's
refractive index. Consequently, the atmosphere introduces losses to
transmitted photons, which are detected at the receiver through a
detection module featuring a limited aperture. The transmitted signal
undergoes degradation due to phenomena like beam wandering, broadening,
deformation, and similar effects. We can examine this scenario by
focusing on a Gaussian beam propagating along the z axis, reaching
the aperture plane positioned at a distance $z={\rm L}$. In
this analysis, we observe that assuming perfect Gaussian beams emitted
by the transmitter is not entirely realistic. Standard telescopes
typically produce beams with intensity distributions that closely
resemble a circular Gaussian profile with some deviations, often caused
by truncation effects at the edges of optical elements. One notable
consequence of these imperfections is the inherent broadening of the
beam due to diffraction. In our model, we can address this phenomenon
by adjusting the parameter representing the initial beam width $\left(\mathcal{W}_{0}\right)$,
thereby accounting for the increased divergence in the far-field resulting
from the imperfect quasi-Gaussian beam. To capture this effect, we
incorporate the transmission of the elliptical beam through a circular
aperture and consider the statistical characteristics of the elliptical
beam as it propagates through turbulence using a Gaussian approximation.
However, it's important to mention certain restriction for simplifications
in our approach, particularly the assumption of isotropic atmospheric
turbulence. For a more detailed formulation, readers are referred
to the Supplemental Material of Ref. \cite{VSV16}. That quasi-Gaussian
beam is directed through a link that spans both the atmosphere and
vacuum, originating from either a transmitter situated in orbit or
a ground station. The link is characterized by non-uniform conditions.
Generally, the varying intensity transmittance of such a signal (received
beam) via a circular aperture of radius $r$ of the receiving telescope
is expressed as follows \cite{VSV12,VSV16}:

\begin{equation}
\begin{array}{lcl}
\eta & = & \int_{\left|\rho\right|^{2}=r^{2}}{\rm d^{2}\boldsymbol{\rho}\left|u\left(\mathbf{\boldsymbol{\rho}},L\right)\right|^{2},}\end{array}\label{eq:Transmittance}
\end{equation}
where $u\left(\mathbf{\boldsymbol{\rho}},{\rm L}\right)$ represents
the beam envelope at the receiver plane, located at a distance ${\rm L}$
from the transmitter, and $\left|u\left(\mathbf{\boldsymbol{\rho}},{\rm L}\right)\right|^{2}$
is the normalized intensity with respect to full $\boldsymbol{\rho}$
plane, where $\boldsymbol{\rho}$ denotes the position vector within
the transverse plane. The vector parameter ${\rm \boldsymbol{v}}$
fully characterizes the state of the beam at the receiver plane (see
Figure \ref{fig:Elliptic_beam_impinge_circular_aperture}),

\begin{equation}
{\rm \boldsymbol{v}}=\left(x_{0},y_{0},\mathcal{W}_{1},\mathcal{W}_{2},\varphi\right),\label{eq:Vector of beam-parameters}
\end{equation}
$x_{0}$, $y_{0}$, $\mathcal{W}_{1/2}$, and $\varphi$ imply the
beam centroid coordinates, the principal semi-axes of the elliptic
beam profile, and the orientation angle of the elliptic beam, respectively.
The transmittance is determined by these beam parameters along with
the radius of the receiving aperture ($r$).

\begin{figure}[h]
\centering{}\includegraphics[scale=0.5]{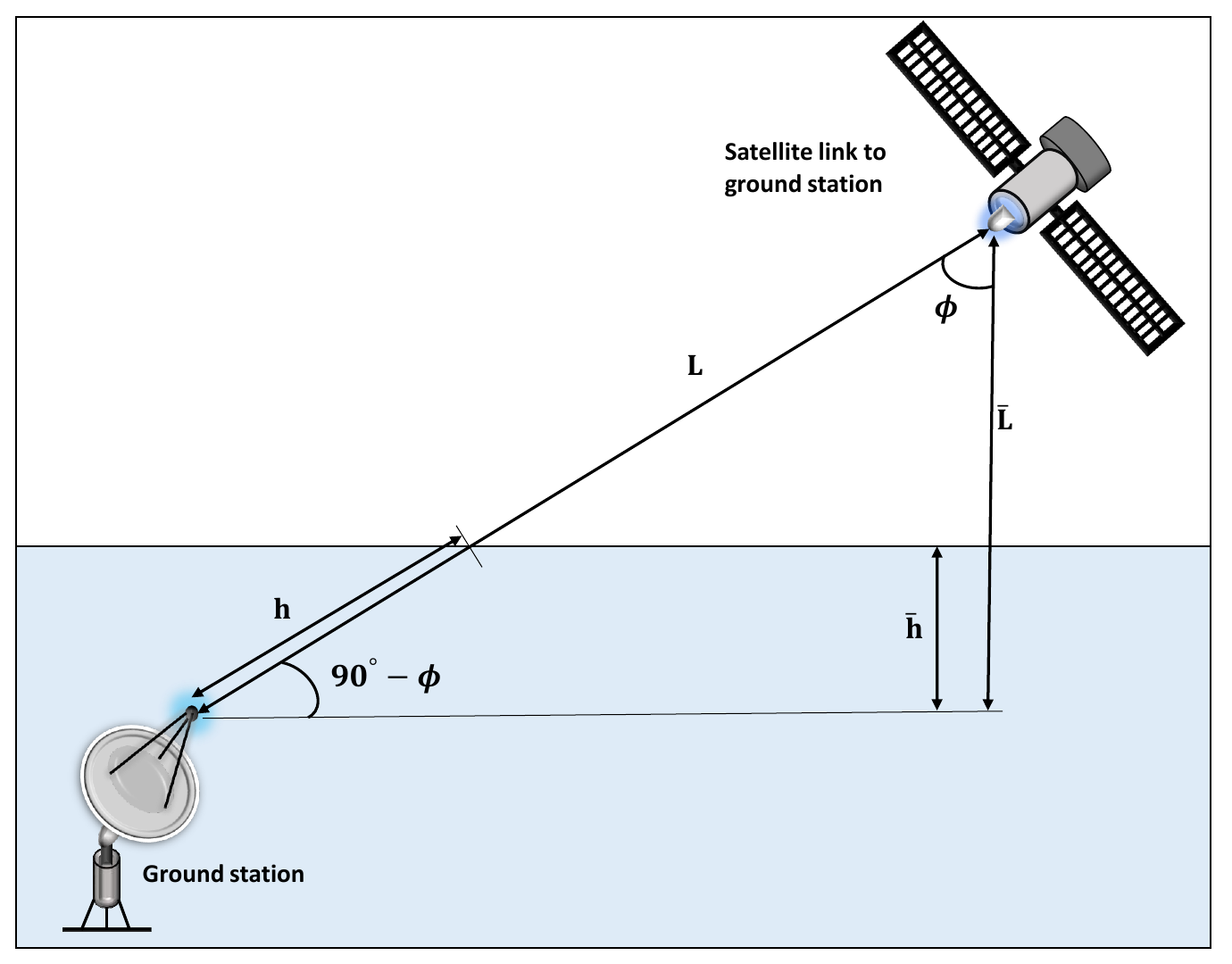}\caption{\label{fig:Non-uniform_free_space_link}(Color online) This figure
depicted that the non-uniform free-space link between the satellite
and the ground station. The diagram highlights key parameters, $\overline{{\rm h}}$
is the thickness of the atmosphere, $\overline{{\rm L}}$ is the altitude
of the satellite, ${\rm h}$ represents length of the propagation
of light inside atmosphere, ${\rm L}$ is the link length between
satellite and ground station, $\phi$ denotes the zenith angle. Up-link
(down-link) configuration represents the transmission of light from
ground station to satellite (satellite to ground station). }
\end{figure}

In general, the atmosphere can be categorized into distinct layers,
each characterized by various physical parameters such as air density,
pressure, temperature, the presence of ionized particles, and more.
The arrangement of these layers varies according to location, particularly
concerning the extent of each layer's thickness. Without loss of generality,
we adopt a simplified model of a satellite-based optical link \cite{LKB19}.
This model entails a uniform atmosphere up to a specific altitude
denoted as $\overline{{\rm h}}$, beyond which a vacuum extends all
the way to the satellite situated at an altitude marked as $\overline{{\rm L}}$,
as illustrated in Figure \ref{fig:Non-uniform_free_space_link}. Rather
than dealing with a continuous range of values characterizing physical
quantities as a function of altitude, this approach involves just
two key parameters. These parameters encompass the value of the physical
quantity within the uniform atmosphere and the effective altitude
range, $\overline{{\rm h}}$. This simplification is likely to be
quite accurate because atmospheric influences are predominantly significant
only within the initial $10$ to $20$ kilometers above the Earth's
surface. This is particularly relevant considering that the standard
orbital height for LEO satellites is above 400 kilometers . In our
analysis, we set the value of $\overline{{\rm L}}$ to $500$ km,
and assume that the zenith angle falls within the range of $\left[0^\circ,80^\circ\right]$.
Under these conditions, the range of the satellite's orbit suitable
for key distribution is approximately ${\rm L}\in\left[500,2000\right]$
km\footnote{The correlation between total link length and zenith angle is, ${\rm L}=\overline{{\rm L}}\sec\phi$.}.
The given context mandates that the effective atmospheric thickness
$\overline{{\rm h}}$ remains constant at 20 km, by the aforementioned
factors. We extend the discussion by maintaining the premise that
the parameters quantifying the influence of atmospheric effects remain
constant (with values greater than $0$) within the atmosphere and
are set to $0$ outside it. In this context, we can make use of the
assumption that,

\begin{equation}
\begin{array}{cl}
{\rm Down-link} & \begin{cases}
C_{n}^{2}\left(z\right) & =C_{n}^{2}\,\text{\textohm}\left(z-\left({\rm L}-{\rm h}\right)\right),\\
n_{0}\left(z\right) & =n_{0}\,\text{\textohm}\left(z-\left({\rm L}-{\rm h}\right)\right),
\end{cases}\\
\\
{\rm Up-link} & \begin{cases}
C_{n}^{2}\left(z\right) & =C_{n}^{2}\,\text{\textohm\ensuremath{\left({\rm h}-z\right)},}\\
n_{0}\left(z\right) & =n_{0}\,\text{\textohm}\left({\rm h}-z\right).
\end{cases}
\end{array}\label{eq:Down-Link and Up-Link condition}
\end{equation}
Here, $C_{n}^{2}$ represents the refractive index structure constant\footnote{Several altitude-dependent models describing the refractive index
structure constant $C_{n}^{2}$ have been documented \cite{V80,HS64,LC06,FSV+10}.
Among these, the parametric fit proposed by Hufnagel and Valley is
widely adopted and faithfully captures the characteristics of $C_{n}^{2}$
in climates characteristic of mid-latitudes \cite{HS64,V80}.}, and $n_{0}$ denotes the density of scattering particles \cite{TP88,T84}.
The function $\text{\textohm}\left(z\right)$ corresponds to the Heaviside
step-function\footnote{The value of this function is zero for negative arguments and one
for positive arguments. This function falls within the broader category
of step functions.}. As stated above, the parameter $z$ signifies the longitudinal coordinate,
while ${\rm L}$ stands for the overall length of the link. Additionally,
${\rm h}$ represents the distance covered within the atmosphere,
as illustrated in the accompanying Figure \ref{fig:Non-uniform_free_space_link}.

Now, let's consider the transmittance, as defined in Eq. (\ref{eq:Transmittance}),
for an elliptic beam that strikes a circular aperture with a radius
of $r$. This transmittance can be expressed as follows \cite{VSV16}:

\begin{equation}
\begin{array}{lcl}
\eta\left(x_{0},y_{0},\mathcal{W}_{1},\mathcal{W}_{2},\varphi\right) & = & \frac{2\,\chi_{{\rm ex}t}}{\pi\mathcal{W}_{1}\mathcal{W}_{2}}\int_{0}^{r}\rho\,{\rm d}\rho\int_{0}^{2\pi}{\rm d}\theta{\rm e^{-2A_{1}\left(\rho cos\theta-\rho_{0}\right)^{2}}}{\rm e^{-2A_{2}\rho^{2}sin^{2}\theta}}e^{-2{\rm A}_{3}\left(\rho{\rm cos}\theta-\rho_{0}\right)\rho{\rm sin}\theta}.\end{array}\label{eq:PDT Equation}
\end{equation}
In this context, $r$ represents the radius of the aperture, while
$\rho$ and $\theta$ denote the polar coordinates of the vector $\boldsymbol{\rho}$,

\[
\begin{array}{lcl}
x & = & \rho{\rm \,cos}\theta,\\
y & = & \rho{\rm \,sin}\theta,
\end{array}
\]
here, $\rho_{0}$ and $\theta_{0}$ represent polar coordinates corresponding
to the vector $\boldsymbol{\rho}_{0}$,

\[
\begin{array}{lcl}
x_{0} & = & \rho_{0}\,{\rm cos}\theta_{0},\\
y_{0} & = & \rho_{0}{\rm \,sin}\theta_{0},
\end{array}
\]
and

\[
\begin{array}{lcl}
{\rm A}_{1} & = & \left(\frac{{\rm cos}^{2}\left(\varphi-\theta_{0}\right)}{\mathcal{W}_{1}^{2}}+\frac{{\rm sin}^{2}\left(\varphi-\theta_{0}\right)}{\mathcal{W}_{2}^{2}}\right),\\
{\rm A}_{2} & = & \left(\frac{{\rm sin}^{2}\left(\varphi-\theta_{0}\right)}{\mathcal{W}_{1}^{2}}+\frac{{\rm cos}^{2}\left(\varphi-\theta_{0}\right)}{\mathcal{W}_{2}^{2}}\right),\\
{\rm A}_{3} & = & \left(\frac{1}{\mathcal{W}_{1}^{2}}-\frac{1}{\mathcal{W}_{2}^{2}}\right){\rm sin\,2\left(\varphi-\theta_{0}\right).}
\end{array}
\]
These expressions can be employed for numerical integration, as described
in Eq. (\ref{eq:PDT Equation}), through the Monte Carlo method or
another effective technique for the same purpose. To simplify the
process of integration using the Monte Carlo method, it requires the
generation of $N$ sets of values for the vector ${\rm \boldsymbol{v}}$
(see Eq. (\ref{eq:Vector of beam-parameters})). It is assumed that
the angle $\left(\varphi-\theta_{0}\right)$ follows a uniform distribution
over the interval $[0,\frac{\pi}{2}]$ and other parameters\footnote{To compute transmittance, first one has to evaluate $\mathcal{W}_{{\rm i}}$
from $\Theta_{{\rm i}}$ using relation $\begin{array}{lcl}
\Theta_{{\rm i}} & = & \ln\left(\frac{\mathcal{W}_{{\rm i}}^{2}}{\mathcal{W}_{{\rm 0}}^{2}}\right),\end{array}$ ${\rm i}=1,2.$ Here, $\mathcal{W}_{0}$ is the beam spot radius
at the transmitter.} ($x_{0},y_{0},\Theta_{{\rm 1}},\Theta_{{\rm 2}}$) follow the normal
distribution \cite{WHW+18}. Substitution of the simulated values
of ${\rm \boldsymbol{v}}$ into Eq. (\ref{eq:PDT Equation}) makes
it feasible to perform the numerical integration. The outcome of this
process also involves the \emph{extinction factor}\footnote{The parameter $\chi_{{\rm ext}}(\phi)$ denotes the extinction losses
caused by atmospheric back-scattering and absorption. It varies depending
on the elevation angle $\left(90^{\circ}-\phi\right)$ or zenith
angle $(\phi)$ \cite{BSH+13,VBB2000}.}\emph{,} $\chi_{{\rm ext}}$, thereby producing $N$ atmospheric transmittance
values, denoted as $\eta\left({\rm \boldsymbol{v}_{i}}\right)$, where
$i$ ranges from $1$ to $N$. The necessary parameters for simulation
are described in Appendix B which are calculated according to our
model. These expressions are different for up-link and down-link configuration
as different expressions mentioned in Eq. (\ref{eq:Down-Link and Up-Link condition})
are used for up-link and down-link configuration.

In the next section, we will evaluate the effectiveness of the HD
protocols selected by us in the satellite-based links. To conduct
this assessment, we need average key rates over the probability distribution
of the transmittance (PDT)\footnote{Some authors followed the relation $\eta_{\delta}=10^{-\frac{\delta}{10}}$
with $\delta=\alpha_{1}{\rm L}$ $[{\rm dB}]$ to represent the channel
transmittance with the form of attenuation, here, ${\rm L}$ total
link length and $\alpha_{1}$ is loss in the channel transmission
${\rm dB/km}$.} computed for different link lengths and configurations. The same
can be expressed as \cite{LKB19},

\begin{equation}
\begin{array}{lclcl}
\bar{R} & = & \intop_{0}^{1}R(\eta)\,P(\eta)\,{\rm d}\eta & = & \stackrel[{\rm i}=1]{N_{bins}}{\sum}R(\eta_{{\rm i}})\,P(\eta_{{\rm i}}),\end{array}\label{eq:Average key-rate}
\end{equation}
where, $\bar{R}$ represents the average key rate, while $R(\eta)$
signifies the key rate corresponding to a specific transmittance value.
The PDT is denoted as $P(\eta)$. To compute the integral average,
the interval $[0,1]$ is divided into $N_{bins}$ bins, each centered
at $\eta_{{\rm i}}$ for $i$ ranging from $1$ to $N_{bins}$, and
is evaluated by combining the weighted sum of the rates. The estimation
of $P(\eta_{{\rm i}})$ relies on random sampling, as explained in
the earlier paragraph.\textcolor{red}{{} }The formulations for the distinct
implementations key rates $R(\eta)$ can be found in Section \ref{subsec:HD-Ext-B92=000026HD-BB84}.

\section{Performance analysis of protocols after simulation \label{sec:III}}

In this section, we elaborately analyze the impact of PDT\footnote{See PDT in Figures 3 and 4 in Ref. \cite{LKB19} after random sampling
of beam parameters ${\rm \boldsymbol{v}}$ for a down-link and an
up-link, respectively.} on key rate after the weighted sum, as well as the probability distribution
of key rate (PDR) concerning the HD-Ext-B92 and HD-BB84 protocols.
The minimum separation between Alice and Bob (i.e., altitude of the
satellite) remains constant at a distance of $\overline{{\rm L}}=500$
km, as the primary focus is on scenarios involving LEO satellites
like the Chinese satellite Micius \cite{LCL+17,YCL+17,RXY+17,YCLL+17}.
We present outcomes of numerical simulation for satellite-based HD-Ext-B92
and HD-BB84 schemes under asymptotic conditions. The simulation incorporates
the experimental parameters outlined in Table \ref{tab:Parameters-associated-with-link-length}
\cite{MFR12,XXL14,LKB19}. The parameters $C_{n}^{2}$,
$n_{0}$, and $h$ are typically determined by fitting experimental
data. However, for the sake of establishing a predictive model, we
parameterize these values in a rational manner. We conduct simulations
under varying atmospheric conditions, encompassing clear, slightly
foggy, and moderately foggy nights, as well as non-windy, moderately
windy, and windy days \cite{LKB19}. A particularly noteworthy aspect
is the comparison between nighttime and daytime operations. In daytime
conditions, elevated temperatures result in stronger winds and heightened
mixing across atmospheric layers, leading to more pronounced turbulence
effects and consequently higher values of $C_{n}^{2}$ compared to
nighttime conditions. Nevertheless, on average, during clear days,
the lower atmosphere exhibits reduced moisture content compared to
nighttime, resulting in diminished beam spreading due to scattering
particles. Conversely, nighttime conditions, characterized by lower
temperatures, yield a less turbulent atmosphere. Additionally, the
formation of haze and mist contributes to higher values of $n_{0}$
compared to daytime conditions. In such scenarios, the impact of scattering
over particulate matter can surpass the effects induced by turbulence. The crucial factors in this scenario include
not only those associated with atmospheric influences but also the
radii of the transmitting and receiving telescopes, along with the
wavelength of the signal. For the satellite in orbit, we opted for
a radius of $r_{{\rm sat}}=15$ cm ($\mathcal{W}_{0}$), while the
ground station telescope has a radius of $r_{{\rm grnd}}=0.5$ m,
and the signal wavelength is $\lambda=785$ nm. Based on Eq. (\ref{eq:Down-Link and Up-Link condition}),
it is evident that a down-link pertains to satellite-to-ground communication,
where atmospheric effects become significant only in the latter part
of the propagation process, i.e., when $z$ exceeds $({\rm L-h})$.
On the other hand, for up-links, these effects are relevant only when
$z$ is below ${\rm h}$.

\begin{table}[h]
\begin{centering}
\begin{tabular}{>{\centering}p{2.5cm}>{\centering}p{4cm}>{\centering}p{5cm}}
\toprule 
Parameter & Value & Short description\tabularnewline
\midrule
$\mathcal{W}_{0}$ & 15 cm, 50 cm & Down-link, up-link\tabularnewline
$r$ & 50 cm, 15 cm & Down-link, up-link\tabularnewline
$\lambda$ & 785 nm & Wavelength of the signal light\tabularnewline
$\beta$ & 0.7 & Parameter in $\chi_{{\rm ext}}(\phi)$\tabularnewline
$\alpha$ & $2\times10^{-6}$ rad & Pointing error\tabularnewline
$\overline{{\rm h}}$ & 20 km & Atmosphere thickness\tabularnewline
$\overline{{\rm L}}$ & 500 km & Minimum altitude (at zenith)\tabularnewline
$n_{0}$ & 0.61 ${\rm m^{-3}}$ & Night-time condition 1\tabularnewline
$n_{0}$ & 0.01 ${\rm m^{-3}}$ & Day-time condition 1\tabularnewline
$n_{0}$ & 3.00 ${\rm m^{-3}}$ & Night-time condition 2\tabularnewline
$n_{0}$ & 0.05 ${\rm m^{-3}}$ & Day-time condition 2\tabularnewline
$n_{0}$ & 6.10 ${\rm m^{-3}}$ & Night-time condition 3\tabularnewline
$n_{0}$ & 0.10 ${\rm m^{-3}}$ & Day-time condition 3\tabularnewline
$C_{n}^{2}$ & $1.12\times10^{-16}$ ${\rm m^{-\frac{2}{3}}}$ & Night-time condition 1\tabularnewline
$C_{n}^{2}$ & $1.64\times10^{-16}$ ${\rm m^{-\frac{2}{3}}}$ & Day-time condition 1\tabularnewline
$C_{n}^{2}$ & $5.50\times10^{-16}$ ${\rm m^{-\frac{2}{3}}}$ & Night-time condition 2\tabularnewline
$C_{n}^{2}$ & $8.00\times10^{-16}$ ${\rm m^{-\frac{2}{3}}}$ & Day-time condition 2\tabularnewline
$C_{n}^{2}$ & $1.10\times10^{-15}$ ${\rm m^{-\frac{2}{3}}}$ & Night-time condition 3\tabularnewline
$C_{n}^{2}$ & $1.60\times10^{-15}$ ${\rm m^{-\frac{2}{3}}}$ & Day-time condition 3\tabularnewline
\bottomrule
\end{tabular}
\par\end{centering}
\caption{\label{tab:Parameters-associated-with-link-length}Parameters associated
with the optical and technical characteristics of the link and different
atmospheric weather conditions.}

\end{table}

From Appendix B, it becomes evident, as expected that the impact of
atmospheric effects is considerably more pronounced in the case of
up-links compared to down-links. The underlying phenomena at play
here, namely beam deflection and broadening, encompass angular effects.
These effects play a role in determining the ultimate size of the
beam, thus influencing the channel losses. Their magnitude is directly
proportional to the distance covered after the initiation of the effect
known as \emph{kick in effect}. For up-links, these effects manifest
near the transmitter, resulting in beam broadening spanning hundreds
of kilometers before detection at the satellite. Conversely, in the
down-link scenario, the majority of the beam's trajectory occurs within
a vacuum, with atmospheric effects coming into play only during the
final fifteen to twenty kilometers before reaching the receiver. A
secondary distinction lies in the origin of fluctuations in the position
of the beam centroid, denoted as $(x_{0},y_{0})$. In up-links, the
atmosphere-induced deflections tend to be significantly more influential
than pointing errors ($\varphi$), which is disregarded. On the other
hand, in down-links, the beam dimensions are already substantially
larger than any turbulent irregularities at the top of the atmosphere.
As a consequence, the resulting beam wandering due to atmospheric
effects can be neglected, rendering pointing errors the dominant contributing
factor.

\begin{figure}[h]
\begin{centering}
\includegraphics[scale=0.5]{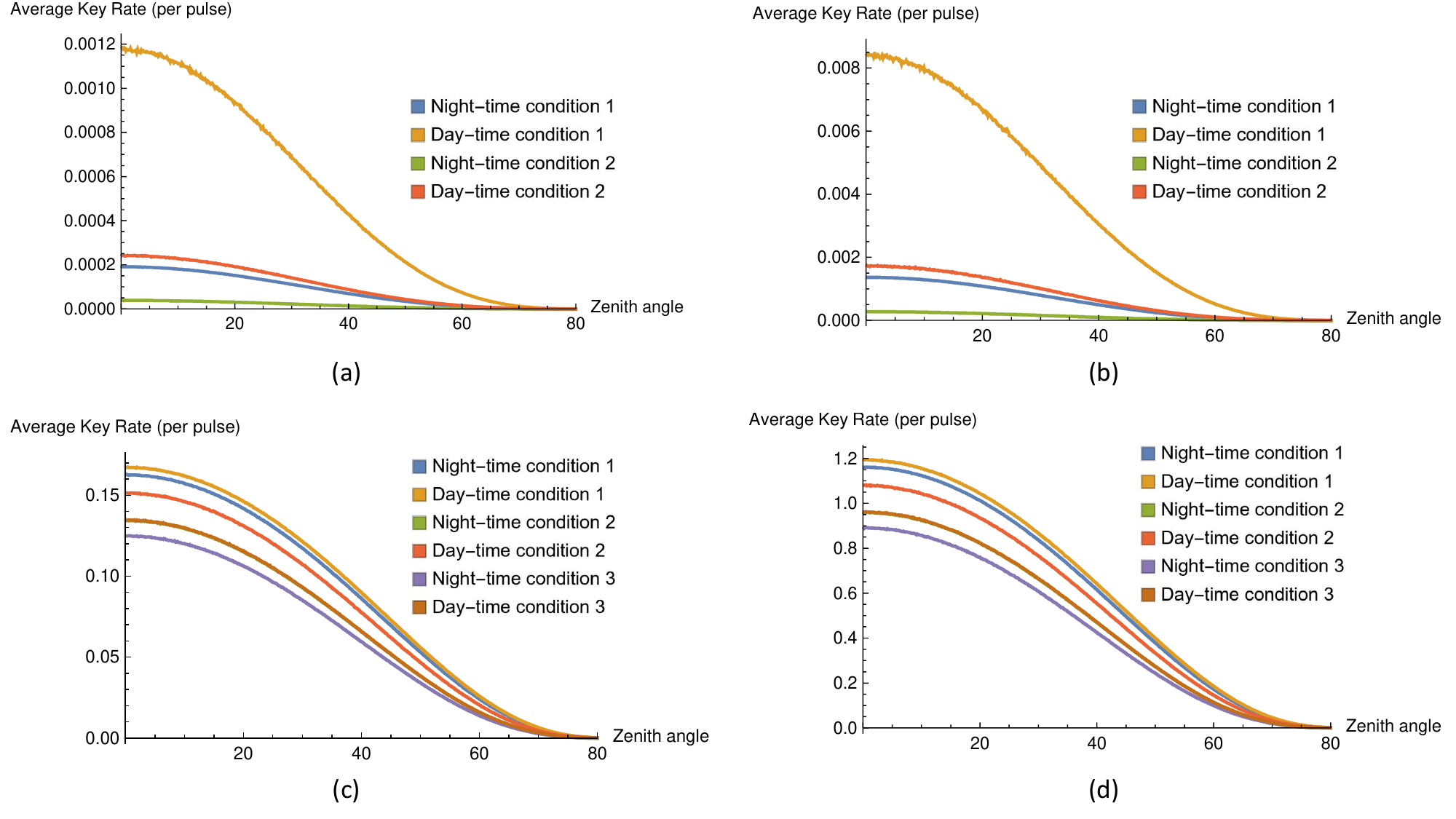}
\par\end{centering}
\caption{\label{fig:AKR_Zenith_Different_Weather}(Color online) Plot of variation
of average key rate (per pulse) with zenith angle in different weather
conditions considering minimal noise (${\rm d}=32$), i.e., day-time
conditions 1, 2 and 3 correspond to not windy, moderate windy and
windy, respectively (described as Day-time condition 1/2/3) and night-time
conditions 1, 2, 3 correspond to clear, slightly foggy and moderate
foggy, respectively (described as Night-time condition 1/2/3). The
upper row corresponds to the up-link scenario and the lower row corresponds
to the down-link scenario: (a) Average key rate generated by HD-Ext-B92
protocol as a function of zenith angle for up-link configuration under
four different weather conditions (Day 1-2 and Night 1-2), (b) Average
key rate generated by HD-BB84 protocol as a function of zenith angle
for up-link configuration under four different weather condition (Day
1-2 and Night 1-2), (c) Average key rate generated by HD-Ext-B92 protocol
as a function of zenith angle for down-link configuration under six
different weather condition (Day 1-2-3 and Night 1-2-3), (d) Average
key rate generated by HD-BB84 protocol as a function of zenith angle
for down-link configuration under six different weather condition
(Day 1-2-3 and Night 1-2-3).}
\end{figure}

Utilizing Equation (\ref{eq:Key-rate equantion}),
which integrates Equations (\ref{eq:Value of S(A|E)}) and (\ref{eq:Value of H(a|b)})
from Appendix A, and Eq. (\ref{eq:Key_rate_HD_BB84}), which represent
as the key rate formula for HD-Ext-B92 and HD-BB84 protocols, along
with PDT in satellite-based communication (refer to Eqs. (\ref{eq:Transmittance})
and (\ref{eq:PDT Equation})), enables the computation of the average
key rate for LEO satellite quantum communication. Employing Eq. (\ref{eq:Average key-rate})
in this calculation facilitates the determination of the probability
distribution of key rates and the assessment of key rate variations
with respect to different parameters in the context of satellite quantum
communication. Now, we aim to investigate the average key rate as
a function of zenith angle, considering minimal noise. Figures \ref{fig:AKR_Zenith_Different_Weather}
illustrate the average key rate using the PDT concerning the angle
relative to the zenith. This analysis is carried out for both up-links
and down-links across various weather conditions for dimension\footnote{The weather data information is used from Ref. \cite{LKB19}. We also
mention the required information in Table \ref{tab:Parameters-associated-with-link-length}.}, ${\rm d}=32$\textcolor{red}{{} }(see Table \ref{tab:Parameters-associated-with-link-length}).
Each data point on the graph is derived from $10,000$ parameter samples
in Eq. (\ref{eq:Vector of beam-parameters}) and computed using Eq.
(\ref{eq:PDT Equation}). In Figures \ref{fig:AKR_Zenith_Different_Weather}
(a) and \ref{fig:AKR_Zenith_Different_Weather} (b), the graphs reveal
that during daytime condition 1, the highest average key rate is yielded
in the zenith position ($\sim0.0012$ and $\sim0.008$) for HD-Ext-B92
and HD-BB84 protocols, respectively, in the up-link configuration.
Notably, the key rate\footnote{For ease of reference, we will refer to the average key rate as the
\textquotedbl key rate\textquotedbl .} is slightly greater for HD-BB84 which corresponds to the expected
result. For the same configuration, the key rate sharply diminishes
under other conditions (Day 2 and Night 1-2). Comparatively, for HD-Ext-B92,
the maximum value of the key rate ($\sim0.0002$) is nearly ten times
lower than that of the HD-BB84 protocol ($\sim0.002$) corresponding
to the day condition 2. A similar comparison holds for night 1/2 conditions.
For these conditions, the key rate becomes approximately zero at zenith
angle $50^{\circ}$. It may be noted that in night-time condition
1, the key rate is lower than in day-time condition 2 for both schemes
within the same configuration. Based on these observations, we can
infer that daytime transmission in the up-link configuration performs
more favorably than nighttime transmission. Due to the very low key
rate during night-time condition 2, we have chosen to negate condition
3, both in night-time and day-time, from the graphical representation. Additionally, in the up-link configuration, the simulation
results reveal a tenfold disparity in key rates between HD-Ext-B92
and HD-BB84 during day-time condition 2. In contrast, during day-time
condition 1, the difference is less pronounced, approximately five
fold. This discrepancy is attributed to the non-windy nature of day-time
condition 1, while day-time condition 2 experiences moderate wind,
resulting in a lower value of $C_{n}^{2}$ for the former condition
compared to the latter. Moreover, the absence of windy conditions
indicates a lower moisture content in the lower atmosphere. Consequently,
the scattering particle density, denoted as $n_{0}$, is lower in
day-time condition 1 compared to day-time condition 2 (see Table \ref{tab:Parameters-associated-with-link-length}).
The down-link configuration is depicted in Figures \ref{fig:AKR_Zenith_Different_Weather}
(c) and \ref{fig:AKR_Zenith_Different_Weather} (d). As previously
discussed, the influence of atmospheric effects is comparatively reduced
in the down-link configuration compared to the up-link configuration.
Consequently, the performance of the link transmittance is superior
for down-link as compared to up-link. This is supported by Figures
\ref{fig:AKR_Zenith_Different_Weather} (c) and \ref{fig:AKR_Zenith_Different_Weather}
(d), which further highlight the enhanced key rate. From these two
figures, the overall plot patterns can be seen to be (sequential arrangement
of plots representing different weather conditions) consistent for
both protocols. The sequence of different weather conditions that
yield higher key rate values follows this order: day-time condition
1, night-time condition 1, day-time condition 2, day-time condition
3, night-time condition 2, and night-time condition 3. Additionally,
it can be seen that similar to the up-link scenario, the daytime conditions
favor channel transmission over the nighttime conditions. This pattern
remains consistent across both scenarios. Of particular interest is
the comparison between operations during night-time and day-time.
During daylight hours, higher temperatures facilitate stronger winds
and heightened mixing across distinct atmospheric layers. This generates
more prominent turbulence effects. However, on average, clear days
witness a reduced moisture content in the lower atmosphere compared
to night-time conditions. Consequently, the scattering of particles
causes less pronounced beam spreading. Conversely, during night-time,
the cooler temperatures result in an atmosphere with lower turbulence
levels, coupled with the formation of mist and haze. In such circumstances,
scattering tends to have a more substantial impact at night-time than
the effects induced by turbulence at day-time. In the down-link scenario,
during day-time condition 1, the highest achievable key rates are
$0.165$ and $1.2$ for HD-Ext-B92 and HD-BB84 protocols, respectively.
Conversely, in night-time condition 3, the highest attainable key
rates are $0.125$ and $0.9$. The key rate ratio, in the down-link
scenario, between the HD-BB84 and HD-Ext-B92 protocols is $7.27$
for the maximum scenario and $7.2$ for the minimum scenario. This
observation substantiates the anticipated outcome that HD-BB84 consistently
outperforms HD-Ext-B92. Furthermore, the key rate decreases significantly
within the zenith angle range of $70^{\circ}$ to $80^{\circ}$
for the down-link scenario, whereas for the up-link scenario, this
reduction begins at a zenith angle of $50^{\circ}$. Intuitively,
down-link transmission exhibits a higher tolerance for larger zenith
angles compared to up-link transmission.

\begin{figure}[h]
\begin{centering}
\includegraphics[scale=0.5]{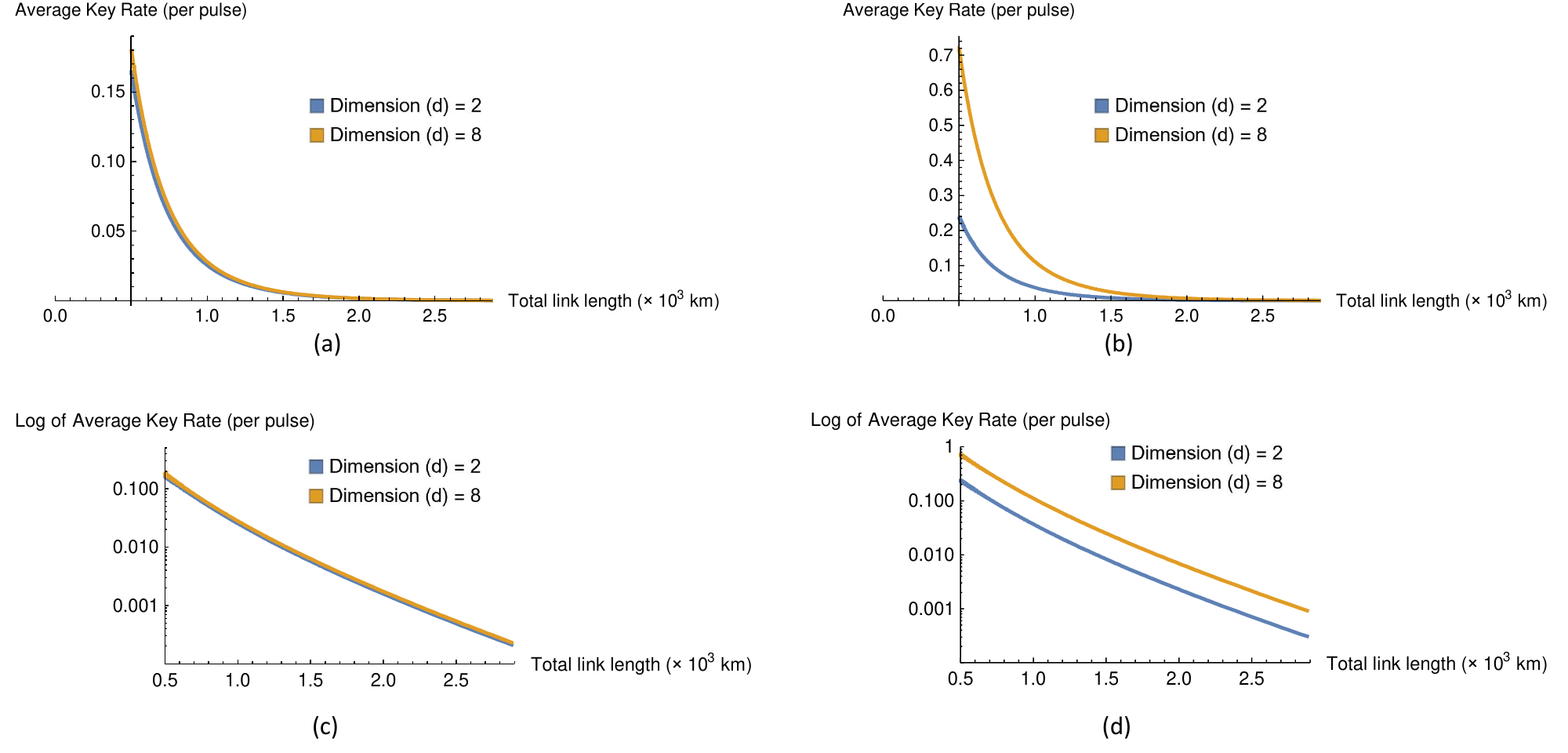}
\par\end{centering}
\caption{\label{fig:AKR_LinkLength_DownLink_Day1}(Color online) Plot of variation
of average key rate (per pulse) with total link length under condition
Day-1 utilizing different dimensions of qudit (${\rm d}=2$ and ${\rm d}=8$):
(a) Average key rate generated by HD-Ext-B92 protocol as a function
of total link length (${\rm L}$) for down-link configuration, (b)
Average key rate generated by HD-BB84 protocol as a function of total
link length (${\rm L}$) for down-link configuration,
(c) and (d) illustrate the same results as depicted in (a) and (b),
respectively, but for a better visualization of the impact of link
length on the average key rate, here a logarithmic scale is used along
the $y$-axis.}
\end{figure}

To obtain the best possible results, hereafter we focus on the down-link
configuration under optimal weather conditions where the average key
rate is highest (cf. Figure \ref{fig:AKR_Zenith_Different_Weather}).
Specifically, we analyze and illustrate the variation of key rate
with total link length (${\rm L}$) in day-time condition 1 within
down-link configuration, assuming an extremely low noise. In this
scenario, the HD-Ext-B92 protocol yields maximum key rates of 0.17
and 0.155 for qudit dimensions 8 and 2, respectively, as illustrated
in Figure \ref{fig:AKR_LinkLength_DownLink_Day1} (a). Notably, the
key rate of the HD-BB84 protocol exhibits notable fluctuations across
different dimensions. As can be seen from Figure \ref{fig:AKR_LinkLength_DownLink_Day1}
(b), for qudit dimensions 8 and 2, the maximum key rates are 0.7 and
0.24, respectively. Furthermore, the key rate decreases
almost linearly for both the HD-QKD protocols and across both dimensions
when plotted on a logarithmic scale. Consequently, it can be inferred
that the decrease in key rate follows an exponential pattern. Specifically,
at a higher zenith angle of $80^\circ$, with a total link distance
of $2900$ km, the key rate of the HD-Ext-B92 protocol is approximately
$10^{-4}$ for both dimensions. In contrast, at the same link distance,
the key rates for HD-BB84 are $10^{-3}$ and $10^{-4}$ for dimensions
$8$ and $2$, respectively. The HD-BB84 protocol outperforms at higher
dimensions, consistent with the findings depicted in the accompanying
figure \ref{fig:AKR_Zenith_Different_Weather}.

\begin{figure}[h]
\begin{centering}
\includegraphics[scale=0.5]{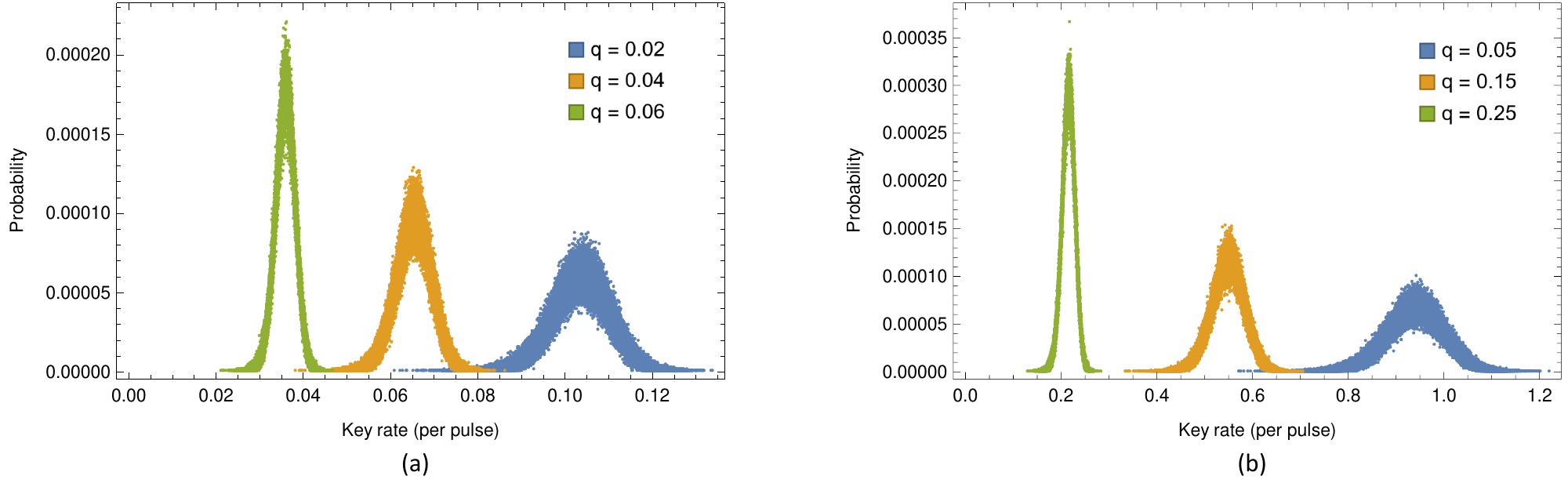}
\par\end{centering}
\caption{\label{fig:Probability_Keyrate_Different_Noise}(Color online) Plot
of the distribution of key-rate variation for different channel noise
parameters (q) at the zenith position under condition Day-1 utilizing
qudit of dimension 32: (a) Probability distribution of key-rate for
HD-Ext-B92 protocol, (b) Probability distribution of key-rate for
HD-BB84 protocol under down-link configuration.}
\end{figure}

\begin{figure}[h]
\begin{centering}
\includegraphics[scale=0.5]{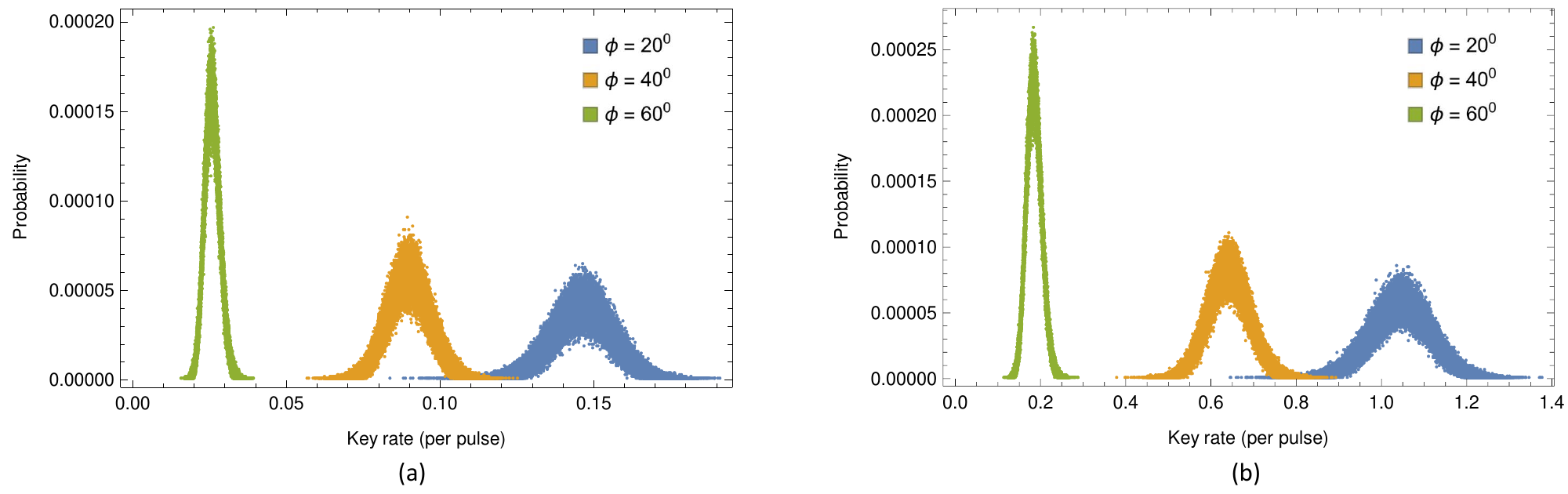}
\par\end{centering}
\caption{\label{fig:Probability_Keyrate_Different_Zenith_Angle}(Color online)
Plot of distribution of key-rate variation for different zenith angles
($\phi$) considering minimal noise, characterized by the parameter
${\rm q}\ll1$, under condition Day-1 utilizing qudit of dimension
32: (a) Probability distribution of key-rate for HD-Ext-B92 protocol,
(b) Probability distribution of key-rate for HD-BB84 protocol under
down-link configuration.}
\end{figure}

In Figure \ref{fig:Probability_Keyrate_Different_Noise}, we present
the PDR with different values of noise parameter (${\rm q}$) at the
zenith position ($\phi=0^{\circ}$) under down-link configuration.
In this context, we employ the optimal performance scenario during
day-time condition 1 utilizing qudit dimension of 32. We have used
a data set of $10^{6}$ beam parameters to simulate the values of
the average key rate and approximate the results to six (five) decimal
places\footnote{This is a good choice of approximation to represent, well-suited for
PDR representation.} to get PDR plots for HD-Ext-B92 (HD-BB84). Within the HD-Ext-B92
protocol, comparing the cases of ${\rm q}=0.02$ and ${\rm q}=0.06$
(in Figure \ref{fig:Probability_Keyrate_Different_Noise} (a)), we
observe a higher key rate for ${\rm q}=0.02$, while the maximum value
of probability of key rate is greater for ${\rm q}=0.06$. The maximum
values of probability are consistently greater with greater values
of noise parameter. Notably, a higher key rate corresponds to a lower
value of probability of occurrence. A specific shape of PDT (as is
the case here) implies that the shape of the PDR would remain the
same with different noise parameters and different zenith angles (or
equivalently with different distances). For example, see that the
shape of the PDR remains same for HD-Ext-B92 protocol and HD-BB84
protocol, although the density of data points are more in the case
of HD-BB84 (see Figure \ref{fig:Probability_Keyrate_Different_Noise}
(a) and (b)). However, this protocol (HD-BB84) exhibits significantly
elevated key rate values as well as higher probabilities compared
to HD-Ext-B92. Subsequently, we also plot the PDR with different zenith
angles in Figure \ref{fig:Probability_Keyrate_Different_Zenith_Angle},
considering extremely low noise characterized by the parameter ${\rm q}\ll1$
at the zenith position under condition Day-1 with the same configuration
(down-link). Notably, the shapes of the PDR curves remain consistent
across both the protocols; however, the data points on the plot appear
more densely concentrated in the HD-BB84 protocol. In this case, we
have utilized a dataset of $10^{6}$ beam parameters to simulate the
values of the average key rate and approximate the results to six
(five) decimal places to get PDR plots for HD-Ext-B92 (HD-BB84). The
peak values of the probability of key rates in the PDR graph for distinct
zenith angles are different for both the protocols. Moreover, for
different zenith angles, the peak values of probability in the PDRs
are consistently greater in HD-BB84 compared to HD-Ext-B92. In conclusion,
we deduce that the PDR curves maintain a uniform shape across varying
zenith angles as PDT considered here has a fixed shape.

\section{Conclusion\label{sec:IV}}

In this paper, we study two protocols for QKD in higher dimensions.
We analyze the key rates of these two higher dimensional protocols
in the context of satellite-based secure quantum communication. To
analyze the effectiveness of these schemes for satellite-based quantum
communication, we employ a robust method known as the elliptic beam
approximation \cite{VSV16}. By employing a generalized
model using this approach, we assess the performance of the HD-Ext-B92
and HD-BB84 protocols. The key rate per pulse and QBER are plotted
against the noise parameter. Notably, our findings reveal that, in
higher dimensions, HD-BB84 outperforms HD-Ext-B92 in terms of both
key rate and noise tolerance. However, HD-BB84 experiences a more
pronounced saturation of QBER in high dimensions. We deduce the key
rate of the HD-Ext-B92 scheme without introducing any additional free
parameters, as opposed to the approach discussed in Ref. \cite{IK21},
and is elaborated in Appendix A. Our analysis comprehensively demonstrates
the impact of link transmittance on the weighted sum of key rate under
nominal noise levels for both the schemes (HD-Ext-B92 and HD-BB84)
under up-link and down-link configurations. Moreover, we delve into
the analysis of PDR across different values of noise parameter (at
the zenith position) and zenith angle (with nominal noise). Remarkably, the PDR
exhibits consistent shapes across all scenarios. It is noteworthy
that the graphical points are denser for HD-BB84; as anticipated this
is because the HD-BB84 protocol makes use of two complete bases. Additionally,
the probability tends to be higher for lower key rate values compared
to higher ones. It may be noted that we employ normal and uniform
distributions to model beam parameters. Alternative distributions
may be employed to account for specific altitudes and atmospheric
conditions. Consequently, variations in key rate could differ in our
analysis, contingent on the consideration of atmospheric effects.
For greater accuracy and interest, utilizing empirical data to obtain
these results is recommended. 

Numerous theoretical studies have been focused on finding the analytical
probability distribution that best aligns with the experimentally
observed transmittance of optical links in free space. The prevalent
distributions employed are the log-normal \cite{LSA01,SNP+13}, Gamma-Gamma
\cite{HAP01}, and Double Weibull \cite{CSK+10} distributions. The
choice among these distributions depends on factors like turbulence
intensity, link distance, and the setup of the transmitting and receiving
telescopes. Conversely, the methodology employed in this study takes
a constructive approach, enabling the determination of the PDT based
on beam characteristics and atmospheric conditions. Further, our work
can be expanded by examining the performance of a cube-sat, such as
utilizing data from an existing satellite with an appropriate payload
(say, from the Chinese satellite \emph{Micius}), while optimizing
the source intensity. This optimization would lead to an enhancement
in the system's key rate and the ability to achieve longer link lengths
(even when tolerating higher zenith angles) \cite{DHC+22,LJ20}. Analysis
of finite key in any quantum communication scheme would be interesting.
Especially consideration of such effects is important in the context
of satellite-based quantum communication because the limited duration
of the connection between the ground station and the satellite would
always lead to a finite key. Thus, future work could involve directing
attention towards finite key analysis in scenarios involving higher
dimensions, as well as assessing key rate performance in relation
to atmospheric transmittance for satellite-based links. In summary,
our investigations into the performance of higher-dimensional QKD
protocols over satellite-based systems may have a substantial impact
on both theoretical and experimental aspects of satellite-based quantum
communication. Thus, the present work definitely establishes the advantages
of using higher dimensional states in satellite-based quantum communication;
but there are challenges associated with the experimental generation
and maintenance of the qudits. In the near future we would like to
address this technical issue and also to find the optimal choice of
dimension that can provide a desired key rate. 

\subsection*{Acknowledgment: }

Authors acknowledge support from the Indian Space Research Organisation
(ISRO) project no: ISRO/RES/3/906/22-23.

\section*{Availability of data and materials}

No additional data is needed for this work.

\section*{Competing interests}

The authors declare that they have no competing interests.

\section*{Funding}

Indian Space Research Organisation (ISRO) project no: ISRO/RES/3/906/22-23.

\section*{Authors' contributions}
SB conceptualized the problem. AD performed most of the calculations. AP and M checked the results. AP, SB, and AD analyzed the results. AP and SB supervised the work and all the authors contributed in writing the paper.

\bibliographystyle{apsrev}
\bibliography{Satellite_first}

\begin{thebibliography}{170}
\expandafter\ifx\csname natexlab\endcsname\relax\def\natexlab#1{#1}\fi
\expandafter\ifx\csname bibnamefont\endcsname\relax
  \def\bibnamefont#1{#1}\fi
\expandafter\ifx\csname bibfnamefont\endcsname\relax
  \def\bibfnamefont#1{#1}\fi
\expandafter\ifx\csname citenamefont\endcsname\relax
  \def\citenamefont#1{#1}\fi
\expandafter\ifx\csname url\endcsname\relax
  \def\url#1{\texttt{#1}}\fi
\expandafter\ifx\csname urlprefix\endcsname\relax\def\urlprefix{URL }\fi
\providecommand{\bibinfo}[2]{#2}
\providecommand{\eprint}[2][]{\url{#2}}

\bibitem[{\citenamefont{Rivest et~al.}(1983)\citenamefont{Rivest, Shamir, and
  Adleman}}]{RSA83}
\bibinfo{author}{\bibfnamefont{R.~L.} \bibnamefont{Rivest}},
  \bibinfo{author}{\bibfnamefont{A.}~\bibnamefont{Shamir}}, \bibnamefont{and}
  \bibinfo{author}{\bibfnamefont{L.~M.} \bibnamefont{Adleman}},
  \emph{\bibinfo{title}{Cryptographic communications system and method}}
  (\bibinfo{year}{1983}), \bibinfo{note}{{US} Patent 4,405,829}.

\bibitem[{\citenamefont{Xu et~al.}(2020)\citenamefont{Xu, Ma, Zhang, Lo, and
  Pan}}]{XMZ+20}
\bibinfo{author}{\bibfnamefont{F.}~\bibnamefont{Xu}},
  \bibinfo{author}{\bibfnamefont{X.}~\bibnamefont{Ma}},
  \bibinfo{author}{\bibfnamefont{Q.}~\bibnamefont{Zhang}},
  \bibinfo{author}{\bibfnamefont{H.-K.} \bibnamefont{Lo}}, \bibnamefont{and}
  \bibinfo{author}{\bibfnamefont{J.-W.} \bibnamefont{Pan}},
  \bibinfo{journal}{Reviews of Modern Physics} \textbf{\bibinfo{volume}{92}},
  \bibinfo{pages}{025002} (\bibinfo{year}{2020}).

\bibitem[{\citenamefont{Bennett and Brassard}(1984)}]{BB84}
\bibinfo{author}{\bibfnamefont{C.~H.} \bibnamefont{Bennett}} \bibnamefont{and}
  \bibinfo{author}{\bibfnamefont{G.}~\bibnamefont{Brassard}},
  \emph{\bibinfo{title}{Quantum cryptography: Public-key distribution and coin
  tossing, {in Proc. IEEE Int. Conf. on Computers, Systems, and Signal
  Processing (Bangalore, India, 1984), pp. 175-179.}}} (\bibinfo{year}{1984}).

\bibitem[{\citenamefont{Ekert}(1991)}]{E91}
\bibinfo{author}{\bibfnamefont{A.~K.} \bibnamefont{Ekert}},
  \bibinfo{journal}{Physical Review Letters} \textbf{\bibinfo{volume}{67}},
  \bibinfo{pages}{661} (\bibinfo{year}{1991}).

\bibitem[{\citenamefont{Bennett}(1992)}]{B92}
\bibinfo{author}{\bibfnamefont{C.~H.} \bibnamefont{Bennett}},
  \bibinfo{journal}{Physical Review Letters} \textbf{\bibinfo{volume}{68}},
  \bibinfo{pages}{3121} (\bibinfo{year}{1992}).

\bibitem[{\citenamefont{Bennett et~al.}(1992)\citenamefont{Bennett, Brassard,
  and Mermin}}]{BBM92}
\bibinfo{author}{\bibfnamefont{C.~H.} \bibnamefont{Bennett}},
  \bibinfo{author}{\bibfnamefont{G.}~\bibnamefont{Brassard}}, \bibnamefont{and}
  \bibinfo{author}{\bibfnamefont{N.~D.} \bibnamefont{Mermin}},
  \bibinfo{journal}{Physical Review Letters} \textbf{\bibinfo{volume}{68}},
  \bibinfo{pages}{557} (\bibinfo{year}{1992}).

\bibitem[{\citenamefont{Scarani et~al.}(2004)\citenamefont{Scarani, Acin,
  Ribordy, and Gisin}}]{SAR+04}
\bibinfo{author}{\bibfnamefont{V.}~\bibnamefont{Scarani}},
  \bibinfo{author}{\bibfnamefont{A.}~\bibnamefont{Acin}},
  \bibinfo{author}{\bibfnamefont{G.}~\bibnamefont{Ribordy}}, \bibnamefont{and}
  \bibinfo{author}{\bibfnamefont{N.}~\bibnamefont{Gisin}},
  \bibinfo{journal}{Physical Review Letters} \textbf{\bibinfo{volume}{92}},
  \bibinfo{pages}{057901} (\bibinfo{year}{2004}).

\bibitem[{\citenamefont{Chatterjee et~al.}(2020)\citenamefont{Chatterjee,
  Joarder, Chatterjee, Sanders, and Sinha}}]{CJC+20}
\bibinfo{author}{\bibfnamefont{R.}~\bibnamefont{Chatterjee}},
  \bibinfo{author}{\bibfnamefont{K.}~\bibnamefont{Joarder}},
  \bibinfo{author}{\bibfnamefont{S.}~\bibnamefont{Chatterjee}},
  \bibinfo{author}{\bibfnamefont{B.~C.} \bibnamefont{Sanders}},
  \bibnamefont{and} \bibinfo{author}{\bibfnamefont{U.}~\bibnamefont{Sinha}},
  \bibinfo{journal}{Physical Review Applied} \textbf{\bibinfo{volume}{14}},
  \bibinfo{pages}{024036} (\bibinfo{year}{2020}).

\bibitem[{\citenamefont{Pathak}(2013)}]{P13}
\bibinfo{author}{\bibfnamefont{A.}~\bibnamefont{Pathak}},
  \emph{\bibinfo{title}{Elements of quantum computation and quantum
  communication}} (\bibinfo{publisher}{CRC Press Boca Raton},
  \bibinfo{year}{2013}).

\bibitem[{\citenamefont{Dutta and Pathak}(2022{\natexlab{a}})}]{DP+22}
\bibinfo{author}{\bibfnamefont{A.}~\bibnamefont{Dutta}} \bibnamefont{and}
  \bibinfo{author}{\bibfnamefont{A.}~\bibnamefont{Pathak}},
  \bibinfo{journal}{arXiv preprint arXiv:2212.13089}
  (\bibinfo{year}{2022}{\natexlab{a}}).

\bibitem[{\citenamefont{Panayi et~al.}(2014)\citenamefont{Panayi, Razavi, Ma,
  and L{\"u}tkenhaus}}]{PRM+14}
\bibinfo{author}{\bibfnamefont{C.}~\bibnamefont{Panayi}},
  \bibinfo{author}{\bibfnamefont{M.}~\bibnamefont{Razavi}},
  \bibinfo{author}{\bibfnamefont{X.}~\bibnamefont{Ma}}, \bibnamefont{and}
  \bibinfo{author}{\bibfnamefont{N.}~\bibnamefont{L{\"u}tkenhaus}},
  \bibinfo{journal}{New Journal of Physics} \textbf{\bibinfo{volume}{16}},
  \bibinfo{pages}{043005} (\bibinfo{year}{2014}).

\bibitem[{\citenamefont{Valivarthi et~al.}(2017)\citenamefont{Valivarthi, Zhou,
  John, Marsili, Verma, Shaw, Nam, Oblak, and Tittel}}]{VZJ+17}
\bibinfo{author}{\bibfnamefont{R.}~\bibnamefont{Valivarthi}},
  \bibinfo{author}{\bibfnamefont{Q.}~\bibnamefont{Zhou}},
  \bibinfo{author}{\bibfnamefont{C.}~\bibnamefont{John}},
  \bibinfo{author}{\bibfnamefont{F.}~\bibnamefont{Marsili}},
  \bibinfo{author}{\bibfnamefont{V.~B.} \bibnamefont{Verma}},
  \bibinfo{author}{\bibfnamefont{M.~D.} \bibnamefont{Shaw}},
  \bibinfo{author}{\bibfnamefont{S.~W.} \bibnamefont{Nam}},
  \bibinfo{author}{\bibfnamefont{D.}~\bibnamefont{Oblak}}, \bibnamefont{and}
  \bibinfo{author}{\bibfnamefont{W.}~\bibnamefont{Tittel}},
  \bibinfo{journal}{Quantum Science and Technology}
  \textbf{\bibinfo{volume}{2}}, \bibinfo{pages}{04LT01} (\bibinfo{year}{2017}).

\bibitem[{\citenamefont{Zhang et~al.}(2018)\citenamefont{Zhang, Zhang, and
  Wang}}]{ZZW18}
\bibinfo{author}{\bibfnamefont{C.-H.} \bibnamefont{Zhang}},
  \bibinfo{author}{\bibfnamefont{C.-M.} \bibnamefont{Zhang}}, \bibnamefont{and}
  \bibinfo{author}{\bibfnamefont{Q.}~\bibnamefont{Wang}},
  \bibinfo{journal}{Communications in Theoretical Physics}
  \textbf{\bibinfo{volume}{70}}, \bibinfo{pages}{331} (\bibinfo{year}{2018}).

\bibitem[{\citenamefont{Horodecki and Stankiewicz}(2020)}]{HS20}
\bibinfo{author}{\bibfnamefont{K.}~\bibnamefont{Horodecki}} \bibnamefont{and}
  \bibinfo{author}{\bibfnamefont{M.}~\bibnamefont{Stankiewicz}},
  \bibinfo{journal}{New Journal of Physics} \textbf{\bibinfo{volume}{22}},
  \bibinfo{pages}{023007} (\bibinfo{year}{2020}).

\bibitem[{\citenamefont{Brougham and Oi}(2022)}]{BO22}
\bibinfo{author}{\bibfnamefont{T.}~\bibnamefont{Brougham}} \bibnamefont{and}
  \bibinfo{author}{\bibfnamefont{D.~K.} \bibnamefont{Oi}},
  \bibinfo{journal}{New Journal of Physics} \textbf{\bibinfo{volume}{24}},
  \bibinfo{pages}{075002} (\bibinfo{year}{2022}).

\bibitem[{\citenamefont{Tajima et~al.}(2017)\citenamefont{Tajima, Kondoh, Ochi,
  Fujiwara, Yoshino, Iizuka, Sakamoto, Tomita, Shimamura, Asami
  et~al.}}]{TKO+17}
\bibinfo{author}{\bibfnamefont{A.}~\bibnamefont{Tajima}},
  \bibinfo{author}{\bibfnamefont{T.}~\bibnamefont{Kondoh}},
  \bibinfo{author}{\bibfnamefont{T.}~\bibnamefont{Ochi}},
  \bibinfo{author}{\bibfnamefont{M.}~\bibnamefont{Fujiwara}},
  \bibinfo{author}{\bibfnamefont{K.}~\bibnamefont{Yoshino}},
  \bibinfo{author}{\bibfnamefont{H.}~\bibnamefont{Iizuka}},
  \bibinfo{author}{\bibfnamefont{T.}~\bibnamefont{Sakamoto}},
  \bibinfo{author}{\bibfnamefont{A.}~\bibnamefont{Tomita}},
  \bibinfo{author}{\bibfnamefont{E.}~\bibnamefont{Shimamura}},
  \bibinfo{author}{\bibfnamefont{S.}~\bibnamefont{Asami}},
  \bibnamefont{et~al.}, \bibinfo{journal}{Quantum Science and Technology}
  \textbf{\bibinfo{volume}{2}}, \bibinfo{pages}{034003} (\bibinfo{year}{2017}).

\bibitem[{\citenamefont{Dutta and Pathak}(2023{\natexlab{a}})}]{DP23}
\bibinfo{author}{\bibfnamefont{A.}~\bibnamefont{Dutta}} \bibnamefont{and}
  \bibinfo{author}{\bibfnamefont{A.}~\bibnamefont{Pathak}},
  \bibinfo{journal}{Quantum Information Processing}
  \textbf{\bibinfo{volume}{22}}, \bibinfo{pages}{13}
  (\bibinfo{year}{2023}{\natexlab{a}}).

\bibitem[{\citenamefont{Dutta and Pathak}(2023{\natexlab{b}})}]{DP+23}
\bibinfo{author}{\bibfnamefont{A.}~\bibnamefont{Dutta}} \bibnamefont{and}
  \bibinfo{author}{\bibfnamefont{A.}~\bibnamefont{Pathak}},
  \bibinfo{journal}{arXiv preprint arXiv:2308.05470}
  (\bibinfo{year}{2023}{\natexlab{b}}).

\bibitem[{\citenamefont{Vajner et~al.}(2022)\citenamefont{Vajner, Rickert, Gao,
  Kaymazlar, and Heindel}}]{VRG+22}
\bibinfo{author}{\bibfnamefont{D.~A.} \bibnamefont{Vajner}},
  \bibinfo{author}{\bibfnamefont{L.}~\bibnamefont{Rickert}},
  \bibinfo{author}{\bibfnamefont{T.}~\bibnamefont{Gao}},
  \bibinfo{author}{\bibfnamefont{K.}~\bibnamefont{Kaymazlar}},
  \bibnamefont{and} \bibinfo{author}{\bibfnamefont{T.}~\bibnamefont{Heindel}},
  \bibinfo{journal}{Advanced Quantum Technologies}
  \textbf{\bibinfo{volume}{5}}, \bibinfo{pages}{2100116}
  (\bibinfo{year}{2022}).

\bibitem[{\citenamefont{Para{\"\i}so et~al.}(2021)\citenamefont{Para{\"\i}so,
  Woodward, Marangon, Lovic, Yuan, and Shields}}]{PWM+21}
\bibinfo{author}{\bibfnamefont{T.~K.} \bibnamefont{Para{\"\i}so}},
  \bibinfo{author}{\bibfnamefont{R.~I.} \bibnamefont{Woodward}},
  \bibinfo{author}{\bibfnamefont{D.~G.} \bibnamefont{Marangon}},
  \bibinfo{author}{\bibfnamefont{V.}~\bibnamefont{Lovic}},
  \bibinfo{author}{\bibfnamefont{Z.}~\bibnamefont{Yuan}}, \bibnamefont{and}
  \bibinfo{author}{\bibfnamefont{A.~J.} \bibnamefont{Shields}},
  \bibinfo{journal}{Advanced Quantum Technologies}
  \textbf{\bibinfo{volume}{4}}, \bibinfo{pages}{2100062}
  (\bibinfo{year}{2021}).

\bibitem[{\citenamefont{Paudel et~al.}(2023)\citenamefont{Paudel, Crawford,
  Lee, Shugayev, Leuenberger, Syamlal, Ohodnicki, Lu, Mollot, and
  Duan}}]{PCL+23}
\bibinfo{author}{\bibfnamefont{H.~P.} \bibnamefont{Paudel}},
  \bibinfo{author}{\bibfnamefont{S.~E.} \bibnamefont{Crawford}},
  \bibinfo{author}{\bibfnamefont{Y.-L.} \bibnamefont{Lee}},
  \bibinfo{author}{\bibfnamefont{R.~A.} \bibnamefont{Shugayev}},
  \bibinfo{author}{\bibfnamefont{M.~N.} \bibnamefont{Leuenberger}},
  \bibinfo{author}{\bibfnamefont{M.}~\bibnamefont{Syamlal}},
  \bibinfo{author}{\bibfnamefont{P.~R.} \bibnamefont{Ohodnicki}},
  \bibinfo{author}{\bibfnamefont{P.}~\bibnamefont{Lu}},
  \bibinfo{author}{\bibfnamefont{D.}~\bibnamefont{Mollot}}, \bibnamefont{and}
  \bibinfo{author}{\bibfnamefont{Y.}~\bibnamefont{Duan}},
  \bibinfo{journal}{Advanced Quantum Technologies}
  \textbf{\bibinfo{volume}{6}}, \bibinfo{pages}{2300096}
  (\bibinfo{year}{2023}).

\bibitem[{\citenamefont{Inagaki et~al.}(2013)\citenamefont{Inagaki, Matsuda,
  Tadanaga, Asobe, and Takesue}}]{IMT+13}
\bibinfo{author}{\bibfnamefont{T.}~\bibnamefont{Inagaki}},
  \bibinfo{author}{\bibfnamefont{N.}~\bibnamefont{Matsuda}},
  \bibinfo{author}{\bibfnamefont{O.}~\bibnamefont{Tadanaga}},
  \bibinfo{author}{\bibfnamefont{M.}~\bibnamefont{Asobe}}, \bibnamefont{and}
  \bibinfo{author}{\bibfnamefont{H.}~\bibnamefont{Takesue}},
  \bibinfo{journal}{Optics Express} \textbf{\bibinfo{volume}{21}},
  \bibinfo{pages}{23241} (\bibinfo{year}{2013}).

\bibitem[{\citenamefont{Korzh et~al.}(2015)\citenamefont{Korzh, Lim, Houlmann,
  Gisin, Li, Nolan, Sanguinetti, Thew, and Zbinden}}]{BLH+15}
\bibinfo{author}{\bibfnamefont{B.}~\bibnamefont{Korzh}},
  \bibinfo{author}{\bibfnamefont{C.~C.~W.} \bibnamefont{Lim}},
  \bibinfo{author}{\bibfnamefont{R.}~\bibnamefont{Houlmann}},
  \bibinfo{author}{\bibfnamefont{N.}~\bibnamefont{Gisin}},
  \bibinfo{author}{\bibfnamefont{M.~J.} \bibnamefont{Li}},
  \bibinfo{author}{\bibfnamefont{D.}~\bibnamefont{Nolan}},
  \bibinfo{author}{\bibfnamefont{B.}~\bibnamefont{Sanguinetti}},
  \bibinfo{author}{\bibfnamefont{R.}~\bibnamefont{Thew}}, \bibnamefont{and}
  \bibinfo{author}{\bibfnamefont{H.}~\bibnamefont{Zbinden}},
  \bibinfo{journal}{Nature Photonics} \textbf{\bibinfo{volume}{9}},
  \bibinfo{pages}{163} (\bibinfo{year}{2015}).

\bibitem[{\citenamefont{Wengerowsky et~al.}(2020)\citenamefont{Wengerowsky,
  Joshi, Steinlechner, Zichi, Liu, Scheidl, Dobrovolskiy, Molen, Los, Zwiller
  et~al.}}]{WJS+20}
\bibinfo{author}{\bibfnamefont{S.}~\bibnamefont{Wengerowsky}},
  \bibinfo{author}{\bibfnamefont{S.~K.} \bibnamefont{Joshi}},
  \bibinfo{author}{\bibfnamefont{F.}~\bibnamefont{Steinlechner}},
  \bibinfo{author}{\bibfnamefont{J.~R.} \bibnamefont{Zichi}},
  \bibinfo{author}{\bibfnamefont{B.}~\bibnamefont{Liu}},
  \bibinfo{author}{\bibfnamefont{T.}~\bibnamefont{Scheidl}},
  \bibinfo{author}{\bibfnamefont{S.~M.} \bibnamefont{Dobrovolskiy}},
  \bibinfo{author}{\bibfnamefont{R.~v.~d.} \bibnamefont{Molen}},
  \bibinfo{author}{\bibfnamefont{J.~W.} \bibnamefont{Los}},
  \bibinfo{author}{\bibfnamefont{V.}~\bibnamefont{Zwiller}},
  \bibnamefont{et~al.}, \bibinfo{journal}{npj Quantum Information}
  \textbf{\bibinfo{volume}{6}}, \bibinfo{pages}{5} (\bibinfo{year}{2020}).

\bibitem[{\citenamefont{Scarani et~al.}(2009)\citenamefont{Scarani,
  Bechmann-Pasquinucci, Cerf, Du{\v{s}}ek, L{\"u}tkenhaus, and Peev}}]{SPC+09}
\bibinfo{author}{\bibfnamefont{V.}~\bibnamefont{Scarani}},
  \bibinfo{author}{\bibfnamefont{H.}~\bibnamefont{Bechmann-Pasquinucci}},
  \bibinfo{author}{\bibfnamefont{N.~J.} \bibnamefont{Cerf}},
  \bibinfo{author}{\bibfnamefont{M.}~\bibnamefont{Du{\v{s}}ek}},
  \bibinfo{author}{\bibfnamefont{N.}~\bibnamefont{L{\"u}tkenhaus}},
  \bibnamefont{and} \bibinfo{author}{\bibfnamefont{M.}~\bibnamefont{Peev}},
  \bibinfo{journal}{Reviews of Modern Physics} \textbf{\bibinfo{volume}{81}},
  \bibinfo{pages}{1301} (\bibinfo{year}{2009}).

\bibitem[{\citenamefont{Briegel et~al.}(1998)\citenamefont{Briegel, D{\"u}r,
  Cirac, and Zoller}}]{BDC+98}
\bibinfo{author}{\bibfnamefont{H.-J.} \bibnamefont{Briegel}},
  \bibinfo{author}{\bibfnamefont{W.}~\bibnamefont{D{\"u}r}},
  \bibinfo{author}{\bibfnamefont{J.~I.} \bibnamefont{Cirac}}, \bibnamefont{and}
  \bibinfo{author}{\bibfnamefont{P.}~\bibnamefont{Zoller}},
  \bibinfo{journal}{Physical Review Letters} \textbf{\bibinfo{volume}{81}},
  \bibinfo{pages}{5932} (\bibinfo{year}{1998}).

\bibitem[{\citenamefont{Sangouard et~al.}(2011)\citenamefont{Sangouard, Simon,
  De~Riedmatten, and Gisin}}]{SSR+11}
\bibinfo{author}{\bibfnamefont{N.}~\bibnamefont{Sangouard}},
  \bibinfo{author}{\bibfnamefont{C.}~\bibnamefont{Simon}},
  \bibinfo{author}{\bibfnamefont{H.}~\bibnamefont{De~Riedmatten}},
  \bibnamefont{and} \bibinfo{author}{\bibfnamefont{N.}~\bibnamefont{Gisin}},
  \bibinfo{journal}{Reviews of Modern Physics} \textbf{\bibinfo{volume}{83}},
  \bibinfo{pages}{33} (\bibinfo{year}{2011}).

\bibitem[{\citenamefont{Pirandola et~al.}(2017)\citenamefont{Pirandola,
  Laurenza, Ottaviani, and Banchi}}]{PLOB17}
\bibinfo{author}{\bibfnamefont{S.}~\bibnamefont{Pirandola}},
  \bibinfo{author}{\bibfnamefont{R.}~\bibnamefont{Laurenza}},
  \bibinfo{author}{\bibfnamefont{C.}~\bibnamefont{Ottaviani}},
  \bibnamefont{and} \bibinfo{author}{\bibfnamefont{L.}~\bibnamefont{Banchi}},
  \bibinfo{journal}{Nature communications} \textbf{\bibinfo{volume}{8}},
  \bibinfo{pages}{15043} (\bibinfo{year}{2017}).

\bibitem[{\citenamefont{Lucamarini et~al.}(2018)\citenamefont{Lucamarini, Yuan,
  Dynes, and Shields}}]{LYD+18}
\bibinfo{author}{\bibfnamefont{M.}~\bibnamefont{Lucamarini}},
  \bibinfo{author}{\bibfnamefont{Z.~L.} \bibnamefont{Yuan}},
  \bibinfo{author}{\bibfnamefont{J.~F.} \bibnamefont{Dynes}}, \bibnamefont{and}
  \bibinfo{author}{\bibfnamefont{A.~J.} \bibnamefont{Shields}},
  \bibinfo{journal}{Nature} \textbf{\bibinfo{volume}{557}},
  \bibinfo{pages}{400} (\bibinfo{year}{2018}).

\bibitem[{\citenamefont{Xie et~al.}(2023{\natexlab{a}})\citenamefont{Xie, Weng,
  Lu, Fu, Wang, Yin, and Chen}}]{XWL+23}
\bibinfo{author}{\bibfnamefont{Y.-M.} \bibnamefont{Xie}},
  \bibinfo{author}{\bibfnamefont{C.-X.} \bibnamefont{Weng}},
  \bibinfo{author}{\bibfnamefont{Y.-S.} \bibnamefont{Lu}},
  \bibinfo{author}{\bibfnamefont{Y.}~\bibnamefont{Fu}},
  \bibinfo{author}{\bibfnamefont{Y.}~\bibnamefont{Wang}},
  \bibinfo{author}{\bibfnamefont{H.-L.} \bibnamefont{Yin}}, \bibnamefont{and}
  \bibinfo{author}{\bibfnamefont{Z.-B.} \bibnamefont{Chen}},
  \bibinfo{journal}{Physical Review A} \textbf{\bibinfo{volume}{107}},
  \bibinfo{pages}{042603} (\bibinfo{year}{2023}{\natexlab{a}}).

\bibitem[{\citenamefont{Xie et~al.}(2022)\citenamefont{Xie, Lu, Weng, Cao, Jia,
  Bao, Wang, Fu, Yin, and Chen}}]{XLW+22}
\bibinfo{author}{\bibfnamefont{Y.-M.} \bibnamefont{Xie}},
  \bibinfo{author}{\bibfnamefont{Y.-S.} \bibnamefont{Lu}},
  \bibinfo{author}{\bibfnamefont{C.-X.} \bibnamefont{Weng}},
  \bibinfo{author}{\bibfnamefont{X.-Y.} \bibnamefont{Cao}},
  \bibinfo{author}{\bibfnamefont{Z.-Y.} \bibnamefont{Jia}},
  \bibinfo{author}{\bibfnamefont{Y.}~\bibnamefont{Bao}},
  \bibinfo{author}{\bibfnamefont{Y.}~\bibnamefont{Wang}},
  \bibinfo{author}{\bibfnamefont{Y.}~\bibnamefont{Fu}},
  \bibinfo{author}{\bibfnamefont{H.-L.} \bibnamefont{Yin}}, \bibnamefont{and}
  \bibinfo{author}{\bibfnamefont{Z.-B.} \bibnamefont{Chen}},
  \bibinfo{journal}{PRX Quantum} \textbf{\bibinfo{volume}{3}},
  \bibinfo{pages}{020315} (\bibinfo{year}{2022}).

\bibitem[{\citenamefont{Zeng et~al.}(2022)\citenamefont{Zeng, Zhou, Wu, and
  Ma}}]{ZZW+22}
\bibinfo{author}{\bibfnamefont{P.}~\bibnamefont{Zeng}},
  \bibinfo{author}{\bibfnamefont{H.}~\bibnamefont{Zhou}},
  \bibinfo{author}{\bibfnamefont{W.}~\bibnamefont{Wu}}, \bibnamefont{and}
  \bibinfo{author}{\bibfnamefont{X.}~\bibnamefont{Ma}},
  \bibinfo{journal}{Nature Communications} \textbf{\bibinfo{volume}{13}},
  \bibinfo{pages}{3903} (\bibinfo{year}{2022}).

\bibitem[{\citenamefont{Xie et~al.}(2023{\natexlab{b}})\citenamefont{Xie, Bai,
  Lu, Weng, Yin, and Chen}}]{XBL+23}
\bibinfo{author}{\bibfnamefont{Y.-M.} \bibnamefont{Xie}},
  \bibinfo{author}{\bibfnamefont{J.-L.} \bibnamefont{Bai}},
  \bibinfo{author}{\bibfnamefont{Y.-S.} \bibnamefont{Lu}},
  \bibinfo{author}{\bibfnamefont{C.-X.} \bibnamefont{Weng}},
  \bibinfo{author}{\bibfnamefont{H.-L.} \bibnamefont{Yin}}, \bibnamefont{and}
  \bibinfo{author}{\bibfnamefont{Z.-B.} \bibnamefont{Chen}},
  \bibinfo{journal}{Physical Review Applied} \textbf{\bibinfo{volume}{19}},
  \bibinfo{pages}{054070} (\bibinfo{year}{2023}{\natexlab{b}}).

\bibitem[{\citenamefont{Chen et~al.}(2021{\natexlab{a}})\citenamefont{Chen,
  Zhang, Liu, Jiang, Zhang, Han, Ma, Hu, Li, Liu et~al.}}]{CZL+21}
\bibinfo{author}{\bibfnamefont{J.-P.} \bibnamefont{Chen}},
  \bibinfo{author}{\bibfnamefont{C.}~\bibnamefont{Zhang}},
  \bibinfo{author}{\bibfnamefont{Y.}~\bibnamefont{Liu}},
  \bibinfo{author}{\bibfnamefont{C.}~\bibnamefont{Jiang}},
  \bibinfo{author}{\bibfnamefont{W.-J.} \bibnamefont{Zhang}},
  \bibinfo{author}{\bibfnamefont{Z.-Y.} \bibnamefont{Han}},
  \bibinfo{author}{\bibfnamefont{S.-Z.} \bibnamefont{Ma}},
  \bibinfo{author}{\bibfnamefont{X.-L.} \bibnamefont{Hu}},
  \bibinfo{author}{\bibfnamefont{Y.-H.} \bibnamefont{Li}},
  \bibinfo{author}{\bibfnamefont{H.}~\bibnamefont{Liu}}, \bibnamefont{et~al.},
  \bibinfo{journal}{Nature Photonics} \textbf{\bibinfo{volume}{15}},
  \bibinfo{pages}{570} (\bibinfo{year}{2021}{\natexlab{a}}).

\bibitem[{\citenamefont{Zhou et~al.}(2023{\natexlab{a}})\citenamefont{Zhou,
  Lin, Xie, Lu, Jing, Yin, and Yuan}}]{ZLX+23}
\bibinfo{author}{\bibfnamefont{L.}~\bibnamefont{Zhou}},
  \bibinfo{author}{\bibfnamefont{J.}~\bibnamefont{Lin}},
  \bibinfo{author}{\bibfnamefont{Y.-M.} \bibnamefont{Xie}},
  \bibinfo{author}{\bibfnamefont{Y.-S.} \bibnamefont{Lu}},
  \bibinfo{author}{\bibfnamefont{Y.}~\bibnamefont{Jing}},
  \bibinfo{author}{\bibfnamefont{H.-L.} \bibnamefont{Yin}}, \bibnamefont{and}
  \bibinfo{author}{\bibfnamefont{Z.}~\bibnamefont{Yuan}},
  \bibinfo{journal}{Physical Review Letters} \textbf{\bibinfo{volume}{130}},
  \bibinfo{pages}{250801} (\bibinfo{year}{2023}{\natexlab{a}}).

\bibitem[{\citenamefont{Zhu et~al.}(2023)\citenamefont{Zhu, Huang, Liu, Zeng,
  Zou, Dai, Tang, Li, You, Wang et~al.}}]{ZHL+23}
\bibinfo{author}{\bibfnamefont{H.-T.} \bibnamefont{Zhu}},
  \bibinfo{author}{\bibfnamefont{Y.}~\bibnamefont{Huang}},
  \bibinfo{author}{\bibfnamefont{H.}~\bibnamefont{Liu}},
  \bibinfo{author}{\bibfnamefont{P.}~\bibnamefont{Zeng}},
  \bibinfo{author}{\bibfnamefont{M.}~\bibnamefont{Zou}},
  \bibinfo{author}{\bibfnamefont{Y.}~\bibnamefont{Dai}},
  \bibinfo{author}{\bibfnamefont{S.}~\bibnamefont{Tang}},
  \bibinfo{author}{\bibfnamefont{H.}~\bibnamefont{Li}},
  \bibinfo{author}{\bibfnamefont{L.}~\bibnamefont{You}},
  \bibinfo{author}{\bibfnamefont{Z.}~\bibnamefont{Wang}}, \bibnamefont{et~al.},
  \bibinfo{journal}{Physical Review Letters} \textbf{\bibinfo{volume}{130}},
  \bibinfo{pages}{030801} (\bibinfo{year}{2023}).

\bibitem[{\citenamefont{Liu et~al.}(2023)\citenamefont{Liu, Zhang, Jiang, Chen,
  Zhang, Pan, Ma, Dong, Xiong, Zhang et~al.}}]{LZJ+23}
\bibinfo{author}{\bibfnamefont{Y.}~\bibnamefont{Liu}},
  \bibinfo{author}{\bibfnamefont{W.-J.} \bibnamefont{Zhang}},
  \bibinfo{author}{\bibfnamefont{C.}~\bibnamefont{Jiang}},
  \bibinfo{author}{\bibfnamefont{J.-P.} \bibnamefont{Chen}},
  \bibinfo{author}{\bibfnamefont{C.}~\bibnamefont{Zhang}},
  \bibinfo{author}{\bibfnamefont{W.-X.} \bibnamefont{Pan}},
  \bibinfo{author}{\bibfnamefont{D.}~\bibnamefont{Ma}},
  \bibinfo{author}{\bibfnamefont{H.}~\bibnamefont{Dong}},
  \bibinfo{author}{\bibfnamefont{J.-M.} \bibnamefont{Xiong}},
  \bibinfo{author}{\bibfnamefont{C.-J.} \bibnamefont{Zhang}},
  \bibnamefont{et~al.}, \bibinfo{journal}{Physical Review Letters}
  \textbf{\bibinfo{volume}{130}}, \bibinfo{pages}{210801}
  (\bibinfo{year}{2023}).

\bibitem[{\citenamefont{VanWiggeren and Roy}(1999)}]{VR99}
\bibinfo{author}{\bibfnamefont{G.~D.} \bibnamefont{VanWiggeren}}
  \bibnamefont{and} \bibinfo{author}{\bibfnamefont{R.}~\bibnamefont{Roy}},
  \bibinfo{journal}{Applied Optics} \textbf{\bibinfo{volume}{38}},
  \bibinfo{pages}{3888} (\bibinfo{year}{1999}).

\bibitem[{\citenamefont{Gordon and Kogelnik}(2000)}]{GK2000}
\bibinfo{author}{\bibfnamefont{J.}~\bibnamefont{Gordon}} \bibnamefont{and}
  \bibinfo{author}{\bibfnamefont{H.}~\bibnamefont{Kogelnik}},
  \bibinfo{journal}{Proceedings of the National Academy of Sciences}
  \textbf{\bibinfo{volume}{97}}, \bibinfo{pages}{4541} (\bibinfo{year}{2000}).

\bibitem[{\citenamefont{Toyoshima et~al.}(2011)\citenamefont{Toyoshima,
  Takenaka, Shoji, Takayama, Takeoka, Fujiwara, Sasaki et~al.}}]{TTS+11}
\bibinfo{author}{\bibfnamefont{M.}~\bibnamefont{Toyoshima}},
  \bibinfo{author}{\bibfnamefont{H.}~\bibnamefont{Takenaka}},
  \bibinfo{author}{\bibfnamefont{Y.}~\bibnamefont{Shoji}},
  \bibinfo{author}{\bibfnamefont{Y.}~\bibnamefont{Takayama}},
  \bibinfo{author}{\bibfnamefont{M.}~\bibnamefont{Takeoka}},
  \bibinfo{author}{\bibfnamefont{M.}~\bibnamefont{Fujiwara}},
  \bibinfo{author}{\bibfnamefont{M.}~\bibnamefont{Sasaki}},
  \bibnamefont{et~al.}, \bibinfo{journal}{International Journal of Optics}
  \textbf{\bibinfo{volume}{2011}} (\bibinfo{year}{2011}).

\bibitem[{\citenamefont{Bedington et~al.}(2017)\citenamefont{Bedington,
  Arrazola, and Ling}}]{BAL17}
\bibinfo{author}{\bibfnamefont{R.}~\bibnamefont{Bedington}},
  \bibinfo{author}{\bibfnamefont{J.~M.} \bibnamefont{Arrazola}},
  \bibnamefont{and} \bibinfo{author}{\bibfnamefont{A.}~\bibnamefont{Ling}},
  \bibinfo{journal}{npj Quantum Information} \textbf{\bibinfo{volume}{3}},
  \bibinfo{pages}{30} (\bibinfo{year}{2017}).

\bibitem[{\citenamefont{Ahmadi et~al.}(2023)\citenamefont{Ahmadi, Schwertfeger,
  Werner, Wiese, Lester, Da~Ros, Krause, Ritter, Abasifard, Cholsuk
  et~al.}}]{ASW+23}
\bibinfo{author}{\bibfnamefont{N.}~\bibnamefont{Ahmadi}},
  \bibinfo{author}{\bibfnamefont{S.}~\bibnamefont{Schwertfeger}},
  \bibinfo{author}{\bibfnamefont{P.}~\bibnamefont{Werner}},
  \bibinfo{author}{\bibfnamefont{L.}~\bibnamefont{Wiese}},
  \bibinfo{author}{\bibfnamefont{J.}~\bibnamefont{Lester}},
  \bibinfo{author}{\bibfnamefont{E.}~\bibnamefont{Da~Ros}},
  \bibinfo{author}{\bibfnamefont{J.}~\bibnamefont{Krause}},
  \bibinfo{author}{\bibfnamefont{S.}~\bibnamefont{Ritter}},
  \bibinfo{author}{\bibfnamefont{M.}~\bibnamefont{Abasifard}},
  \bibinfo{author}{\bibfnamefont{C.}~\bibnamefont{Cholsuk}},
  \bibnamefont{et~al.}, \bibinfo{journal}{Advanced Quantum Technologies} p.
  \bibinfo{pages}{2300343} (\bibinfo{year}{2023}).

\bibitem[{\citenamefont{Ursin et~al.}(2009)\citenamefont{Ursin, Jennewein,
  Kofler, Perdigues, Cacciapuoti, de~Matos, Aspelmeyer, Valencia, Scheidl, Acin
  et~al.}}]{UJK+09}
\bibinfo{author}{\bibfnamefont{R.}~\bibnamefont{Ursin}},
  \bibinfo{author}{\bibfnamefont{T.}~\bibnamefont{Jennewein}},
  \bibinfo{author}{\bibfnamefont{J.}~\bibnamefont{Kofler}},
  \bibinfo{author}{\bibfnamefont{J.~M.} \bibnamefont{Perdigues}},
  \bibinfo{author}{\bibfnamefont{L.}~\bibnamefont{Cacciapuoti}},
  \bibinfo{author}{\bibfnamefont{C.~J.} \bibnamefont{de~Matos}},
  \bibinfo{author}{\bibfnamefont{M.}~\bibnamefont{Aspelmeyer}},
  \bibinfo{author}{\bibfnamefont{A.}~\bibnamefont{Valencia}},
  \bibinfo{author}{\bibfnamefont{T.}~\bibnamefont{Scheidl}},
  \bibinfo{author}{\bibfnamefont{A.}~\bibnamefont{Acin}}, \bibnamefont{et~al.},
  \bibinfo{journal}{Europhysics News} \textbf{\bibinfo{volume}{40}},
  \bibinfo{pages}{26} (\bibinfo{year}{2009}).

\bibitem[{\citenamefont{Scheidl et~al.}(2013)\citenamefont{Scheidl, Wille, and
  Ursin}}]{SWU13}
\bibinfo{author}{\bibfnamefont{T.}~\bibnamefont{Scheidl}},
  \bibinfo{author}{\bibfnamefont{E.}~\bibnamefont{Wille}}, \bibnamefont{and}
  \bibinfo{author}{\bibfnamefont{R.}~\bibnamefont{Ursin}},
  \bibinfo{journal}{New Journal of Physics} \textbf{\bibinfo{volume}{15}},
  \bibinfo{pages}{043008} (\bibinfo{year}{2013}).

\bibitem[{\citenamefont{Sharma and Banerjee}(2019)}]{SB19}
\bibinfo{author}{\bibfnamefont{V.}~\bibnamefont{Sharma}} \bibnamefont{and}
  \bibinfo{author}{\bibfnamefont{S.}~\bibnamefont{Banerjee}},
  \bibinfo{journal}{Quantum Information Processing}
  \textbf{\bibinfo{volume}{18}}, \bibinfo{pages}{67} (\bibinfo{year}{2019}).

\bibitem[{\citenamefont{Korotkova et~al.}(2005)\citenamefont{Korotkova, Salem,
  Dogariu, and Wolf}}]{KSD+05}
\bibinfo{author}{\bibfnamefont{O.}~\bibnamefont{Korotkova}},
  \bibinfo{author}{\bibfnamefont{M.}~\bibnamefont{Salem}},
  \bibinfo{author}{\bibfnamefont{A.}~\bibnamefont{Dogariu}}, \bibnamefont{and}
  \bibinfo{author}{\bibfnamefont{E.}~\bibnamefont{Wolf}},
  \bibinfo{journal}{Waves in Random and Complex Media}
  \textbf{\bibinfo{volume}{15}}, \bibinfo{pages}{353} (\bibinfo{year}{2005}).

\bibitem[{\citenamefont{Zhang et~al.}(2017)\citenamefont{Zhang, Ding, and
  Dang}}]{ZDD17}
\bibinfo{author}{\bibfnamefont{J.}~\bibnamefont{Zhang}},
  \bibinfo{author}{\bibfnamefont{S.}~\bibnamefont{Ding}}, \bibnamefont{and}
  \bibinfo{author}{\bibfnamefont{A.}~\bibnamefont{Dang}},
  \bibinfo{journal}{Applied Optics} \textbf{\bibinfo{volume}{56}},
  \bibinfo{pages}{5145} (\bibinfo{year}{2017}).

\bibitem[{\citenamefont{Zhu et~al.}(2021)\citenamefont{Zhu, Janasik, Fyffe,
  Hay, Zhou, Kantor, Winder, Boyd, Leuchs, and Shi}}]{ZJF+21}
\bibinfo{author}{\bibfnamefont{Z.}~\bibnamefont{Zhu}},
  \bibinfo{author}{\bibfnamefont{M.}~\bibnamefont{Janasik}},
  \bibinfo{author}{\bibfnamefont{A.}~\bibnamefont{Fyffe}},
  \bibinfo{author}{\bibfnamefont{D.}~\bibnamefont{Hay}},
  \bibinfo{author}{\bibfnamefont{Y.}~\bibnamefont{Zhou}},
  \bibinfo{author}{\bibfnamefont{B.}~\bibnamefont{Kantor}},
  \bibinfo{author}{\bibfnamefont{T.}~\bibnamefont{Winder}},
  \bibinfo{author}{\bibfnamefont{R.~W.} \bibnamefont{Boyd}},
  \bibinfo{author}{\bibfnamefont{G.}~\bibnamefont{Leuchs}}, \bibnamefont{and}
  \bibinfo{author}{\bibfnamefont{Z.}~\bibnamefont{Shi}},
  \bibinfo{journal}{Nature Communications} \textbf{\bibinfo{volume}{12}},
  \bibinfo{pages}{1666} (\bibinfo{year}{2021}).

\bibitem[{\citenamefont{Pirandola}(2021{\natexlab{a}})}]{P21}
\bibinfo{author}{\bibfnamefont{S.}~\bibnamefont{Pirandola}},
  \bibinfo{journal}{Physical Review Research} \textbf{\bibinfo{volume}{3}},
  \bibinfo{pages}{013279} (\bibinfo{year}{2021}{\natexlab{a}}).

\bibitem[{\citenamefont{Pirandola}(2021{\natexlab{b}})}]{P+21}
\bibinfo{author}{\bibfnamefont{S.}~\bibnamefont{Pirandola}},
  \bibinfo{journal}{Physical Review Research} \textbf{\bibinfo{volume}{3}},
  \bibinfo{pages}{023130} (\bibinfo{year}{2021}{\natexlab{b}}).

\bibitem[{\citenamefont{Xavier et~al.}(2009)\citenamefont{Xavier, Walenta,
  De~Faria, Tempor{\~a}o, Gisin, Zbinden, and Von~der Weid}}]{XWF+09}
\bibinfo{author}{\bibfnamefont{G.}~\bibnamefont{Xavier}},
  \bibinfo{author}{\bibfnamefont{N.}~\bibnamefont{Walenta}},
  \bibinfo{author}{\bibfnamefont{G.~V.} \bibnamefont{De~Faria}},
  \bibinfo{author}{\bibfnamefont{G.}~\bibnamefont{Tempor{\~a}o}},
  \bibinfo{author}{\bibfnamefont{N.}~\bibnamefont{Gisin}},
  \bibinfo{author}{\bibfnamefont{H.}~\bibnamefont{Zbinden}}, \bibnamefont{and}
  \bibinfo{author}{\bibfnamefont{J.}~\bibnamefont{Von~der Weid}},
  \bibinfo{journal}{New Journal of Physics} \textbf{\bibinfo{volume}{11}},
  \bibinfo{pages}{045015} (\bibinfo{year}{2009}).

\bibitem[{\citenamefont{Ding et~al.}(2017)\citenamefont{Ding, Chen, Chen, Wang,
  Wang, Yin, Guo, Han et~al.}}]{DCC+17}
\bibinfo{author}{\bibfnamefont{Y.-Y.} \bibnamefont{Ding}},
  \bibinfo{author}{\bibfnamefont{W.}~\bibnamefont{Chen}},
  \bibinfo{author}{\bibfnamefont{H.}~\bibnamefont{Chen}},
  \bibinfo{author}{\bibfnamefont{C.}~\bibnamefont{Wang}},
  \bibinfo{author}{\bibfnamefont{S.}~\bibnamefont{Wang}},
  \bibinfo{author}{\bibfnamefont{Z.-Q.} \bibnamefont{Yin}},
  \bibinfo{author}{\bibfnamefont{G.-C.} \bibnamefont{Guo}},
  \bibinfo{author}{\bibfnamefont{Z.-F.} \bibnamefont{Han}},
  \bibnamefont{et~al.}, \bibinfo{journal}{Optics Letters}
  \textbf{\bibinfo{volume}{42}}, \bibinfo{pages}{1023} (\bibinfo{year}{2017}).

\bibitem[{\citenamefont{Li et~al.}(2018)\citenamefont{Li, Gao, Li, Xue, Wang,
  Lu, Xiang, Zhao, Yan, Chen et~al.}}]{LGL+18}
\bibinfo{author}{\bibfnamefont{D.-D.} \bibnamefont{Li}},
  \bibinfo{author}{\bibfnamefont{S.}~\bibnamefont{Gao}},
  \bibinfo{author}{\bibfnamefont{G.-C.} \bibnamefont{Li}},
  \bibinfo{author}{\bibfnamefont{L.}~\bibnamefont{Xue}},
  \bibinfo{author}{\bibfnamefont{L.-W.} \bibnamefont{Wang}},
  \bibinfo{author}{\bibfnamefont{C.-B.} \bibnamefont{Lu}},
  \bibinfo{author}{\bibfnamefont{Y.}~\bibnamefont{Xiang}},
  \bibinfo{author}{\bibfnamefont{Z.-Y.} \bibnamefont{Zhao}},
  \bibinfo{author}{\bibfnamefont{L.-C.} \bibnamefont{Yan}},
  \bibinfo{author}{\bibfnamefont{Z.-Y.} \bibnamefont{Chen}},
  \bibnamefont{et~al.}, \bibinfo{journal}{Optics Express}
  \textbf{\bibinfo{volume}{26}}, \bibinfo{pages}{22793} (\bibinfo{year}{2018}).

\bibitem[{\citenamefont{Neumann et~al.}(2022)\citenamefont{Neumann, Buchner,
  Bulla, Bohmann, and Ursin}}]{NBB+22}
\bibinfo{author}{\bibfnamefont{S.~P.} \bibnamefont{Neumann}},
  \bibinfo{author}{\bibfnamefont{A.}~\bibnamefont{Buchner}},
  \bibinfo{author}{\bibfnamefont{L.}~\bibnamefont{Bulla}},
  \bibinfo{author}{\bibfnamefont{M.}~\bibnamefont{Bohmann}}, \bibnamefont{and}
  \bibinfo{author}{\bibfnamefont{R.}~\bibnamefont{Ursin}},
  \bibinfo{journal}{Nature Communications} \textbf{\bibinfo{volume}{13}},
  \bibinfo{pages}{6134} (\bibinfo{year}{2022}).

\bibitem[{\citenamefont{Lee et~al.}(2022)\citenamefont{Lee, Mohammadi, Babcock,
  Higgins, Podmore, and Jennewein}}]{LKB+22}
\bibinfo{author}{\bibfnamefont{Y.~S.} \bibnamefont{Lee}},
  \bibinfo{author}{\bibfnamefont{K.}~\bibnamefont{Mohammadi}},
  \bibinfo{author}{\bibfnamefont{L.}~\bibnamefont{Babcock}},
  \bibinfo{author}{\bibfnamefont{B.~L.} \bibnamefont{Higgins}},
  \bibinfo{author}{\bibfnamefont{H.}~\bibnamefont{Podmore}}, \bibnamefont{and}
  \bibinfo{author}{\bibfnamefont{T.}~\bibnamefont{Jennewein}},
  \bibinfo{journal}{Review of Scientific Instruments}
  \textbf{\bibinfo{volume}{93}} (\bibinfo{year}{2022}).

\bibitem[{\citenamefont{Chatterjee et~al.}(2023)\citenamefont{Chatterjee,
  Goswami, Chatterjee, and Sinha}}]{CGC+23}
\bibinfo{author}{\bibfnamefont{S.}~\bibnamefont{Chatterjee}},
  \bibinfo{author}{\bibfnamefont{K.}~\bibnamefont{Goswami}},
  \bibinfo{author}{\bibfnamefont{R.}~\bibnamefont{Chatterjee}},
  \bibnamefont{and} \bibinfo{author}{\bibfnamefont{U.}~\bibnamefont{Sinha}},
  \bibinfo{journal}{Communications Physics} \textbf{\bibinfo{volume}{6}},
  \bibinfo{pages}{116} (\bibinfo{year}{2023}).

\bibitem[{\citenamefont{Rideout et~al.}(2012)\citenamefont{Rideout, Jennewein,
  Amelino-Camelia, Demarie, Higgins, Kempf, Kent, Laflamme, Ma, Mann
  et~al.}}]{RJC+12}
\bibinfo{author}{\bibfnamefont{D.}~\bibnamefont{Rideout}},
  \bibinfo{author}{\bibfnamefont{T.}~\bibnamefont{Jennewein}},
  \bibinfo{author}{\bibfnamefont{G.}~\bibnamefont{Amelino-Camelia}},
  \bibinfo{author}{\bibfnamefont{T.~F.} \bibnamefont{Demarie}},
  \bibinfo{author}{\bibfnamefont{B.~L.} \bibnamefont{Higgins}},
  \bibinfo{author}{\bibfnamefont{A.}~\bibnamefont{Kempf}},
  \bibinfo{author}{\bibfnamefont{A.}~\bibnamefont{Kent}},
  \bibinfo{author}{\bibfnamefont{R.}~\bibnamefont{Laflamme}},
  \bibinfo{author}{\bibfnamefont{X.}~\bibnamefont{Ma}},
  \bibinfo{author}{\bibfnamefont{R.~B.} \bibnamefont{Mann}},
  \bibnamefont{et~al.}, \bibinfo{journal}{Classical and Quantum Gravity}
  \textbf{\bibinfo{volume}{29}}, \bibinfo{pages}{224011}
  (\bibinfo{year}{2012}).

\bibitem[{\citenamefont{Joshi et~al.}(2018)\citenamefont{Joshi, Pienaar, Ralph,
  Cacciapuoti, McCutcheon, Rarity, Giggenbach, Lim, Makarov, Fuentes
  et~al.}}]{JPR+18}
\bibinfo{author}{\bibfnamefont{S.~K.} \bibnamefont{Joshi}},
  \bibinfo{author}{\bibfnamefont{J.}~\bibnamefont{Pienaar}},
  \bibinfo{author}{\bibfnamefont{T.~C.} \bibnamefont{Ralph}},
  \bibinfo{author}{\bibfnamefont{L.}~\bibnamefont{Cacciapuoti}},
  \bibinfo{author}{\bibfnamefont{W.}~\bibnamefont{McCutcheon}},
  \bibinfo{author}{\bibfnamefont{J.}~\bibnamefont{Rarity}},
  \bibinfo{author}{\bibfnamefont{D.}~\bibnamefont{Giggenbach}},
  \bibinfo{author}{\bibfnamefont{J.~G.} \bibnamefont{Lim}},
  \bibinfo{author}{\bibfnamefont{V.}~\bibnamefont{Makarov}},
  \bibinfo{author}{\bibfnamefont{I.}~\bibnamefont{Fuentes}},
  \bibnamefont{et~al.}, \bibinfo{journal}{New Journal of Physics}
  \textbf{\bibinfo{volume}{20}}, \bibinfo{pages}{063016}
  (\bibinfo{year}{2018}).

\bibitem[{\citenamefont{Giovannetti
  et~al.}(2001{\natexlab{a}})\citenamefont{Giovannetti, Lloyd, and
  Maccone}}]{GLM01}
\bibinfo{author}{\bibfnamefont{V.}~\bibnamefont{Giovannetti}},
  \bibinfo{author}{\bibfnamefont{S.}~\bibnamefont{Lloyd}}, \bibnamefont{and}
  \bibinfo{author}{\bibfnamefont{L.}~\bibnamefont{Maccone}},
  \bibinfo{journal}{Nature} \textbf{\bibinfo{volume}{412}},
  \bibinfo{pages}{417} (\bibinfo{year}{2001}{\natexlab{a}}).

\bibitem[{\citenamefont{Giovannetti
  et~al.}(2001{\natexlab{b}})\citenamefont{Giovannetti, Lloyd, Maccone, and
  Wong}}]{GLM+01}
\bibinfo{author}{\bibfnamefont{V.}~\bibnamefont{Giovannetti}},
  \bibinfo{author}{\bibfnamefont{S.}~\bibnamefont{Lloyd}},
  \bibinfo{author}{\bibfnamefont{L.}~\bibnamefont{Maccone}}, \bibnamefont{and}
  \bibinfo{author}{\bibfnamefont{F.}~\bibnamefont{Wong}},
  \bibinfo{journal}{Physical Review Letters} \textbf{\bibinfo{volume}{87}},
  \bibinfo{pages}{117902} (\bibinfo{year}{2001}{\natexlab{b}}).

\bibitem[{\citenamefont{Ho et~al.}(2009)\citenamefont{Ho, Lamas-Linares, and
  Kurtsiefer}}]{HLK09}
\bibinfo{author}{\bibfnamefont{C.}~\bibnamefont{Ho}},
  \bibinfo{author}{\bibfnamefont{A.}~\bibnamefont{Lamas-Linares}},
  \bibnamefont{and}
  \bibinfo{author}{\bibfnamefont{C.}~\bibnamefont{Kurtsiefer}},
  \bibinfo{journal}{New Journal of Physics} \textbf{\bibinfo{volume}{11}},
  \bibinfo{pages}{045011} (\bibinfo{year}{2009}).

\bibitem[{\citenamefont{Ahmadi et~al.}(2014)\citenamefont{Ahmadi, Bruschi,
  Sab{\'\i}n, Adesso, and Fuentes}}]{ABS+14}
\bibinfo{author}{\bibfnamefont{M.}~\bibnamefont{Ahmadi}},
  \bibinfo{author}{\bibfnamefont{D.~E.} \bibnamefont{Bruschi}},
  \bibinfo{author}{\bibfnamefont{C.}~\bibnamefont{Sab{\'\i}n}},
  \bibinfo{author}{\bibfnamefont{G.}~\bibnamefont{Adesso}}, \bibnamefont{and}
  \bibinfo{author}{\bibfnamefont{I.}~\bibnamefont{Fuentes}},
  \bibinfo{journal}{Scientific Reports} \textbf{\bibinfo{volume}{4}},
  \bibinfo{pages}{4996} (\bibinfo{year}{2014}).

\bibitem[{\citenamefont{Ursin et~al.}(2007)\citenamefont{Ursin, Tiefenbacher,
  Schmitt-Manderbach, Weier, Scheidl, Lindenthal, Blauensteiner, Jennewein,
  Perdigues, Trojek et~al.}}]{UTM+07}
\bibinfo{author}{\bibfnamefont{R.}~\bibnamefont{Ursin}},
  \bibinfo{author}{\bibfnamefont{F.}~\bibnamefont{Tiefenbacher}},
  \bibinfo{author}{\bibfnamefont{T.}~\bibnamefont{Schmitt-Manderbach}},
  \bibinfo{author}{\bibfnamefont{H.}~\bibnamefont{Weier}},
  \bibinfo{author}{\bibfnamefont{T.}~\bibnamefont{Scheidl}},
  \bibinfo{author}{\bibfnamefont{M.}~\bibnamefont{Lindenthal}},
  \bibinfo{author}{\bibfnamefont{B.}~\bibnamefont{Blauensteiner}},
  \bibinfo{author}{\bibfnamefont{T.}~\bibnamefont{Jennewein}},
  \bibinfo{author}{\bibfnamefont{J.}~\bibnamefont{Perdigues}},
  \bibinfo{author}{\bibfnamefont{P.}~\bibnamefont{Trojek}},
  \bibnamefont{et~al.}, \bibinfo{journal}{Nature Physics}
  \textbf{\bibinfo{volume}{3}}, \bibinfo{pages}{481} (\bibinfo{year}{2007}).

\bibitem[{\citenamefont{Villoresi et~al.}(2008)\citenamefont{Villoresi,
  Jennewein, Tamburini, Aspelmeyer, Bonato, Ursin, Pernechele, Luceri, Bianco,
  Zeilinger et~al.}}]{VJT+08}
\bibinfo{author}{\bibfnamefont{P.}~\bibnamefont{Villoresi}},
  \bibinfo{author}{\bibfnamefont{T.}~\bibnamefont{Jennewein}},
  \bibinfo{author}{\bibfnamefont{F.}~\bibnamefont{Tamburini}},
  \bibinfo{author}{\bibfnamefont{M.}~\bibnamefont{Aspelmeyer}},
  \bibinfo{author}{\bibfnamefont{C.}~\bibnamefont{Bonato}},
  \bibinfo{author}{\bibfnamefont{R.}~\bibnamefont{Ursin}},
  \bibinfo{author}{\bibfnamefont{C.}~\bibnamefont{Pernechele}},
  \bibinfo{author}{\bibfnamefont{V.}~\bibnamefont{Luceri}},
  \bibinfo{author}{\bibfnamefont{G.}~\bibnamefont{Bianco}},
  \bibinfo{author}{\bibfnamefont{A.}~\bibnamefont{Zeilinger}},
  \bibnamefont{et~al.}, \bibinfo{journal}{New Journal of Physics}
  \textbf{\bibinfo{volume}{10}}, \bibinfo{pages}{033038}
  (\bibinfo{year}{2008}).

\bibitem[{\citenamefont{Fedrizzi et~al.}(2009)\citenamefont{Fedrizzi, Ursin,
  Herbst, Nespoli, Prevedel, Scheidl, Tiefenbacher, Jennewein, and
  Zeilinger}}]{FUH+09}
\bibinfo{author}{\bibfnamefont{A.}~\bibnamefont{Fedrizzi}},
  \bibinfo{author}{\bibfnamefont{R.}~\bibnamefont{Ursin}},
  \bibinfo{author}{\bibfnamefont{T.}~\bibnamefont{Herbst}},
  \bibinfo{author}{\bibfnamefont{M.}~\bibnamefont{Nespoli}},
  \bibinfo{author}{\bibfnamefont{R.}~\bibnamefont{Prevedel}},
  \bibinfo{author}{\bibfnamefont{T.}~\bibnamefont{Scheidl}},
  \bibinfo{author}{\bibfnamefont{F.}~\bibnamefont{Tiefenbacher}},
  \bibinfo{author}{\bibfnamefont{T.}~\bibnamefont{Jennewein}},
  \bibnamefont{and}
  \bibinfo{author}{\bibfnamefont{A.}~\bibnamefont{Zeilinger}},
  \bibinfo{journal}{Nature Physics} \textbf{\bibinfo{volume}{5}},
  \bibinfo{pages}{389} (\bibinfo{year}{2009}).

\bibitem[{\citenamefont{Yin et~al.}(2012)\citenamefont{Yin, Ren, Lu, Cao, Yong,
  Wu, Liu, Liao, Zhou, Jiang et~al.}}]{YRL+12}
\bibinfo{author}{\bibfnamefont{J.}~\bibnamefont{Yin}},
  \bibinfo{author}{\bibfnamefont{J.-G.} \bibnamefont{Ren}},
  \bibinfo{author}{\bibfnamefont{H.}~\bibnamefont{Lu}},
  \bibinfo{author}{\bibfnamefont{Y.}~\bibnamefont{Cao}},
  \bibinfo{author}{\bibfnamefont{H.-L.} \bibnamefont{Yong}},
  \bibinfo{author}{\bibfnamefont{Y.-P.} \bibnamefont{Wu}},
  \bibinfo{author}{\bibfnamefont{C.}~\bibnamefont{Liu}},
  \bibinfo{author}{\bibfnamefont{S.-K.} \bibnamefont{Liao}},
  \bibinfo{author}{\bibfnamefont{F.}~\bibnamefont{Zhou}},
  \bibinfo{author}{\bibfnamefont{Y.}~\bibnamefont{Jiang}},
  \bibnamefont{et~al.}, \bibinfo{journal}{Nature}
  \textbf{\bibinfo{volume}{488}}, \bibinfo{pages}{185} (\bibinfo{year}{2012}).

\bibitem[{\citenamefont{Wang et~al.}(2013)\citenamefont{Wang, Yang, Liao,
  Zhang, Shen, Hu, Wu, Yang, Jiang, Tang et~al.}}]{WYL+13}
\bibinfo{author}{\bibfnamefont{J.-Y.} \bibnamefont{Wang}},
  \bibinfo{author}{\bibfnamefont{B.}~\bibnamefont{Yang}},
  \bibinfo{author}{\bibfnamefont{S.-K.} \bibnamefont{Liao}},
  \bibinfo{author}{\bibfnamefont{L.}~\bibnamefont{Zhang}},
  \bibinfo{author}{\bibfnamefont{Q.}~\bibnamefont{Shen}},
  \bibinfo{author}{\bibfnamefont{X.-F.} \bibnamefont{Hu}},
  \bibinfo{author}{\bibfnamefont{J.-C.} \bibnamefont{Wu}},
  \bibinfo{author}{\bibfnamefont{S.-J.} \bibnamefont{Yang}},
  \bibinfo{author}{\bibfnamefont{H.}~\bibnamefont{Jiang}},
  \bibinfo{author}{\bibfnamefont{Y.-L.} \bibnamefont{Tang}},
  \bibnamefont{et~al.}, \bibinfo{journal}{Nature Photonics}
  \textbf{\bibinfo{volume}{7}}, \bibinfo{pages}{387} (\bibinfo{year}{2013}).

\bibitem[{\citenamefont{Nauerth et~al.}(2013)\citenamefont{Nauerth, Moll, Rau,
  Fuchs, Horwath, Frick, and Weinfurter}}]{NMR+13}
\bibinfo{author}{\bibfnamefont{S.}~\bibnamefont{Nauerth}},
  \bibinfo{author}{\bibfnamefont{F.}~\bibnamefont{Moll}},
  \bibinfo{author}{\bibfnamefont{M.}~\bibnamefont{Rau}},
  \bibinfo{author}{\bibfnamefont{C.}~\bibnamefont{Fuchs}},
  \bibinfo{author}{\bibfnamefont{J.}~\bibnamefont{Horwath}},
  \bibinfo{author}{\bibfnamefont{S.}~\bibnamefont{Frick}}, \bibnamefont{and}
  \bibinfo{author}{\bibfnamefont{H.}~\bibnamefont{Weinfurter}},
  \bibinfo{journal}{Nature Photonics} \textbf{\bibinfo{volume}{7}},
  \bibinfo{pages}{382} (\bibinfo{year}{2013}).

\bibitem[{\citenamefont{Cao et~al.}(2013)\citenamefont{Cao, Liang, Yin, Yong,
  Zhou, Wu, Ren, Li, Pan, Yang et~al.}}]{CLY+13}
\bibinfo{author}{\bibfnamefont{Y.}~\bibnamefont{Cao}},
  \bibinfo{author}{\bibfnamefont{H.}~\bibnamefont{Liang}},
  \bibinfo{author}{\bibfnamefont{J.}~\bibnamefont{Yin}},
  \bibinfo{author}{\bibfnamefont{H.-L.} \bibnamefont{Yong}},
  \bibinfo{author}{\bibfnamefont{F.}~\bibnamefont{Zhou}},
  \bibinfo{author}{\bibfnamefont{Y.-P.} \bibnamefont{Wu}},
  \bibinfo{author}{\bibfnamefont{J.-G.} \bibnamefont{Ren}},
  \bibinfo{author}{\bibfnamefont{Y.-H.} \bibnamefont{Li}},
  \bibinfo{author}{\bibfnamefont{G.-S.} \bibnamefont{Pan}},
  \bibinfo{author}{\bibfnamefont{T.}~\bibnamefont{Yang}}, \bibnamefont{et~al.},
  \bibinfo{journal}{Optics Express} \textbf{\bibinfo{volume}{21}},
  \bibinfo{pages}{27260} (\bibinfo{year}{2013}).

\bibitem[{\citenamefont{Pugh et~al.}(2017)\citenamefont{Pugh, Kaiser, Bourgoin,
  Jin, Sultana, Agne, Anisimova, Makarov, Choi, Higgins et~al.}}]{PKB+17}
\bibinfo{author}{\bibfnamefont{C.~J.} \bibnamefont{Pugh}},
  \bibinfo{author}{\bibfnamefont{S.}~\bibnamefont{Kaiser}},
  \bibinfo{author}{\bibfnamefont{J.-P.} \bibnamefont{Bourgoin}},
  \bibinfo{author}{\bibfnamefont{J.}~\bibnamefont{Jin}},
  \bibinfo{author}{\bibfnamefont{N.}~\bibnamefont{Sultana}},
  \bibinfo{author}{\bibfnamefont{S.}~\bibnamefont{Agne}},
  \bibinfo{author}{\bibfnamefont{E.}~\bibnamefont{Anisimova}},
  \bibinfo{author}{\bibfnamefont{V.}~\bibnamefont{Makarov}},
  \bibinfo{author}{\bibfnamefont{E.}~\bibnamefont{Choi}},
  \bibinfo{author}{\bibfnamefont{B.~L.} \bibnamefont{Higgins}},
  \bibnamefont{et~al.}, \bibinfo{journal}{Quantum Science and Technology}
  \textbf{\bibinfo{volume}{2}}, \bibinfo{pages}{024009} (\bibinfo{year}{2017}).

\bibitem[{\citenamefont{Bonato et~al.}(2009)\citenamefont{Bonato, Tomaello,
  Da~Deppo, Naletto, and Villoresi}}]{BTD09}
\bibinfo{author}{\bibfnamefont{C.}~\bibnamefont{Bonato}},
  \bibinfo{author}{\bibfnamefont{A.}~\bibnamefont{Tomaello}},
  \bibinfo{author}{\bibfnamefont{V.}~\bibnamefont{Da~Deppo}},
  \bibinfo{author}{\bibfnamefont{G.}~\bibnamefont{Naletto}}, \bibnamefont{and}
  \bibinfo{author}{\bibfnamefont{P.}~\bibnamefont{Villoresi}},
  \bibinfo{journal}{New Journal of Physics} \textbf{\bibinfo{volume}{11}},
  \bibinfo{pages}{045017} (\bibinfo{year}{2009}).

\bibitem[{\citenamefont{Trinh et~al.}(2022)\citenamefont{Trinh,
  Carrasco-Casado, Takenaka, Fujiwara, Kitamura, Sasaki, and
  Toyoshima}}]{TCT+22}
\bibinfo{author}{\bibfnamefont{P.~V.} \bibnamefont{Trinh}},
  \bibinfo{author}{\bibfnamefont{A.}~\bibnamefont{Carrasco-Casado}},
  \bibinfo{author}{\bibfnamefont{H.}~\bibnamefont{Takenaka}},
  \bibinfo{author}{\bibfnamefont{M.}~\bibnamefont{Fujiwara}},
  \bibinfo{author}{\bibfnamefont{M.}~\bibnamefont{Kitamura}},
  \bibinfo{author}{\bibfnamefont{M.}~\bibnamefont{Sasaki}}, \bibnamefont{and}
  \bibinfo{author}{\bibfnamefont{M.}~\bibnamefont{Toyoshima}},
  \bibinfo{journal}{Communications Physics} \textbf{\bibinfo{volume}{5}},
  \bibinfo{pages}{225} (\bibinfo{year}{2022}).

\bibitem[{\citenamefont{Vasylyev et~al.}(2019)\citenamefont{Vasylyev, Vogel,
  and Moll}}]{VVM19}
\bibinfo{author}{\bibfnamefont{D.}~\bibnamefont{Vasylyev}},
  \bibinfo{author}{\bibfnamefont{W.}~\bibnamefont{Vogel}}, \bibnamefont{and}
  \bibinfo{author}{\bibfnamefont{F.}~\bibnamefont{Moll}},
  \bibinfo{journal}{Physical Review A} \textbf{\bibinfo{volume}{99}},
  \bibinfo{pages}{053830} (\bibinfo{year}{2019}).

\bibitem[{\citenamefont{Bourgoin et~al.}(2015)\citenamefont{Bourgoin, Higgins,
  Gigov, Holloway, Pugh, Kaiser, Cranmer, and Jennewein}}]{BHG+15}
\bibinfo{author}{\bibfnamefont{J.-P.} \bibnamefont{Bourgoin}},
  \bibinfo{author}{\bibfnamefont{B.~L.} \bibnamefont{Higgins}},
  \bibinfo{author}{\bibfnamefont{N.}~\bibnamefont{Gigov}},
  \bibinfo{author}{\bibfnamefont{C.}~\bibnamefont{Holloway}},
  \bibinfo{author}{\bibfnamefont{C.~J.} \bibnamefont{Pugh}},
  \bibinfo{author}{\bibfnamefont{S.}~\bibnamefont{Kaiser}},
  \bibinfo{author}{\bibfnamefont{M.}~\bibnamefont{Cranmer}}, \bibnamefont{and}
  \bibinfo{author}{\bibfnamefont{T.}~\bibnamefont{Jennewein}},
  \bibinfo{journal}{Optics Express} \textbf{\bibinfo{volume}{23}},
  \bibinfo{pages}{33437} (\bibinfo{year}{2015}).

\bibitem[{\citenamefont{Liu et~al.}(2020)\citenamefont{Liu, Tian, Gu, Fan, Ni,
  Yang, Zhang, Hu, Guo, Cao et~al.}}]{LTG+20}
\bibinfo{author}{\bibfnamefont{H.-Y.} \bibnamefont{Liu}},
  \bibinfo{author}{\bibfnamefont{X.-H.} \bibnamefont{Tian}},
  \bibinfo{author}{\bibfnamefont{C.}~\bibnamefont{Gu}},
  \bibinfo{author}{\bibfnamefont{P.}~\bibnamefont{Fan}},
  \bibinfo{author}{\bibfnamefont{X.}~\bibnamefont{Ni}},
  \bibinfo{author}{\bibfnamefont{R.}~\bibnamefont{Yang}},
  \bibinfo{author}{\bibfnamefont{J.-N.} \bibnamefont{Zhang}},
  \bibinfo{author}{\bibfnamefont{M.}~\bibnamefont{Hu}},
  \bibinfo{author}{\bibfnamefont{J.}~\bibnamefont{Guo}},
  \bibinfo{author}{\bibfnamefont{X.}~\bibnamefont{Cao}}, \bibnamefont{et~al.},
  \bibinfo{journal}{National Science Review} \textbf{\bibinfo{volume}{7}},
  \bibinfo{pages}{921} (\bibinfo{year}{2020}).

\bibitem[{\citenamefont{Jennewein and Higgins}(2013)}]{JH13}
\bibinfo{author}{\bibfnamefont{T.}~\bibnamefont{Jennewein}} \bibnamefont{and}
  \bibinfo{author}{\bibfnamefont{B.}~\bibnamefont{Higgins}},
  \bibinfo{journal}{Physics World} \textbf{\bibinfo{volume}{26}},
  \bibinfo{pages}{52} (\bibinfo{year}{2013}).

\bibitem[{\citenamefont{Khan et~al.}(2018)\citenamefont{Khan, Heim, Neuzner,
  and Marquardt}}]{KHN+18}
\bibinfo{author}{\bibfnamefont{I.}~\bibnamefont{Khan}},
  \bibinfo{author}{\bibfnamefont{B.}~\bibnamefont{Heim}},
  \bibinfo{author}{\bibfnamefont{A.}~\bibnamefont{Neuzner}}, \bibnamefont{and}
  \bibinfo{author}{\bibfnamefont{C.}~\bibnamefont{Marquardt}},
  \bibinfo{journal}{Optics and Photonics News} \textbf{\bibinfo{volume}{29}},
  \bibinfo{pages}{26} (\bibinfo{year}{2018}).

\bibitem[{\citenamefont{Yin et~al.}(2017{\natexlab{a}})\citenamefont{Yin, Cao,
  Li, Liao, Zhang, Ren, Cai, Liu, Li, Dai et~al.}}]{YCLL+17}
\bibinfo{author}{\bibfnamefont{J.}~\bibnamefont{Yin}},
  \bibinfo{author}{\bibfnamefont{Y.}~\bibnamefont{Cao}},
  \bibinfo{author}{\bibfnamefont{Y.-H.} \bibnamefont{Li}},
  \bibinfo{author}{\bibfnamefont{S.-K.} \bibnamefont{Liao}},
  \bibinfo{author}{\bibfnamefont{L.}~\bibnamefont{Zhang}},
  \bibinfo{author}{\bibfnamefont{J.-G.} \bibnamefont{Ren}},
  \bibinfo{author}{\bibfnamefont{W.-Q.} \bibnamefont{Cai}},
  \bibinfo{author}{\bibfnamefont{W.-Y.} \bibnamefont{Liu}},
  \bibinfo{author}{\bibfnamefont{B.}~\bibnamefont{Li}},
  \bibinfo{author}{\bibfnamefont{H.}~\bibnamefont{Dai}}, \bibnamefont{et~al.},
  \bibinfo{journal}{Science} \textbf{\bibinfo{volume}{356}},
  \bibinfo{pages}{1140} (\bibinfo{year}{2017}{\natexlab{a}}).

\bibitem[{\citenamefont{Liao et~al.}(2017{\natexlab{a}})\citenamefont{Liao,
  Cai, Liu, Zhang, Li, Ren, Yin, Shen, Cao, Li et~al.}}]{LCL+17}
\bibinfo{author}{\bibfnamefont{S.-K.} \bibnamefont{Liao}},
  \bibinfo{author}{\bibfnamefont{W.-Q.} \bibnamefont{Cai}},
  \bibinfo{author}{\bibfnamefont{W.-Y.} \bibnamefont{Liu}},
  \bibinfo{author}{\bibfnamefont{L.}~\bibnamefont{Zhang}},
  \bibinfo{author}{\bibfnamefont{Y.}~\bibnamefont{Li}},
  \bibinfo{author}{\bibfnamefont{J.-G.} \bibnamefont{Ren}},
  \bibinfo{author}{\bibfnamefont{J.}~\bibnamefont{Yin}},
  \bibinfo{author}{\bibfnamefont{Q.}~\bibnamefont{Shen}},
  \bibinfo{author}{\bibfnamefont{Y.}~\bibnamefont{Cao}},
  \bibinfo{author}{\bibfnamefont{Z.-P.} \bibnamefont{Li}},
  \bibnamefont{et~al.}, \bibinfo{journal}{Nature}
  \textbf{\bibinfo{volume}{549}}, \bibinfo{pages}{43}
  (\bibinfo{year}{2017}{\natexlab{a}}).

\bibitem[{\citenamefont{Takenaka et~al.}(2017)\citenamefont{Takenaka,
  Carrasco-Casado, Fujiwara, Kitamura, Sasaki, and Toyoshima}}]{TCF+17}
\bibinfo{author}{\bibfnamefont{H.}~\bibnamefont{Takenaka}},
  \bibinfo{author}{\bibfnamefont{A.}~\bibnamefont{Carrasco-Casado}},
  \bibinfo{author}{\bibfnamefont{M.}~\bibnamefont{Fujiwara}},
  \bibinfo{author}{\bibfnamefont{M.}~\bibnamefont{Kitamura}},
  \bibinfo{author}{\bibfnamefont{M.}~\bibnamefont{Sasaki}}, \bibnamefont{and}
  \bibinfo{author}{\bibfnamefont{M.}~\bibnamefont{Toyoshima}},
  \bibinfo{journal}{Nature Photonics} \textbf{\bibinfo{volume}{11}},
  \bibinfo{pages}{502} (\bibinfo{year}{2017}).

\bibitem[{\citenamefont{Liao et~al.}(2017{\natexlab{b}})\citenamefont{Liao,
  Lin, Ren, Liu, Qiang, Yin, Li, Shen, Zhang, Liang et~al.}}]{LLR+17}
\bibinfo{author}{\bibfnamefont{S.-K.} \bibnamefont{Liao}},
  \bibinfo{author}{\bibfnamefont{J.}~\bibnamefont{Lin}},
  \bibinfo{author}{\bibfnamefont{J.-G.} \bibnamefont{Ren}},
  \bibinfo{author}{\bibfnamefont{W.-Y.} \bibnamefont{Liu}},
  \bibinfo{author}{\bibfnamefont{J.}~\bibnamefont{Qiang}},
  \bibinfo{author}{\bibfnamefont{J.}~\bibnamefont{Yin}},
  \bibinfo{author}{\bibfnamefont{Y.}~\bibnamefont{Li}},
  \bibinfo{author}{\bibfnamefont{Q.}~\bibnamefont{Shen}},
  \bibinfo{author}{\bibfnamefont{L.}~\bibnamefont{Zhang}},
  \bibinfo{author}{\bibfnamefont{X.-F.} \bibnamefont{Liang}},
  \bibnamefont{et~al.}, \bibinfo{journal}{Chinese Physics Letters}
  \textbf{\bibinfo{volume}{34}}, \bibinfo{pages}{090302}
  (\bibinfo{year}{2017}{\natexlab{b}}).

\bibitem[{\citenamefont{Yin et~al.}(2020)\citenamefont{Yin, Li, Liao, Yang,
  Cao, Zhang, Ren, Cai, Liu, Li et~al.}}]{YLL+20}
\bibinfo{author}{\bibfnamefont{J.}~\bibnamefont{Yin}},
  \bibinfo{author}{\bibfnamefont{Y.-H.} \bibnamefont{Li}},
  \bibinfo{author}{\bibfnamefont{S.-K.} \bibnamefont{Liao}},
  \bibinfo{author}{\bibfnamefont{M.}~\bibnamefont{Yang}},
  \bibinfo{author}{\bibfnamefont{Y.}~\bibnamefont{Cao}},
  \bibinfo{author}{\bibfnamefont{L.}~\bibnamefont{Zhang}},
  \bibinfo{author}{\bibfnamefont{J.-G.} \bibnamefont{Ren}},
  \bibinfo{author}{\bibfnamefont{W.-Q.} \bibnamefont{Cai}},
  \bibinfo{author}{\bibfnamefont{W.-Y.} \bibnamefont{Liu}},
  \bibinfo{author}{\bibfnamefont{S.-L.} \bibnamefont{Li}},
  \bibnamefont{et~al.}, \bibinfo{journal}{Nature}
  \textbf{\bibinfo{volume}{582}}, \bibinfo{pages}{501} (\bibinfo{year}{2020}).

\bibitem[{\citenamefont{Yin et~al.}(2017{\natexlab{b}})\citenamefont{Yin, Cao,
  Li, Ren, Liao, Zhang, Cai, Liu, Li, Dai et~al.}}]{YCL+17}
\bibinfo{author}{\bibfnamefont{J.}~\bibnamefont{Yin}},
  \bibinfo{author}{\bibfnamefont{Y.}~\bibnamefont{Cao}},
  \bibinfo{author}{\bibfnamefont{Y.-H.} \bibnamefont{Li}},
  \bibinfo{author}{\bibfnamefont{J.-G.} \bibnamefont{Ren}},
  \bibinfo{author}{\bibfnamefont{S.-K.} \bibnamefont{Liao}},
  \bibinfo{author}{\bibfnamefont{L.}~\bibnamefont{Zhang}},
  \bibinfo{author}{\bibfnamefont{W.-Q.} \bibnamefont{Cai}},
  \bibinfo{author}{\bibfnamefont{W.-Y.} \bibnamefont{Liu}},
  \bibinfo{author}{\bibfnamefont{B.}~\bibnamefont{Li}},
  \bibinfo{author}{\bibfnamefont{H.}~\bibnamefont{Dai}}, \bibnamefont{et~al.},
  \bibinfo{journal}{Physical Review Letters} \textbf{\bibinfo{volume}{119}},
  \bibinfo{pages}{200501} (\bibinfo{year}{2017}{\natexlab{b}}).

\bibitem[{\citenamefont{Ren et~al.}(2017)\citenamefont{Ren, Xu, Yong, Zhang,
  Liao, Yin, Liu, Cai, Yang, Li et~al.}}]{RXY+17}
\bibinfo{author}{\bibfnamefont{J.-G.} \bibnamefont{Ren}},
  \bibinfo{author}{\bibfnamefont{P.}~\bibnamefont{Xu}},
  \bibinfo{author}{\bibfnamefont{H.-L.} \bibnamefont{Yong}},
  \bibinfo{author}{\bibfnamefont{L.}~\bibnamefont{Zhang}},
  \bibinfo{author}{\bibfnamefont{S.-K.} \bibnamefont{Liao}},
  \bibinfo{author}{\bibfnamefont{J.}~\bibnamefont{Yin}},
  \bibinfo{author}{\bibfnamefont{W.-Y.} \bibnamefont{Liu}},
  \bibinfo{author}{\bibfnamefont{W.-Q.} \bibnamefont{Cai}},
  \bibinfo{author}{\bibfnamefont{M.}~\bibnamefont{Yang}},
  \bibinfo{author}{\bibfnamefont{L.}~\bibnamefont{Li}}, \bibnamefont{et~al.},
  \bibinfo{journal}{Nature} \textbf{\bibinfo{volume}{549}}, \bibinfo{pages}{70}
  (\bibinfo{year}{2017}).

\bibitem[{\citenamefont{Wehner et~al.}(2018)\citenamefont{Wehner, Elkouss, and
  Hanson}}]{WEH18}
\bibinfo{author}{\bibfnamefont{S.}~\bibnamefont{Wehner}},
  \bibinfo{author}{\bibfnamefont{D.}~\bibnamefont{Elkouss}}, \bibnamefont{and}
  \bibinfo{author}{\bibfnamefont{R.}~\bibnamefont{Hanson}},
  \bibinfo{journal}{Science} \textbf{\bibinfo{volume}{362}},
  \bibinfo{pages}{eaam9288} (\bibinfo{year}{2018}).

\bibitem[{\citenamefont{Yin et~al.}(2023)\citenamefont{Yin, Fu, Li, Weng, Li,
  Gu, Lu, Huang, and Chen}}]{YFL+23}
\bibinfo{author}{\bibfnamefont{H.-L.} \bibnamefont{Yin}},
  \bibinfo{author}{\bibfnamefont{Y.}~\bibnamefont{Fu}},
  \bibinfo{author}{\bibfnamefont{C.-L.} \bibnamefont{Li}},
  \bibinfo{author}{\bibfnamefont{C.-X.} \bibnamefont{Weng}},
  \bibinfo{author}{\bibfnamefont{B.-H.} \bibnamefont{Li}},
  \bibinfo{author}{\bibfnamefont{J.}~\bibnamefont{Gu}},
  \bibinfo{author}{\bibfnamefont{Y.-S.} \bibnamefont{Lu}},
  \bibinfo{author}{\bibfnamefont{S.}~\bibnamefont{Huang}}, \bibnamefont{and}
  \bibinfo{author}{\bibfnamefont{Z.-B.} \bibnamefont{Chen}},
  \bibinfo{journal}{National Science Review} \textbf{\bibinfo{volume}{10}},
  \bibinfo{pages}{nwac228} (\bibinfo{year}{2023}).

\bibitem[{\citenamefont{Cao et~al.}(2024)\citenamefont{Cao, Li, Wang, Fu, Yin,
  and Chen}}]{CLW+24}
\bibinfo{author}{\bibfnamefont{X.-Y.} \bibnamefont{Cao}},
  \bibinfo{author}{\bibfnamefont{B.-H.} \bibnamefont{Li}},
  \bibinfo{author}{\bibfnamefont{Y.}~\bibnamefont{Wang}},
  \bibinfo{author}{\bibfnamefont{Y.}~\bibnamefont{Fu}},
  \bibinfo{author}{\bibfnamefont{H.-L.} \bibnamefont{Yin}}, \bibnamefont{and}
  \bibinfo{author}{\bibfnamefont{Z.-B.} \bibnamefont{Chen}},
  \bibinfo{journal}{Science Advances} \textbf{\bibinfo{volume}{10}},
  \bibinfo{pages}{eadk3258} (\bibinfo{year}{2024}).

\bibitem[{\citenamefont{Tang et~al.}(2016{\natexlab{a}})\citenamefont{Tang,
  Chandrasekara, Tan, Cheng, Sha, Hiang, Oi, and Ling}}]{TCT+16}
\bibinfo{author}{\bibfnamefont{Z.}~\bibnamefont{Tang}},
  \bibinfo{author}{\bibfnamefont{R.}~\bibnamefont{Chandrasekara}},
  \bibinfo{author}{\bibfnamefont{Y.~C.} \bibnamefont{Tan}},
  \bibinfo{author}{\bibfnamefont{C.}~\bibnamefont{Cheng}},
  \bibinfo{author}{\bibfnamefont{L.}~\bibnamefont{Sha}},
  \bibinfo{author}{\bibfnamefont{G.~C.} \bibnamefont{Hiang}},
  \bibinfo{author}{\bibfnamefont{D.~K.} \bibnamefont{Oi}}, \bibnamefont{and}
  \bibinfo{author}{\bibfnamefont{A.}~\bibnamefont{Ling}},
  \bibinfo{journal}{Physical Review Applied} \textbf{\bibinfo{volume}{5}},
  \bibinfo{pages}{054022} (\bibinfo{year}{2016}{\natexlab{a}}).

\bibitem[{\citenamefont{Tang et~al.}(2016{\natexlab{b}})\citenamefont{Tang,
  Chandrasekara, Tan, Cheng, Durak, and Ling}}]{TCTC+16}
\bibinfo{author}{\bibfnamefont{Z.}~\bibnamefont{Tang}},
  \bibinfo{author}{\bibfnamefont{R.}~\bibnamefont{Chandrasekara}},
  \bibinfo{author}{\bibfnamefont{Y.~C.} \bibnamefont{Tan}},
  \bibinfo{author}{\bibfnamefont{C.}~\bibnamefont{Cheng}},
  \bibinfo{author}{\bibfnamefont{K.}~\bibnamefont{Durak}}, \bibnamefont{and}
  \bibinfo{author}{\bibfnamefont{A.}~\bibnamefont{Ling}},
  \bibinfo{journal}{Scientific Reports} \textbf{\bibinfo{volume}{6}},
  \bibinfo{pages}{25603} (\bibinfo{year}{2016}{\natexlab{b}}).

\bibitem[{\citenamefont{Steinlechner et~al.}(2016)\citenamefont{Steinlechner,
  de~Vries, Fleischmann, Wille, Beckert, and Ursin}}]{SVF+16}
\bibinfo{author}{\bibfnamefont{F.}~\bibnamefont{Steinlechner}},
  \bibinfo{author}{\bibfnamefont{O.}~\bibnamefont{de~Vries}},
  \bibinfo{author}{\bibfnamefont{N.}~\bibnamefont{Fleischmann}},
  \bibinfo{author}{\bibfnamefont{E.}~\bibnamefont{Wille}},
  \bibinfo{author}{\bibfnamefont{E.}~\bibnamefont{Beckert}}, \bibnamefont{and}
  \bibinfo{author}{\bibfnamefont{R.}~\bibnamefont{Ursin}}, in
  \emph{\bibinfo{booktitle}{2016 Conference on Lasers and Electro-Optics
  (CLEO)}} (\bibinfo{organization}{IEEE}, \bibinfo{year}{2016}), pp.
  \bibinfo{pages}{1--2}.

\bibitem[{\citenamefont{Peng et~al.}(2005)\citenamefont{Peng, Yang, Bao, Zhang,
  Jin, Feng, Yang, Yang, Yin, Zhang et~al.}}]{PYB+05}
\bibinfo{author}{\bibfnamefont{C.-Z.} \bibnamefont{Peng}},
  \bibinfo{author}{\bibfnamefont{T.}~\bibnamefont{Yang}},
  \bibinfo{author}{\bibfnamefont{X.-H.} \bibnamefont{Bao}},
  \bibinfo{author}{\bibfnamefont{J.}~\bibnamefont{Zhang}},
  \bibinfo{author}{\bibfnamefont{X.-M.} \bibnamefont{Jin}},
  \bibinfo{author}{\bibfnamefont{F.-Y.} \bibnamefont{Feng}},
  \bibinfo{author}{\bibfnamefont{B.}~\bibnamefont{Yang}},
  \bibinfo{author}{\bibfnamefont{J.}~\bibnamefont{Yang}},
  \bibinfo{author}{\bibfnamefont{J.}~\bibnamefont{Yin}},
  \bibinfo{author}{\bibfnamefont{Q.}~\bibnamefont{Zhang}},
  \bibnamefont{et~al.}, \bibinfo{journal}{Physical Review Letters}
  \textbf{\bibinfo{volume}{94}}, \bibinfo{pages}{150501}
  (\bibinfo{year}{2005}).

\bibitem[{\citenamefont{Marcikic et~al.}(2006)\citenamefont{Marcikic,
  Lamas-Linares, and Kurtsiefer}}]{MLK+06}
\bibinfo{author}{\bibfnamefont{I.}~\bibnamefont{Marcikic}},
  \bibinfo{author}{\bibfnamefont{A.}~\bibnamefont{Lamas-Linares}},
  \bibnamefont{and}
  \bibinfo{author}{\bibfnamefont{C.}~\bibnamefont{Kurtsiefer}},
  \bibinfo{journal}{Applied Physics Letters} \textbf{\bibinfo{volume}{89}}
  (\bibinfo{year}{2006}).

\bibitem[{\citenamefont{Erven et~al.}(2008)\citenamefont{Erven, Couteau,
  Laflamme, and Weihs}}]{ECL+08}
\bibinfo{author}{\bibfnamefont{C.}~\bibnamefont{Erven}},
  \bibinfo{author}{\bibfnamefont{C.}~\bibnamefont{Couteau}},
  \bibinfo{author}{\bibfnamefont{R.}~\bibnamefont{Laflamme}}, \bibnamefont{and}
  \bibinfo{author}{\bibfnamefont{G.}~\bibnamefont{Weihs}},
  \bibinfo{journal}{Optics Express} \textbf{\bibinfo{volume}{16}},
  \bibinfo{pages}{16840} (\bibinfo{year}{2008}).

\bibitem[{\citenamefont{Scheidl et~al.}(2009)\citenamefont{Scheidl, Ursin,
  Fedrizzi, Ramelow, Ma, Herbst, Prevedel, Ratschbacher, Kofler, Jennewein
  et~al.}}]{SUF+09}
\bibinfo{author}{\bibfnamefont{T.}~\bibnamefont{Scheidl}},
  \bibinfo{author}{\bibfnamefont{R.}~\bibnamefont{Ursin}},
  \bibinfo{author}{\bibfnamefont{A.}~\bibnamefont{Fedrizzi}},
  \bibinfo{author}{\bibfnamefont{S.}~\bibnamefont{Ramelow}},
  \bibinfo{author}{\bibfnamefont{X.-S.} \bibnamefont{Ma}},
  \bibinfo{author}{\bibfnamefont{T.}~\bibnamefont{Herbst}},
  \bibinfo{author}{\bibfnamefont{R.}~\bibnamefont{Prevedel}},
  \bibinfo{author}{\bibfnamefont{L.}~\bibnamefont{Ratschbacher}},
  \bibinfo{author}{\bibfnamefont{J.}~\bibnamefont{Kofler}},
  \bibinfo{author}{\bibfnamefont{T.}~\bibnamefont{Jennewein}},
  \bibnamefont{et~al.}, \bibinfo{journal}{New Journal of Physics}
  \textbf{\bibinfo{volume}{11}}, \bibinfo{pages}{085002}
  (\bibinfo{year}{2009}).

\bibitem[{\citenamefont{Ecker et~al.}(2021)\citenamefont{Ecker, Liu,
  Handsteiner, Fink, Rauch, Steinlechner, Scheidl, Zeilinger, and
  Ursin}}]{ELH+21}
\bibinfo{author}{\bibfnamefont{S.}~\bibnamefont{Ecker}},
  \bibinfo{author}{\bibfnamefont{B.}~\bibnamefont{Liu}},
  \bibinfo{author}{\bibfnamefont{J.}~\bibnamefont{Handsteiner}},
  \bibinfo{author}{\bibfnamefont{M.}~\bibnamefont{Fink}},
  \bibinfo{author}{\bibfnamefont{D.}~\bibnamefont{Rauch}},
  \bibinfo{author}{\bibfnamefont{F.}~\bibnamefont{Steinlechner}},
  \bibinfo{author}{\bibfnamefont{T.}~\bibnamefont{Scheidl}},
  \bibinfo{author}{\bibfnamefont{A.}~\bibnamefont{Zeilinger}},
  \bibnamefont{and} \bibinfo{author}{\bibfnamefont{R.}~\bibnamefont{Ursin}},
  \bibinfo{journal}{npj Quantum Information} \textbf{\bibinfo{volume}{7}},
  \bibinfo{pages}{5} (\bibinfo{year}{2021}).

\bibitem[{\citenamefont{Schmitt-Manderbach
  et~al.}(2007)\citenamefont{Schmitt-Manderbach, Weier, F{\"u}rst, Ursin,
  Tiefenbacher, Scheidl, Perdigues, Sodnik, Kurtsiefer, Rarity
  et~al.}}]{MWF+07}
\bibinfo{author}{\bibfnamefont{T.}~\bibnamefont{Schmitt-Manderbach}},
  \bibinfo{author}{\bibfnamefont{H.}~\bibnamefont{Weier}},
  \bibinfo{author}{\bibfnamefont{M.}~\bibnamefont{F{\"u}rst}},
  \bibinfo{author}{\bibfnamefont{R.}~\bibnamefont{Ursin}},
  \bibinfo{author}{\bibfnamefont{F.}~\bibnamefont{Tiefenbacher}},
  \bibinfo{author}{\bibfnamefont{T.}~\bibnamefont{Scheidl}},
  \bibinfo{author}{\bibfnamefont{J.}~\bibnamefont{Perdigues}},
  \bibinfo{author}{\bibfnamefont{Z.}~\bibnamefont{Sodnik}},
  \bibinfo{author}{\bibfnamefont{C.}~\bibnamefont{Kurtsiefer}},
  \bibinfo{author}{\bibfnamefont{J.~G.} \bibnamefont{Rarity}},
  \bibnamefont{et~al.}, \bibinfo{journal}{Physical Review Letters}
  \textbf{\bibinfo{volume}{98}}, \bibinfo{pages}{010504}
  (\bibinfo{year}{2007}).

\bibitem[{\citenamefont{Dubey et~al.}(2024)\citenamefont{Dubey, Bhole, Dutta,
  Behera, Losu, Pandeeti, Metkar, Banerjee, and Pathak}}]{DBD+24}
\bibinfo{author}{\bibfnamefont{U.}~\bibnamefont{Dubey}},
  \bibinfo{author}{\bibfnamefont{P.}~\bibnamefont{Bhole}},
  \bibinfo{author}{\bibfnamefont{A.}~\bibnamefont{Dutta}},
  \bibinfo{author}{\bibfnamefont{D.~P.} \bibnamefont{Behera}},
  \bibinfo{author}{\bibfnamefont{V.}~\bibnamefont{Losu}},
  \bibinfo{author}{\bibfnamefont{G.~S.~D.} \bibnamefont{Pandeeti}},
  \bibinfo{author}{\bibfnamefont{A.~R.} \bibnamefont{Metkar}},
  \bibinfo{author}{\bibfnamefont{A.}~\bibnamefont{Banerjee}}, \bibnamefont{and}
  \bibinfo{author}{\bibfnamefont{A.}~\bibnamefont{Pathak}},
  \bibinfo{journal}{Physics Open} p. \bibinfo{pages}{100210}
  (\bibinfo{year}{2024}).

\bibitem[{\citenamefont{Liao et~al.}(2018)\citenamefont{Liao, Cai, Handsteiner,
  Liu, Yin, Zhang, Rauch, Fink, Ren, Liu et~al.}}]{LCH+18}
\bibinfo{author}{\bibfnamefont{S.-K.} \bibnamefont{Liao}},
  \bibinfo{author}{\bibfnamefont{W.-Q.} \bibnamefont{Cai}},
  \bibinfo{author}{\bibfnamefont{J.}~\bibnamefont{Handsteiner}},
  \bibinfo{author}{\bibfnamefont{B.}~\bibnamefont{Liu}},
  \bibinfo{author}{\bibfnamefont{J.}~\bibnamefont{Yin}},
  \bibinfo{author}{\bibfnamefont{L.}~\bibnamefont{Zhang}},
  \bibinfo{author}{\bibfnamefont{D.}~\bibnamefont{Rauch}},
  \bibinfo{author}{\bibfnamefont{M.}~\bibnamefont{Fink}},
  \bibinfo{author}{\bibfnamefont{J.-G.} \bibnamefont{Ren}},
  \bibinfo{author}{\bibfnamefont{W.-Y.} \bibnamefont{Liu}},
  \bibnamefont{et~al.}, \bibinfo{journal}{Physical Review Letters}
  \textbf{\bibinfo{volume}{120}}, \bibinfo{pages}{030501}
  (\bibinfo{year}{2018}).

\bibitem[{\citenamefont{Shenoy-Hejamadi
  et~al.}(2017)\citenamefont{Shenoy-Hejamadi, Pathak, and
  Radhakrishna}}]{SPR17}
\bibinfo{author}{\bibfnamefont{A.}~\bibnamefont{Shenoy-Hejamadi}},
  \bibinfo{author}{\bibfnamefont{A.}~\bibnamefont{Pathak}}, \bibnamefont{and}
  \bibinfo{author}{\bibfnamefont{S.}~\bibnamefont{Radhakrishna}},
  \bibinfo{journal}{Quanta} \textbf{\bibinfo{volume}{6}}, \bibinfo{pages}{1}
  (\bibinfo{year}{2017}).

\bibitem[{\citenamefont{Pirandola et~al.}(2020)\citenamefont{Pirandola,
  Andersen, Banchi, Berta, Bunandar, Colbeck, Englund, Gehring, Lupo, Ottaviani
  et~al.}}]{PAB+20}
\bibinfo{author}{\bibfnamefont{S.}~\bibnamefont{Pirandola}},
  \bibinfo{author}{\bibfnamefont{U.~L.} \bibnamefont{Andersen}},
  \bibinfo{author}{\bibfnamefont{L.}~\bibnamefont{Banchi}},
  \bibinfo{author}{\bibfnamefont{M.}~\bibnamefont{Berta}},
  \bibinfo{author}{\bibfnamefont{D.}~\bibnamefont{Bunandar}},
  \bibinfo{author}{\bibfnamefont{R.}~\bibnamefont{Colbeck}},
  \bibinfo{author}{\bibfnamefont{D.}~\bibnamefont{Englund}},
  \bibinfo{author}{\bibfnamefont{T.}~\bibnamefont{Gehring}},
  \bibinfo{author}{\bibfnamefont{C.}~\bibnamefont{Lupo}},
  \bibinfo{author}{\bibfnamefont{C.}~\bibnamefont{Ottaviani}},
  \bibnamefont{et~al.}, \bibinfo{journal}{Advances in Optics and Photonics}
  \textbf{\bibinfo{volume}{12}}, \bibinfo{pages}{1012} (\bibinfo{year}{2020}).

\bibitem[{\citenamefont{Lucamarini et~al.}(2009)\citenamefont{Lucamarini,
  Di~Giuseppe, and Tamaki}}]{LGT09}
\bibinfo{author}{\bibfnamefont{M.}~\bibnamefont{Lucamarini}},
  \bibinfo{author}{\bibfnamefont{G.}~\bibnamefont{Di~Giuseppe}},
  \bibnamefont{and} \bibinfo{author}{\bibfnamefont{K.}~\bibnamefont{Tamaki}},
  \bibinfo{journal}{Physical Review A} \textbf{\bibinfo{volume}{80}},
  \bibinfo{pages}{032327} (\bibinfo{year}{2009}).

\bibitem[{\citenamefont{Zhou et~al.}(2022)\citenamefont{Zhou, Yuan, Wang, Ling,
  Fu, Fang, Wang, Liu, Porfyrakis, Briggs et~al.}}]{ZYW+22}
\bibinfo{author}{\bibfnamefont{S.}~\bibnamefont{Zhou}},
  \bibinfo{author}{\bibfnamefont{J.}~\bibnamefont{Yuan}},
  \bibinfo{author}{\bibfnamefont{Z.-Y.} \bibnamefont{Wang}},
  \bibinfo{author}{\bibfnamefont{K.}~\bibnamefont{Ling}},
  \bibinfo{author}{\bibfnamefont{P.-X.} \bibnamefont{Fu}},
  \bibinfo{author}{\bibfnamefont{Y.-H.} \bibnamefont{Fang}},
  \bibinfo{author}{\bibfnamefont{Y.-X.} \bibnamefont{Wang}},
  \bibinfo{author}{\bibfnamefont{Z.}~\bibnamefont{Liu}},
  \bibinfo{author}{\bibfnamefont{K.}~\bibnamefont{Porfyrakis}},
  \bibinfo{author}{\bibfnamefont{G.~A.~D.} \bibnamefont{Briggs}},
  \bibnamefont{et~al.}, \bibinfo{journal}{Angewandte Chemie International
  Edition} \textbf{\bibinfo{volume}{61}}, \bibinfo{pages}{e202115263}
  (\bibinfo{year}{2022}).

\bibitem[{\citenamefont{Meier et~al.}(2023)\citenamefont{Meier, de~Melo,
  Lamsal, Bersano, Segura~Carrillo, Harter, Omanakuttan, Mitra, Deutsch,
  Boshier et~al.}}]{MML+23}
\bibinfo{author}{\bibfnamefont{E.}~\bibnamefont{Meier}},
  \bibinfo{author}{\bibfnamefont{L.}~\bibnamefont{de~Melo}},
  \bibinfo{author}{\bibfnamefont{H.}~\bibnamefont{Lamsal}},
  \bibinfo{author}{\bibfnamefont{T.}~\bibnamefont{Bersano}},
  \bibinfo{author}{\bibfnamefont{E.}~\bibnamefont{Segura~Carrillo}},
  \bibinfo{author}{\bibfnamefont{A.}~\bibnamefont{Harter}},
  \bibinfo{author}{\bibfnamefont{S.}~\bibnamefont{Omanakuttan}},
  \bibinfo{author}{\bibfnamefont{A.}~\bibnamefont{Mitra}},
  \bibinfo{author}{\bibfnamefont{I.}~\bibnamefont{Deutsch}},
  \bibinfo{author}{\bibfnamefont{M.}~\bibnamefont{Boshier}},
  \bibnamefont{et~al.}, \bibinfo{journal}{Bulletin of the American Physical
  Society}  (\bibinfo{year}{2023}).

\bibitem[{\citenamefont{Cozzolino et~al.}(2019)\citenamefont{Cozzolino, Da~Lio,
  Bacco, and Oxenl{\o}we}}]{CLB+19}
\bibinfo{author}{\bibfnamefont{D.}~\bibnamefont{Cozzolino}},
  \bibinfo{author}{\bibfnamefont{B.}~\bibnamefont{Da~Lio}},
  \bibinfo{author}{\bibfnamefont{D.}~\bibnamefont{Bacco}}, \bibnamefont{and}
  \bibinfo{author}{\bibfnamefont{L.~K.} \bibnamefont{Oxenl{\o}we}},
  \bibinfo{journal}{Advanced Quantum Technologies}
  \textbf{\bibinfo{volume}{2}}, \bibinfo{pages}{1900038}
  (\bibinfo{year}{2019}).

\bibitem[{\citenamefont{Malpani et~al.}(2019)\citenamefont{Malpani, Alam,
  Thapliyal, Pathak, Narayanan, and Banerjee}}]{PAT+19}
\bibinfo{author}{\bibfnamefont{P.}~\bibnamefont{Malpani}},
  \bibinfo{author}{\bibfnamefont{N.}~\bibnamefont{Alam}},
  \bibinfo{author}{\bibfnamefont{K.}~\bibnamefont{Thapliyal}},
  \bibinfo{author}{\bibfnamefont{A.}~\bibnamefont{Pathak}},
  \bibinfo{author}{\bibfnamefont{V.}~\bibnamefont{Narayanan}},
  \bibnamefont{and} \bibinfo{author}{\bibfnamefont{S.}~\bibnamefont{Banerjee}},
  \bibinfo{journal}{Annalen der Physik} \textbf{\bibinfo{volume}{531}},
  \bibinfo{pages}{1800318} (\bibinfo{year}{2019}).

\bibitem[{\citenamefont{Bechmann-Pasquinucci and Tittel}(2000)}]{PT200}
\bibinfo{author}{\bibfnamefont{H.}~\bibnamefont{Bechmann-Pasquinucci}}
  \bibnamefont{and} \bibinfo{author}{\bibfnamefont{W.}~\bibnamefont{Tittel}},
  \bibinfo{journal}{Physical Review A} \textbf{\bibinfo{volume}{61}},
  \bibinfo{pages}{062308} (\bibinfo{year}{2000}).

\bibitem[{\citenamefont{Sheridan and Scarani}(2010)}]{SS10}
\bibinfo{author}{\bibfnamefont{L.}~\bibnamefont{Sheridan}} \bibnamefont{and}
  \bibinfo{author}{\bibfnamefont{V.}~\bibnamefont{Scarani}},
  \bibinfo{journal}{Physical Review A} \textbf{\bibinfo{volume}{82}},
  \bibinfo{pages}{030301} (\bibinfo{year}{2010}).

\bibitem[{\citenamefont{Vlachou et~al.}(2018)\citenamefont{Vlachou, Krawec,
  Mateus, Paunkovi{\'c}, and Souto}}]{VKM+18}
\bibinfo{author}{\bibfnamefont{C.}~\bibnamefont{Vlachou}},
  \bibinfo{author}{\bibfnamefont{W.}~\bibnamefont{Krawec}},
  \bibinfo{author}{\bibfnamefont{P.}~\bibnamefont{Mateus}},
  \bibinfo{author}{\bibfnamefont{N.}~\bibnamefont{Paunkovi{\'c}}},
  \bibnamefont{and} \bibinfo{author}{\bibfnamefont{A.}~\bibnamefont{Souto}},
  \bibinfo{journal}{Quantum Information Processing}
  \textbf{\bibinfo{volume}{17}}, \bibinfo{pages}{1} (\bibinfo{year}{2018}).

\bibitem[{\citenamefont{Sharma et~al.}(2016)\citenamefont{Sharma, Thapliyal,
  Pathak, and Banerjee}}]{STP+16}
\bibinfo{author}{\bibfnamefont{V.}~\bibnamefont{Sharma}},
  \bibinfo{author}{\bibfnamefont{K.}~\bibnamefont{Thapliyal}},
  \bibinfo{author}{\bibfnamefont{A.}~\bibnamefont{Pathak}}, \bibnamefont{and}
  \bibinfo{author}{\bibfnamefont{S.}~\bibnamefont{Banerjee}},
  \bibinfo{journal}{Quantum Information Processing}
  \textbf{\bibinfo{volume}{15}}, \bibinfo{pages}{4681} (\bibinfo{year}{2016}).

\bibitem[{\citenamefont{Iqbal and Krawec}(2021)}]{IK21}
\bibinfo{author}{\bibfnamefont{H.}~\bibnamefont{Iqbal}} \bibnamefont{and}
  \bibinfo{author}{\bibfnamefont{W.~O.} \bibnamefont{Krawec}},
  \bibinfo{journal}{Quantum Information Processing}
  \textbf{\bibinfo{volume}{20}}, \bibinfo{pages}{344} (\bibinfo{year}{2021}).

\bibitem[{\citenamefont{Vasylyev et~al.}(2016)\citenamefont{Vasylyev, Semenov,
  and Vogel}}]{VSV16}
\bibinfo{author}{\bibfnamefont{D.}~\bibnamefont{Vasylyev}},
  \bibinfo{author}{\bibfnamefont{A.}~\bibnamefont{Semenov}}, \bibnamefont{and}
  \bibinfo{author}{\bibfnamefont{W.}~\bibnamefont{Vogel}},
  \bibinfo{journal}{Physical Review Letters} \textbf{\bibinfo{volume}{117}},
  \bibinfo{pages}{090501} (\bibinfo{year}{2016}).

\bibitem[{\citenamefont{Vasylyev et~al.}(2017)\citenamefont{Vasylyev, Semenov,
  Vogel, G{\"u}nthner, Thurn, Bayraktar, and Marquardt}}]{VSV+17}
\bibinfo{author}{\bibfnamefont{D.}~\bibnamefont{Vasylyev}},
  \bibinfo{author}{\bibfnamefont{A.}~\bibnamefont{Semenov}},
  \bibinfo{author}{\bibfnamefont{W.}~\bibnamefont{Vogel}},
  \bibinfo{author}{\bibfnamefont{K.}~\bibnamefont{G{\"u}nthner}},
  \bibinfo{author}{\bibfnamefont{A.}~\bibnamefont{Thurn}},
  \bibinfo{author}{\bibfnamefont{{\"O}.}~\bibnamefont{Bayraktar}},
  \bibnamefont{and}
  \bibinfo{author}{\bibfnamefont{C.}~\bibnamefont{Marquardt}},
  \bibinfo{journal}{Physical Review A} \textbf{\bibinfo{volume}{96}},
  \bibinfo{pages}{043856} (\bibinfo{year}{2017}).

\bibitem[{\citenamefont{Liorni et~al.}(2019)\citenamefont{Liorni, Kampermann,
  and Bru{\ss}}}]{LKB19}
\bibinfo{author}{\bibfnamefont{C.}~\bibnamefont{Liorni}},
  \bibinfo{author}{\bibfnamefont{H.}~\bibnamefont{Kampermann}},
  \bibnamefont{and} \bibinfo{author}{\bibfnamefont{D.}~\bibnamefont{Bru{\ss}}},
  \bibinfo{journal}{New Journal of Physics} \textbf{\bibinfo{volume}{21}},
  \bibinfo{pages}{093055} (\bibinfo{year}{2019}).

\bibitem[{\citenamefont{Bourgoin et~al.}(2013)\citenamefont{Bourgoin,
  Meyer-Scott, Higgins, Helou, Erven, Huebel, Kumar, Hudson, D'Souza, Girard
  et~al.}}]{BSH+13}
\bibinfo{author}{\bibfnamefont{J.}~\bibnamefont{Bourgoin}},
  \bibinfo{author}{\bibfnamefont{E.}~\bibnamefont{Meyer-Scott}},
  \bibinfo{author}{\bibfnamefont{B.~L.} \bibnamefont{Higgins}},
  \bibinfo{author}{\bibfnamefont{B.}~\bibnamefont{Helou}},
  \bibinfo{author}{\bibfnamefont{C.}~\bibnamefont{Erven}},
  \bibinfo{author}{\bibfnamefont{H.}~\bibnamefont{Huebel}},
  \bibinfo{author}{\bibfnamefont{B.}~\bibnamefont{Kumar}},
  \bibinfo{author}{\bibfnamefont{D.}~\bibnamefont{Hudson}},
  \bibinfo{author}{\bibfnamefont{I.}~\bibnamefont{D'Souza}},
  \bibinfo{author}{\bibfnamefont{R.}~\bibnamefont{Girard}},
  \bibnamefont{et~al.}, \bibinfo{journal}{New Journal of Physics}
  \textbf{\bibinfo{volume}{15}}, \bibinfo{pages}{023006}
  (\bibinfo{year}{2013}).

\bibitem[{\citenamefont{Tamaki et~al.}(2003)\citenamefont{Tamaki, Koashi, and
  Imoto}}]{TKI03}
\bibinfo{author}{\bibfnamefont{K.}~\bibnamefont{Tamaki}},
  \bibinfo{author}{\bibfnamefont{M.}~\bibnamefont{Koashi}}, \bibnamefont{and}
  \bibinfo{author}{\bibfnamefont{N.}~\bibnamefont{Imoto}},
  \bibinfo{journal}{Physical Review Letters} \textbf{\bibinfo{volume}{90}},
  \bibinfo{pages}{167904} (\bibinfo{year}{2003}).

\bibitem[{\citenamefont{Matsumoto}(2013)}]{M13}
\bibinfo{author}{\bibfnamefont{R.}~\bibnamefont{Matsumoto}}, in
  \emph{\bibinfo{booktitle}{2013 IEEE International Symposium on Information
  Theory}} (\bibinfo{organization}{IEEE}, \bibinfo{year}{2013}), pp.
  \bibinfo{pages}{351--353}.

\bibitem[{\citenamefont{Amer and Krawec}(2020)}]{AK20}
\bibinfo{author}{\bibfnamefont{O.}~\bibnamefont{Amer}} \bibnamefont{and}
  \bibinfo{author}{\bibfnamefont{W.~O.} \bibnamefont{Krawec}}, in
  \emph{\bibinfo{booktitle}{2020 IEEE International Symposium on Information
  Theory (ISIT)}} (\bibinfo{organization}{IEEE}, \bibinfo{year}{2020}), pp.
  \bibinfo{pages}{1944--1948}.

\bibitem[{\citenamefont{Hu et~al.}(2020{\natexlab{a}})\citenamefont{Hu, Zhang,
  Liu, Cai, Ye, Guo, Xing, Huang, Huang, Li et~al.}}]{HZL+20}
\bibinfo{author}{\bibfnamefont{X.-M.} \bibnamefont{Hu}},
  \bibinfo{author}{\bibfnamefont{C.}~\bibnamefont{Zhang}},
  \bibinfo{author}{\bibfnamefont{B.-H.} \bibnamefont{Liu}},
  \bibinfo{author}{\bibfnamefont{Y.}~\bibnamefont{Cai}},
  \bibinfo{author}{\bibfnamefont{X.-J.} \bibnamefont{Ye}},
  \bibinfo{author}{\bibfnamefont{Y.}~\bibnamefont{Guo}},
  \bibinfo{author}{\bibfnamefont{W.-B.} \bibnamefont{Xing}},
  \bibinfo{author}{\bibfnamefont{C.-X.} \bibnamefont{Huang}},
  \bibinfo{author}{\bibfnamefont{Y.-F.} \bibnamefont{Huang}},
  \bibinfo{author}{\bibfnamefont{C.-F.} \bibnamefont{Li}},
  \bibnamefont{et~al.}, \bibinfo{journal}{Physical Review Letters}
  \textbf{\bibinfo{volume}{125}}, \bibinfo{pages}{230501}
  (\bibinfo{year}{2020}{\natexlab{a}}).

\bibitem[{\citenamefont{Bouwmeester et~al.}(1997)\citenamefont{Bouwmeester,
  Pan, Mattle, Eibl, Weinfurter, and Zeilinger}}]{BPM+97}
\bibinfo{author}{\bibfnamefont{D.}~\bibnamefont{Bouwmeester}},
  \bibinfo{author}{\bibfnamefont{J.-W.} \bibnamefont{Pan}},
  \bibinfo{author}{\bibfnamefont{K.}~\bibnamefont{Mattle}},
  \bibinfo{author}{\bibfnamefont{M.}~\bibnamefont{Eibl}},
  \bibinfo{author}{\bibfnamefont{H.}~\bibnamefont{Weinfurter}},
  \bibnamefont{and}
  \bibinfo{author}{\bibfnamefont{A.}~\bibnamefont{Zeilinger}},
  \bibinfo{journal}{Nature} \textbf{\bibinfo{volume}{390}},
  \bibinfo{pages}{575} (\bibinfo{year}{1997}).

\bibitem[{\citenamefont{Fattal et~al.}(2004)\citenamefont{Fattal, Diamanti,
  Inoue, and Yamamoto}}]{FDI+04}
\bibinfo{author}{\bibfnamefont{D.}~\bibnamefont{Fattal}},
  \bibinfo{author}{\bibfnamefont{E.}~\bibnamefont{Diamanti}},
  \bibinfo{author}{\bibfnamefont{K.}~\bibnamefont{Inoue}}, \bibnamefont{and}
  \bibinfo{author}{\bibfnamefont{Y.}~\bibnamefont{Yamamoto}},
  \bibinfo{journal}{Physical Review Letters} \textbf{\bibinfo{volume}{92}},
  \bibinfo{pages}{037904} (\bibinfo{year}{2004}).

\bibitem[{\citenamefont{Olmschenk et~al.}(2009)\citenamefont{Olmschenk,
  Matsukevich, Maunz, Hayes, Duan, and Monroe}}]{OMM+09}
\bibinfo{author}{\bibfnamefont{S.}~\bibnamefont{Olmschenk}},
  \bibinfo{author}{\bibfnamefont{D.}~\bibnamefont{Matsukevich}},
  \bibinfo{author}{\bibfnamefont{P.}~\bibnamefont{Maunz}},
  \bibinfo{author}{\bibfnamefont{D.}~\bibnamefont{Hayes}},
  \bibinfo{author}{\bibfnamefont{L.-M.} \bibnamefont{Duan}}, \bibnamefont{and}
  \bibinfo{author}{\bibfnamefont{C.}~\bibnamefont{Monroe}},
  \bibinfo{journal}{Science} \textbf{\bibinfo{volume}{323}},
  \bibinfo{pages}{486} (\bibinfo{year}{2009}).

\bibitem[{\citenamefont{Furusawa et~al.}(1998)\citenamefont{Furusawa,
  S{\o}rensen, Braunstein, Fuchs, Kimble, and Polzik}}]{FSB+98}
\bibinfo{author}{\bibfnamefont{A.}~\bibnamefont{Furusawa}},
  \bibinfo{author}{\bibfnamefont{J.~L.} \bibnamefont{S{\o}rensen}},
  \bibinfo{author}{\bibfnamefont{S.~L.} \bibnamefont{Braunstein}},
  \bibinfo{author}{\bibfnamefont{C.~A.} \bibnamefont{Fuchs}},
  \bibinfo{author}{\bibfnamefont{H.~J.} \bibnamefont{Kimble}},
  \bibnamefont{and} \bibinfo{author}{\bibfnamefont{E.~S.}
  \bibnamefont{Polzik}}, \bibinfo{journal}{science}
  \textbf{\bibinfo{volume}{282}}, \bibinfo{pages}{706} (\bibinfo{year}{1998}).

\bibitem[{\citenamefont{Yonezawa et~al.}(2007)\citenamefont{Yonezawa,
  Braunstein, and Furusawa}}]{YBF07}
\bibinfo{author}{\bibfnamefont{H.}~\bibnamefont{Yonezawa}},
  \bibinfo{author}{\bibfnamefont{S.~L.} \bibnamefont{Braunstein}},
  \bibnamefont{and} \bibinfo{author}{\bibfnamefont{A.}~\bibnamefont{Furusawa}},
  \bibinfo{journal}{Physical Review Letters} \textbf{\bibinfo{volume}{99}},
  \bibinfo{pages}{110503} (\bibinfo{year}{2007}).

\bibitem[{\citenamefont{Lee et~al.}(2011)\citenamefont{Lee, Benichi, Takeno,
  Takeda, Webb, Huntington, and Furusawa}}]{LBT+11}
\bibinfo{author}{\bibfnamefont{N.}~\bibnamefont{Lee}},
  \bibinfo{author}{\bibfnamefont{H.}~\bibnamefont{Benichi}},
  \bibinfo{author}{\bibfnamefont{Y.}~\bibnamefont{Takeno}},
  \bibinfo{author}{\bibfnamefont{S.}~\bibnamefont{Takeda}},
  \bibinfo{author}{\bibfnamefont{J.}~\bibnamefont{Webb}},
  \bibinfo{author}{\bibfnamefont{E.}~\bibnamefont{Huntington}},
  \bibnamefont{and} \bibinfo{author}{\bibfnamefont{A.}~\bibnamefont{Furusawa}},
  \bibinfo{journal}{Science} \textbf{\bibinfo{volume}{332}},
  \bibinfo{pages}{330} (\bibinfo{year}{2011}).

\bibitem[{\citenamefont{Dada et~al.}(2011)\citenamefont{Dada, Leach, Buller,
  Padgett, and Andersson}}]{DLB+11}
\bibinfo{author}{\bibfnamefont{A.~C.} \bibnamefont{Dada}},
  \bibinfo{author}{\bibfnamefont{J.}~\bibnamefont{Leach}},
  \bibinfo{author}{\bibfnamefont{G.~S.} \bibnamefont{Buller}},
  \bibinfo{author}{\bibfnamefont{M.~J.} \bibnamefont{Padgett}},
  \bibnamefont{and}
  \bibinfo{author}{\bibfnamefont{E.}~\bibnamefont{Andersson}},
  \bibinfo{journal}{Nature Physics} \textbf{\bibinfo{volume}{7}},
  \bibinfo{pages}{677} (\bibinfo{year}{2011}).

\bibitem[{\citenamefont{Martin et~al.}(2017)\citenamefont{Martin, Guerreiro,
  Tiranov, Designolle, Fr{\"o}wis, Brunner, Huber, and Gisin}}]{MGT+17}
\bibinfo{author}{\bibfnamefont{A.}~\bibnamefont{Martin}},
  \bibinfo{author}{\bibfnamefont{T.}~\bibnamefont{Guerreiro}},
  \bibinfo{author}{\bibfnamefont{A.}~\bibnamefont{Tiranov}},
  \bibinfo{author}{\bibfnamefont{S.}~\bibnamefont{Designolle}},
  \bibinfo{author}{\bibfnamefont{F.}~\bibnamefont{Fr{\"o}wis}},
  \bibinfo{author}{\bibfnamefont{N.}~\bibnamefont{Brunner}},
  \bibinfo{author}{\bibfnamefont{M.}~\bibnamefont{Huber}}, \bibnamefont{and}
  \bibinfo{author}{\bibfnamefont{N.}~\bibnamefont{Gisin}},
  \bibinfo{journal}{Physical Review Letters} \textbf{\bibinfo{volume}{118}},
  \bibinfo{pages}{110501} (\bibinfo{year}{2017}).

\bibitem[{\citenamefont{Kues et~al.}(2017)\citenamefont{Kues, Reimer, Roztocki,
  Cort{\'e}s, Sciara, Wetzel, Zhang, Cino, Chu, Little et~al.}}]{KRR+17}
\bibinfo{author}{\bibfnamefont{M.}~\bibnamefont{Kues}},
  \bibinfo{author}{\bibfnamefont{C.}~\bibnamefont{Reimer}},
  \bibinfo{author}{\bibfnamefont{P.}~\bibnamefont{Roztocki}},
  \bibinfo{author}{\bibfnamefont{L.~R.} \bibnamefont{Cort{\'e}s}},
  \bibinfo{author}{\bibfnamefont{S.}~\bibnamefont{Sciara}},
  \bibinfo{author}{\bibfnamefont{B.}~\bibnamefont{Wetzel}},
  \bibinfo{author}{\bibfnamefont{Y.}~\bibnamefont{Zhang}},
  \bibinfo{author}{\bibfnamefont{A.}~\bibnamefont{Cino}},
  \bibinfo{author}{\bibfnamefont{S.~T.} \bibnamefont{Chu}},
  \bibinfo{author}{\bibfnamefont{B.~E.} \bibnamefont{Little}},
  \bibnamefont{et~al.}, \bibinfo{journal}{Nature}
  \textbf{\bibinfo{volume}{546}}, \bibinfo{pages}{622} (\bibinfo{year}{2017}).

\bibitem[{\citenamefont{Hu et~al.}(2020{\natexlab{b}})\citenamefont{Hu, Xing,
  Liu, Huang, Li, Guo, Erker, and Huber}}]{HXL+20}
\bibinfo{author}{\bibfnamefont{X.-M.} \bibnamefont{Hu}},
  \bibinfo{author}{\bibfnamefont{W.-B.} \bibnamefont{Xing}},
  \bibinfo{author}{\bibfnamefont{B.-H.} \bibnamefont{Liu}},
  \bibinfo{author}{\bibfnamefont{Y.-F.} \bibnamefont{Huang}},
  \bibinfo{author}{\bibfnamefont{C.-F.} \bibnamefont{Li}},
  \bibinfo{author}{\bibfnamefont{G.-C.} \bibnamefont{Guo}},
  \bibinfo{author}{\bibfnamefont{P.}~\bibnamefont{Erker}}, \bibnamefont{and}
  \bibinfo{author}{\bibfnamefont{M.}~\bibnamefont{Huber}},
  \bibinfo{journal}{Physical Review Letters} \textbf{\bibinfo{volume}{125}},
  \bibinfo{pages}{090503} (\bibinfo{year}{2020}{\natexlab{b}}).

\bibitem[{\citenamefont{Valencia et~al.}(2020)\citenamefont{Valencia,
  Srivastav, Pivoluska, Huber, Friis, McCutcheon, and Malik}}]{VSP+20}
\bibinfo{author}{\bibfnamefont{N.~H.} \bibnamefont{Valencia}},
  \bibinfo{author}{\bibfnamefont{V.}~\bibnamefont{Srivastav}},
  \bibinfo{author}{\bibfnamefont{M.}~\bibnamefont{Pivoluska}},
  \bibinfo{author}{\bibfnamefont{M.}~\bibnamefont{Huber}},
  \bibinfo{author}{\bibfnamefont{N.}~\bibnamefont{Friis}},
  \bibinfo{author}{\bibfnamefont{W.}~\bibnamefont{McCutcheon}},
  \bibnamefont{and} \bibinfo{author}{\bibfnamefont{M.}~\bibnamefont{Malik}},
  \bibinfo{journal}{Quantum} \textbf{\bibinfo{volume}{4}}, \bibinfo{pages}{376}
  (\bibinfo{year}{2020}).

\bibitem[{\citenamefont{Lu and Pan}(2021)}]{LP21}
\bibinfo{author}{\bibfnamefont{C.-Y.} \bibnamefont{Lu}} \bibnamefont{and}
  \bibinfo{author}{\bibfnamefont{J.-W.} \bibnamefont{Pan}},
  \bibinfo{journal}{Nature Nanotechnology} \textbf{\bibinfo{volume}{16}},
  \bibinfo{pages}{1294} (\bibinfo{year}{2021}).

\bibitem[{\citenamefont{Thomas and Senellart}(2021)}]{TS21}
\bibinfo{author}{\bibfnamefont{S.}~\bibnamefont{Thomas}} \bibnamefont{and}
  \bibinfo{author}{\bibfnamefont{P.}~\bibnamefont{Senellart}},
  \bibinfo{journal}{Nature Nanotechnology} \textbf{\bibinfo{volume}{16}},
  \bibinfo{pages}{367} (\bibinfo{year}{2021}).

\bibitem[{\citenamefont{Hwang}(2003)}]{H03}
\bibinfo{author}{\bibfnamefont{W.-Y.} \bibnamefont{Hwang}},
  \bibinfo{journal}{Physical Review Letters} \textbf{\bibinfo{volume}{91}},
  \bibinfo{pages}{057901} (\bibinfo{year}{2003}).

\bibitem[{\citenamefont{Lo et~al.}(2005)\citenamefont{Lo, Ma, and
  Chen}}]{LMC05}
\bibinfo{author}{\bibfnamefont{H.-K.} \bibnamefont{Lo}},
  \bibinfo{author}{\bibfnamefont{X.}~\bibnamefont{Ma}}, \bibnamefont{and}
  \bibinfo{author}{\bibfnamefont{K.}~\bibnamefont{Chen}},
  \bibinfo{journal}{Physical Review Letters} \textbf{\bibinfo{volume}{94}},
  \bibinfo{pages}{230504} (\bibinfo{year}{2005}).

\bibitem[{\citenamefont{Zhou et~al.}(2018)\citenamefont{Zhou, Huang, and
  Guo}}]{ZHG18}
\bibinfo{author}{\bibfnamefont{J.}~\bibnamefont{Zhou}},
  \bibinfo{author}{\bibfnamefont{D.}~\bibnamefont{Huang}}, \bibnamefont{and}
  \bibinfo{author}{\bibfnamefont{Y.}~\bibnamefont{Guo}},
  \bibinfo{journal}{Physical Review A} \textbf{\bibinfo{volume}{98}},
  \bibinfo{pages}{042303} (\bibinfo{year}{2018}).

\bibitem[{\citenamefont{Zhou et~al.}(2023{\natexlab{b}})\citenamefont{Zhou, Wu,
  Feng, Li, Shi, and Shi}}]{ZWF+23}
\bibinfo{author}{\bibfnamefont{J.}~\bibnamefont{Zhou}},
  \bibinfo{author}{\bibfnamefont{L.}~\bibnamefont{Wu}},
  \bibinfo{author}{\bibfnamefont{Y.}~\bibnamefont{Feng}},
  \bibinfo{author}{\bibfnamefont{H.}~\bibnamefont{Li}},
  \bibinfo{author}{\bibfnamefont{J.}~\bibnamefont{Shi}}, \bibnamefont{and}
  \bibinfo{author}{\bibfnamefont{R.}~\bibnamefont{Shi}},
  \bibinfo{journal}{Quantum Information Processing}
  \textbf{\bibinfo{volume}{22}}, \bibinfo{pages}{356}
  (\bibinfo{year}{2023}{\natexlab{b}}).

\bibitem[{\citenamefont{Chen et~al.}(2021{\natexlab{b}})\citenamefont{Chen,
  Deng, Gan, Chen, and Islam}}]{CDG+21}
\bibinfo{author}{\bibfnamefont{C.-M.} \bibnamefont{Chen}},
  \bibinfo{author}{\bibfnamefont{X.}~\bibnamefont{Deng}},
  \bibinfo{author}{\bibfnamefont{W.}~\bibnamefont{Gan}},
  \bibinfo{author}{\bibfnamefont{J.}~\bibnamefont{Chen}}, \bibnamefont{and}
  \bibinfo{author}{\bibfnamefont{S.~H.} \bibnamefont{Islam}},
  \bibinfo{journal}{The Journal of Supercomputing}
  \textbf{\bibinfo{volume}{77}}, \bibinfo{pages}{9046}
  (\bibinfo{year}{2021}{\natexlab{b}}).

\bibitem[{\citenamefont{Feng et~al.}(2021)\citenamefont{Feng, Lv, Liu, and
  Zhang}}]{FLL+21}
\bibinfo{author}{\bibfnamefont{B.}~\bibnamefont{Feng}},
  \bibinfo{author}{\bibfnamefont{C.}~\bibnamefont{Lv}},
  \bibinfo{author}{\bibfnamefont{J.}~\bibnamefont{Liu}}, \bibnamefont{and}
  \bibinfo{author}{\bibfnamefont{T.}~\bibnamefont{Zhang}}, in
  \emph{\bibinfo{booktitle}{Journal of Physics: Conference Series}}
  (\bibinfo{organization}{IOP Publishing}, \bibinfo{year}{2021}), vol.
  \bibinfo{volume}{1757}, p. \bibinfo{pages}{012111}.

\bibitem[{\citenamefont{Bai et~al.}(2021)\citenamefont{Bai, Zhang, Jiang, Wu,
  Shi, Lin, and Pei}}]{BZJ+21}
\bibinfo{author}{\bibfnamefont{E.}~\bibnamefont{Bai}},
  \bibinfo{author}{\bibfnamefont{Y.}~\bibnamefont{Zhang}},
  \bibinfo{author}{\bibfnamefont{X.}~\bibnamefont{Jiang}},
  \bibinfo{author}{\bibfnamefont{Y.}~\bibnamefont{Wu}},
  \bibinfo{author}{\bibfnamefont{Z.}~\bibnamefont{Shi}},
  \bibinfo{author}{\bibfnamefont{X.}~\bibnamefont{Lin}}, \bibnamefont{and}
  \bibinfo{author}{\bibfnamefont{Z.}~\bibnamefont{Pei}}, in
  \emph{\bibinfo{booktitle}{2021 IEEE International Conference on Power
  Electronics, Computer Applications (ICPECA)}} (\bibinfo{organization}{IEEE},
  \bibinfo{year}{2021}), pp. \bibinfo{pages}{571--574}.

\bibitem[{\citenamefont{Zhang and Mao}(2020)}]{ZM20}
\bibinfo{author}{\bibfnamefont{P.}~\bibnamefont{Zhang}} \bibnamefont{and}
  \bibinfo{author}{\bibfnamefont{X.}~\bibnamefont{Mao}}, in
  \emph{\bibinfo{booktitle}{Journal of Physics: Conference Series}}
  (\bibinfo{organization}{IOP Publishing}, \bibinfo{year}{2020}), vol.
  \bibinfo{volume}{1621}, p. \bibinfo{pages}{012017}.

\bibitem[{\citenamefont{Devetak and Winter}(2005)}]{DW05}
\bibinfo{author}{\bibfnamefont{I.}~\bibnamefont{Devetak}} \bibnamefont{and}
  \bibinfo{author}{\bibfnamefont{A.}~\bibnamefont{Winter}},
  \bibinfo{journal}{Proceedings of The Royal Society A: Mathematical, Physical
  and Engineering Sciences} \textbf{\bibinfo{volume}{461}},
  \bibinfo{pages}{207} (\bibinfo{year}{2005}).

\bibitem[{\citenamefont{Renner et~al.}(2005)\citenamefont{Renner, Gisin, and
  Kraus}}]{RGK_05}
\bibinfo{author}{\bibfnamefont{R.}~\bibnamefont{Renner}},
  \bibinfo{author}{\bibfnamefont{N.}~\bibnamefont{Gisin}}, \bibnamefont{and}
  \bibinfo{author}{\bibfnamefont{B.}~\bibnamefont{Kraus}},
  \bibinfo{journal}{Physical Review A} \textbf{\bibinfo{volume}{72}},
  \bibinfo{pages}{012332} (\bibinfo{year}{2005}).

\bibitem[{\citenamefont{Krawec}(2016)}]{K16}
\bibinfo{author}{\bibfnamefont{W.~O.} \bibnamefont{Krawec}},
  \bibinfo{journal}{arXiv preprint arXiv:1608.07728}  (\bibinfo{year}{2016}).

\bibitem[{\citenamefont{Cerf et~al.}(2002)\citenamefont{Cerf, Bourennane,
  Karlsson, and Gisin}}]{CBK+02}
\bibinfo{author}{\bibfnamefont{N.~J.} \bibnamefont{Cerf}},
  \bibinfo{author}{\bibfnamefont{M.}~\bibnamefont{Bourennane}},
  \bibinfo{author}{\bibfnamefont{A.}~\bibnamefont{Karlsson}}, \bibnamefont{and}
  \bibinfo{author}{\bibfnamefont{N.}~\bibnamefont{Gisin}},
  \bibinfo{journal}{Physical Review Letters} \textbf{\bibinfo{volume}{88}},
  \bibinfo{pages}{127902} (\bibinfo{year}{2002}).

\bibitem[{\citenamefont{Berta et~al.}(2010)\citenamefont{Berta, Christandl,
  Colbeck, Renes, and Renner}}]{BCC+10}
\bibinfo{author}{\bibfnamefont{M.}~\bibnamefont{Berta}},
  \bibinfo{author}{\bibfnamefont{M.}~\bibnamefont{Christandl}},
  \bibinfo{author}{\bibfnamefont{R.}~\bibnamefont{Colbeck}},
  \bibinfo{author}{\bibfnamefont{J.~M.} \bibnamefont{Renes}}, \bibnamefont{and}
  \bibinfo{author}{\bibfnamefont{R.}~\bibnamefont{Renner}},
  \bibinfo{journal}{Nature Physics} \textbf{\bibinfo{volume}{6}},
  \bibinfo{pages}{659} (\bibinfo{year}{2010}).

\bibitem[{\citenamefont{Kraus}(1987)}]{K87}
\bibinfo{author}{\bibfnamefont{K.}~\bibnamefont{Kraus}},
  \bibinfo{journal}{Physical Review D} \textbf{\bibinfo{volume}{35}},
  \bibinfo{pages}{3070} (\bibinfo{year}{1987}).

\bibitem[{\citenamefont{Maassen and Uffink}(1988)}]{MU88}
\bibinfo{author}{\bibfnamefont{H.}~\bibnamefont{Maassen}} \bibnamefont{and}
  \bibinfo{author}{\bibfnamefont{J.~B.} \bibnamefont{Uffink}},
  \bibinfo{journal}{Physical Review Letters} \textbf{\bibinfo{volume}{60}},
  \bibinfo{pages}{1103} (\bibinfo{year}{1988}).

\bibitem[{\citenamefont{Christandl et~al.}(2004)\citenamefont{Christandl,
  Renner, and Ekert}}]{CRE_04}
\bibinfo{author}{\bibfnamefont{M.}~\bibnamefont{Christandl}},
  \bibinfo{author}{\bibfnamefont{R.}~\bibnamefont{Renner}}, \bibnamefont{and}
  \bibinfo{author}{\bibfnamefont{A.}~\bibnamefont{Ekert}},
  \bibinfo{journal}{arXiv preprint quant-ph/0402131}  (\bibinfo{year}{2004}).

\bibitem[{\citenamefont{Capmany et~al.}(2009)\citenamefont{Capmany,
  Ortigosa-Blanch, Mora, Ruiz-Alba, Amaya, and Martinez}}]{BMA+09}
\bibinfo{author}{\bibfnamefont{J.}~\bibnamefont{Capmany}},
  \bibinfo{author}{\bibfnamefont{A.}~\bibnamefont{Ortigosa-Blanch}},
  \bibinfo{author}{\bibfnamefont{J.}~\bibnamefont{Mora}},
  \bibinfo{author}{\bibfnamefont{A.}~\bibnamefont{Ruiz-Alba}},
  \bibinfo{author}{\bibfnamefont{W.}~\bibnamefont{Amaya}}, \bibnamefont{and}
  \bibinfo{author}{\bibfnamefont{A.}~\bibnamefont{Martinez}},
  \bibinfo{journal}{IEEE Journal of Selected Topics in Quantum Electronics}
  \textbf{\bibinfo{volume}{15}}, \bibinfo{pages}{1607} (\bibinfo{year}{2009}).

\bibitem[{\citenamefont{Barnett et~al.}(1993)\citenamefont{Barnett, Huttner,
  and Phoenix}}]{BHP93}
\bibinfo{author}{\bibfnamefont{S.~M.} \bibnamefont{Barnett}},
  \bibinfo{author}{\bibfnamefont{B.}~\bibnamefont{Huttner}}, \bibnamefont{and}
  \bibinfo{author}{\bibfnamefont{S.~J.} \bibnamefont{Phoenix}},
  \bibinfo{journal}{Journal of Modern Optics} \textbf{\bibinfo{volume}{40}},
  \bibinfo{pages}{2501} (\bibinfo{year}{1993}).

\bibitem[{\citenamefont{Watanabe et~al.}(2008)\citenamefont{Watanabe,
  Matsumoto, and Uyematsu}}]{WMU08}
\bibinfo{author}{\bibfnamefont{S.}~\bibnamefont{Watanabe}},
  \bibinfo{author}{\bibfnamefont{R.}~\bibnamefont{Matsumoto}},
  \bibnamefont{and} \bibinfo{author}{\bibfnamefont{T.}~\bibnamefont{Uyematsu}},
  \bibinfo{journal}{Physical Review A} \textbf{\bibinfo{volume}{78}},
  \bibinfo{pages}{042316} (\bibinfo{year}{2008}).

\bibitem[{\citenamefont{Matsumoto and Watanabe}(2008)}]{MW08}
\bibinfo{author}{\bibfnamefont{R.}~\bibnamefont{Matsumoto}} \bibnamefont{and}
  \bibinfo{author}{\bibfnamefont{S.}~\bibnamefont{Watanabe}},
  \bibinfo{journal}{IEICE Transactions on Fundamentals of Electronics,
  Communications and Computer Sciences} \textbf{\bibinfo{volume}{91}},
  \bibinfo{pages}{2870} (\bibinfo{year}{2008}).

\bibitem[{\citenamefont{Tamaki et~al.}(2014)\citenamefont{Tamaki, Curty, Kato,
  Lo, and Azuma}}]{TCK+14}
\bibinfo{author}{\bibfnamefont{K.}~\bibnamefont{Tamaki}},
  \bibinfo{author}{\bibfnamefont{M.}~\bibnamefont{Curty}},
  \bibinfo{author}{\bibfnamefont{G.}~\bibnamefont{Kato}},
  \bibinfo{author}{\bibfnamefont{H.-K.} \bibnamefont{Lo}}, \bibnamefont{and}
  \bibinfo{author}{\bibfnamefont{K.}~\bibnamefont{Azuma}},
  \bibinfo{journal}{Physical Review A} \textbf{\bibinfo{volume}{90}},
  \bibinfo{pages}{052314} (\bibinfo{year}{2014}).

\bibitem[{\citenamefont{Vasylyev et~al.}(2012)\citenamefont{Vasylyev, Semenov,
  and Vogel}}]{VSV12}
\bibinfo{author}{\bibfnamefont{D.~Y.} \bibnamefont{Vasylyev}},
  \bibinfo{author}{\bibfnamefont{A.}~\bibnamefont{Semenov}}, \bibnamefont{and}
  \bibinfo{author}{\bibfnamefont{W.}~\bibnamefont{Vogel}},
  \bibinfo{journal}{Physical Review Letters} \textbf{\bibinfo{volume}{108}},
  \bibinfo{pages}{220501} (\bibinfo{year}{2012}).

\bibitem[{\citenamefont{Valley}(1980)}]{V80}
\bibinfo{author}{\bibfnamefont{G.~C.} \bibnamefont{Valley}},
  \bibinfo{journal}{Applied Optics} \textbf{\bibinfo{volume}{19}},
  \bibinfo{pages}{574} (\bibinfo{year}{1980}).

\bibitem[{\citenamefont{Hufnagel and Stanley}(1964)}]{HS64}
\bibinfo{author}{\bibfnamefont{R.}~\bibnamefont{Hufnagel}} \bibnamefont{and}
  \bibinfo{author}{\bibfnamefont{N.}~\bibnamefont{Stanley}},
  \bibinfo{journal}{JOSA} \textbf{\bibinfo{volume}{54}}, \bibinfo{pages}{52}
  (\bibinfo{year}{1964}).

\bibitem[{\citenamefont{Lawson and Carrano}(2006)}]{LC06}
\bibinfo{author}{\bibfnamefont{J.~K.} \bibnamefont{Lawson}} \bibnamefont{and}
  \bibinfo{author}{\bibfnamefont{C.~J.} \bibnamefont{Carrano}}, in
  \emph{\bibinfo{booktitle}{Atmospheric Optical Modeling, Measurement, and
  Simulation II}} (\bibinfo{organization}{SPIE}, \bibinfo{year}{2006}), vol.
  \bibinfo{volume}{6303}, pp. \bibinfo{pages}{38--49}.

\bibitem[{\citenamefont{Frehlich et~al.}(2010)\citenamefont{Frehlich, Sharman,
  Vandenberghe, Yu, Liu, Knievel, and Jumper}}]{FSV+10}
\bibinfo{author}{\bibfnamefont{R.}~\bibnamefont{Frehlich}},
  \bibinfo{author}{\bibfnamefont{R.}~\bibnamefont{Sharman}},
  \bibinfo{author}{\bibfnamefont{F.}~\bibnamefont{Vandenberghe}},
  \bibinfo{author}{\bibfnamefont{W.}~\bibnamefont{Yu}},
  \bibinfo{author}{\bibfnamefont{Y.}~\bibnamefont{Liu}},
  \bibinfo{author}{\bibfnamefont{J.}~\bibnamefont{Knievel}}, \bibnamefont{and}
  \bibinfo{author}{\bibfnamefont{G.}~\bibnamefont{Jumper}},
  \bibinfo{journal}{Journal of Applied Meteorology and Climatology}
  \textbf{\bibinfo{volume}{49}}, \bibinfo{pages}{1742} (\bibinfo{year}{2010}).

\bibitem[{\citenamefont{Tomasi and Paccagnella}(1988)}]{TP88}
\bibinfo{author}{\bibfnamefont{C.}~\bibnamefont{Tomasi}} \bibnamefont{and}
  \bibinfo{author}{\bibfnamefont{T.}~\bibnamefont{Paccagnella}},
  \bibinfo{journal}{Pure and Applied Geophysics}
  \textbf{\bibinfo{volume}{127}}, \bibinfo{pages}{93} (\bibinfo{year}{1988}).

\bibitem[{\citenamefont{Tomasi}(1984)}]{T84}
\bibinfo{author}{\bibfnamefont{C.}~\bibnamefont{Tomasi}},
  \bibinfo{journal}{Journal of Geophysical Research: Atmospheres}
  \textbf{\bibinfo{volume}{89}}, \bibinfo{pages}{2563} (\bibinfo{year}{1984}).

\bibitem[{\citenamefont{Wang et~al.}(2018)\citenamefont{Wang, Huang, Wang, and
  Zeng}}]{WHW+18}
\bibinfo{author}{\bibfnamefont{S.}~\bibnamefont{Wang}},
  \bibinfo{author}{\bibfnamefont{P.}~\bibnamefont{Huang}},
  \bibinfo{author}{\bibfnamefont{T.}~\bibnamefont{Wang}}, \bibnamefont{and}
  \bibinfo{author}{\bibfnamefont{G.}~\bibnamefont{Zeng}}, \bibinfo{journal}{New
  Journal of Physics} \textbf{\bibinfo{volume}{20}}, \bibinfo{pages}{083037}
  (\bibinfo{year}{2018}).

\bibitem[{\citenamefont{Vargas et~al.}(2000)\citenamefont{Vargas, Ben{\'\i}tez,
  and Bajo}}]{VBB2000}
\bibinfo{author}{\bibfnamefont{M.~J.} \bibnamefont{Vargas}},
  \bibinfo{author}{\bibfnamefont{P.~M.} \bibnamefont{Ben{\'\i}tez}},
  \bibnamefont{and} \bibinfo{author}{\bibfnamefont{F.~S.} \bibnamefont{Bajo}},
  \bibinfo{journal}{European Journal of Physics} \textbf{\bibinfo{volume}{21}},
  \bibinfo{pages}{245} (\bibinfo{year}{2000}).

\bibitem[{\citenamefont{Ma et~al.}(2012)\citenamefont{Ma, Fung, and
  Razavi}}]{MFR12}
\bibinfo{author}{\bibfnamefont{X.}~\bibnamefont{Ma}},
  \bibinfo{author}{\bibfnamefont{C.-H.~F.} \bibnamefont{Fung}},
  \bibnamefont{and} \bibinfo{author}{\bibfnamefont{M.}~\bibnamefont{Razavi}},
  \bibinfo{journal}{Physical Review A} \textbf{\bibinfo{volume}{86}},
  \bibinfo{pages}{052305} (\bibinfo{year}{2012}).

\bibitem[{\citenamefont{Xu et~al.}(2014)\citenamefont{Xu, Xu, and Lo}}]{XXL14}
\bibinfo{author}{\bibfnamefont{F.}~\bibnamefont{Xu}},
  \bibinfo{author}{\bibfnamefont{H.}~\bibnamefont{Xu}}, \bibnamefont{and}
  \bibinfo{author}{\bibfnamefont{H.-K.} \bibnamefont{Lo}},
  \bibinfo{journal}{Physical Review A} \textbf{\bibinfo{volume}{89}},
  \bibinfo{pages}{052333} (\bibinfo{year}{2014}).

\bibitem[{\citenamefont{Limpert et~al.}(2001)\citenamefont{Limpert, Stahel, and
  Abbt}}]{LSA01}
\bibinfo{author}{\bibfnamefont{E.}~\bibnamefont{Limpert}},
  \bibinfo{author}{\bibfnamefont{W.~A.} \bibnamefont{Stahel}},
  \bibnamefont{and} \bibinfo{author}{\bibfnamefont{M.}~\bibnamefont{Abbt}},
  \bibinfo{journal}{BioScience} \textbf{\bibinfo{volume}{51}},
  \bibinfo{pages}{341} (\bibinfo{year}{2001}).

\bibitem[{\citenamefont{Stassinakis et~al.}(2013)\citenamefont{Stassinakis,
  Nistazakis, Peppas, and Tombras}}]{SNP+13}
\bibinfo{author}{\bibfnamefont{A.}~\bibnamefont{Stassinakis}},
  \bibinfo{author}{\bibfnamefont{H.}~\bibnamefont{Nistazakis}},
  \bibinfo{author}{\bibfnamefont{K.}~\bibnamefont{Peppas}}, \bibnamefont{and}
  \bibinfo{author}{\bibfnamefont{G.}~\bibnamefont{Tombras}},
  \bibinfo{journal}{Optics \& Laser Technology} \textbf{\bibinfo{volume}{54}},
  \bibinfo{pages}{329} (\bibinfo{year}{2013}).

\bibitem[{\citenamefont{Al-Habash et~al.}(2001)\citenamefont{Al-Habash,
  Andrews, and Phillips}}]{HAP01}
\bibinfo{author}{\bibfnamefont{M.}~\bibnamefont{Al-Habash}},
  \bibinfo{author}{\bibfnamefont{L.~C.} \bibnamefont{Andrews}},
  \bibnamefont{and} \bibinfo{author}{\bibfnamefont{R.~L.}
  \bibnamefont{Phillips}}, \bibinfo{journal}{Optical Engineering}
  \textbf{\bibinfo{volume}{40}}, \bibinfo{pages}{1554} (\bibinfo{year}{2001}).

\bibitem[{\citenamefont{Chatzidiamantis
  et~al.}(2010)\citenamefont{Chatzidiamantis, Sandalidis, Karagiannidis,
  Kotsopoulos, and Matthaiou}}]{CSK+10}
\bibinfo{author}{\bibfnamefont{N.~D.} \bibnamefont{Chatzidiamantis}},
  \bibinfo{author}{\bibfnamefont{H.~G.} \bibnamefont{Sandalidis}},
  \bibinfo{author}{\bibfnamefont{G.~K.} \bibnamefont{Karagiannidis}},
  \bibinfo{author}{\bibfnamefont{S.~A.} \bibnamefont{Kotsopoulos}},
  \bibnamefont{and}
  \bibinfo{author}{\bibfnamefont{M.}~\bibnamefont{Matthaiou}}, in
  \emph{\bibinfo{booktitle}{2010 17th International Conference on
  Telecommunications}} (\bibinfo{organization}{IEEE}, \bibinfo{year}{2010}),
  pp. \bibinfo{pages}{487--492}.

\bibitem[{\citenamefont{Dong et~al.}(2022)\citenamefont{Dong, Huang, Cui, and
  Jiao}}]{DHC+22}
\bibinfo{author}{\bibfnamefont{Q.}~\bibnamefont{Dong}},
  \bibinfo{author}{\bibfnamefont{G.}~\bibnamefont{Huang}},
  \bibinfo{author}{\bibfnamefont{W.}~\bibnamefont{Cui}}, \bibnamefont{and}
  \bibinfo{author}{\bibfnamefont{R.}~\bibnamefont{Jiao}},
  \bibinfo{journal}{Quantum Science and Technology}
  \textbf{\bibinfo{volume}{7}}, \bibinfo{pages}{015014} (\bibinfo{year}{2022}).

\bibitem[{\citenamefont{Liang and Jiao}(2020)}]{LJ20}
\bibinfo{author}{\bibfnamefont{W.}~\bibnamefont{Liang}} \bibnamefont{and}
  \bibinfo{author}{\bibfnamefont{R.}~\bibnamefont{Jiao}}, \bibinfo{journal}{New
  Journal of Physics} \textbf{\bibinfo{volume}{22}}, \bibinfo{pages}{083074}
  (\bibinfo{year}{2020}).

\bibitem[{\citenamefont{Dutta and Pathak}(2022{\natexlab{b}})}]{DP22}
\bibinfo{author}{\bibfnamefont{A.}~\bibnamefont{Dutta}} \bibnamefont{and}
  \bibinfo{author}{\bibfnamefont{A.}~\bibnamefont{Pathak}},
  \bibinfo{journal}{Quantum Information Processing}
  \textbf{\bibinfo{volume}{21}}, \bibinfo{pages}{369}
  (\bibinfo{year}{2022}{\natexlab{b}}).

\end{thebibliography}

\appendix

\section*{Appendix A \label{sec:Appendix-A}}

We recap the security analysis proposed in Ref. \cite{IK21} and show
our important modification in the investigation of the minimum value
key rate (per pulse) for HD-Ext-B92 protocol. We elaborate the theorem
\cite{K16} which provides the lower bound of the conditional von
Neumann entropy of classical-quantum state $\rho_{aE}$ in Hilbert
space\footnote{Alice's register and Eve's quantum memory are represented in Hilbert
space $\mathcal{H}_{a}$ and $\mathcal{H}_{E}$, respectively.} $\mathcal{H}_{a}\otimes\mathcal{H}_{E}$.

\emph{Theorem }Let $\mathcal{H}_{a}$ and $\mathcal{H}_{E}$ are finite-dimensional
Hilbert space and consider the following state of Alice and Eve in
the form of density matrix,

\begin{equation}
\rho_{aE}=\frac{1}{M}\left(|0\rangle\langle0|_{a}\otimes\left[\sum_{x=1}^{{\rm d}}|E_{x}^{0}\rangle\langle E_{x}^{0}|\right]+|1\rangle\langle1|_{a}\otimes\left[\sum_{x=1}^{{\rm d}}|E_{x}^{1}\rangle\langle E_{x}^{1}|\right]\right),\label{eq:Classical-Quantum state of ALice and Bob}
\end{equation}
where $M(>0)$ is normalization factor, ${\rm d}$ has finite value,
and each state\footnote{Eve's states are not necessarily normalized, nor orthogonal; it might
be that $|E_{x}^{y}\rangle\equiv0$ also.} $|E_{x}^{y}\rangle\in\mathcal{H}_{E}$. Assuming $K_{x}^{y}=\langle E_{x}^{y}|E_{x}^{y}\rangle\geq0$,
then,

\[
S\left(a|E\right)_{\rho_{aE}}\geq\sum_{x=1}^{{\rm d}}\left(\frac{K_{x}^{0}+K_{x}^{1}}{M}\right)S_{x},
\]
where
\[
S_{x}=\begin{cases}
h\left(\frac{K_{x}^{0}}{K_{x}^{0}\,+\,K_{x}^{1}}\right)-h\left(\delta_{x}\right) & {\rm if}\,\,K_{x}^{0}\wedge K_{x}^{1}\geq0,\\
0 & {\rm otherwise},
\end{cases}
\]
and
\[
\delta_{x}=\frac{1}{2}+\frac{\sqrt{\left(K_{x}^{0}-K_{x}^{1}\right)^{2}+4{\rm Re^{2}}\langle E_{x}^{0}|E_{x}^{1}\rangle}}{2\left(K_{x}^{0}+K_{x}^{1}\right)}.
\]
This \emph{Theorem}{}
serves as the foundation for our analysis, facilitating the derivation
of the key rate equation for the protocol discussed throughout the
remainder of this appendix.

The action of Eve's unitary operation $\mathcal{E}_{TE}$ on Alice's
transmitted state $|\Upsilon\rangle_{T}$ and Eve's ancilla state
$|\chi\rangle_{E}$ is described in the following,

\[
\mathcal{E}_{TE}|\Upsilon\rangle_{T}\otimes|\chi\rangle_{E}=\sum_{c=1}^{d}|c,E_{c}^{\Upsilon}\rangle_{TE},
\]
and

\[
\begin{array}{lcl}
\mathcal{E}_{TE}|\psi\rangle_{T}\otimes|\chi\rangle_{E} & = & \mathcal{E}_{TE}\frac{1}{\sqrt{2}}\left(|m\rangle+|n\rangle\right)_{T}\otimes|\chi\rangle_{E}\\
 & = & \frac{1}{\sqrt{2}}\stackrel[c=1]{d}{\sum}|c\rangle_{T}\otimes|F_{c}\rangle_{E},
\end{array}
\]
where $|F_{c}\rangle_{E}:=|E_{c}^{m}\rangle_{E}+|E_{c}^{n}\rangle_{E}$,
and $E_{c}^{\Upsilon}$ is an arbitrary state in Eve's ancillary basis
when Alice's transmitted state before and after Eve's operation are
$|\Upsilon\rangle_{T}$ and $|c\rangle_{T}$, respectively. As $\mathcal{E}_{TE}$
is a unitary operation the relation holds as $\stackrel[c=1]{d}{\sum}\langle E_{c}^{\Upsilon}|E_{c}^{\Upsilon}\rangle=1$.
After Eve's unitary operation on the classical-quantum state, $\rho_{aT}=\frac{1}{2}\left(|0\rangle\langle0|_{a}\otimes|m\rangle\langle m|_{T}+|1\rangle\langle1|_{a}\otimes|\psi\rangle\langle\psi|_{T}\right)$
is as following,

\begin{equation}
\begin{array}{lcl}
\rho_{aTE} & = & \mathcal{E}_{TE}\left(\rho_{aT}\right)\\
 & = & \frac{1}{2}\left[|0\rangle\langle0|_{a}\otimes P\left(\stackrel[c=1]{d}{\sum}|c,E_{c}^{m}\rangle_{TE}\right)+|1\rangle\langle1|_{a}\otimes P\left(\frac{1}{\sqrt{2}}\stackrel[c=1]{d}{\sum}|c,F_{c}\rangle_{TE}\right)\right],
\end{array}\label{eq:Alice-Transmitted-Eve's state}
\end{equation}
where $P\left(|\upsilon\rangle\right)=|\upsilon\rangle\langle\upsilon|$
is projection operator. After receiving the transmitted register $T$
Bob will apply the measurement operators $M_{0}=I_{a}\otimes\left(I-|\psi\rangle\langle\psi|\right)_{T}\otimes I_{E}$
and $M_{1}=I_{a}\otimes\left(I-|m\rangle\langle m|\right)_{T}\otimes I_{E}$
on $T$. Using Eq. (\ref{eq:Alice-Transmitted-Eve's state}) we can
write density state after Bob's operations,

\begin{equation}
\begin{array}{lcl}
\rho_{aTE}^{0} & = & M_{0}\left(\rho_{aTE}\right)M_{0}^{\dagger}\\
 & = & \frac{1}{2}\left[|0\rangle\langle0|\otimes P\left\{ \underset{c\ne m,c\ne n}{\sum}|c,E_{c}^{m}\rangle+\frac{1}{2}\,|m,E_{m}^{m}-E_{n}^{m}\rangle-\frac{1}{2}\,|n,E_{m}^{m}-E_{n}^{m}\rangle\right\} \right.\\
 & + & \left.|1\rangle\langle1|\otimes P\left\{ \frac{1}{\sqrt{2}}\left(\underset{c\ne m,c\ne n}{\sum}|c,F_{c}\rangle+\frac{1}{2}\,|m,F_{m}-F_{n}\rangle-\frac{1}{2}\,|n,F_{m}-F_{n}\rangle\right)\right\} \right]_{aTE},
\end{array}\label{eq:Alice-Transmit-Eve when Bob gets 0}
\end{equation}
and
\begin{equation}
\begin{array}{lcl}
\rho_{aTE}^{1} & = & M_{1}\left(\rho_{aTE}\right)M_{1}^{\dagger}\\
 & = & \frac{1}{2}\left[|0\rangle\langle0|\otimes P\left(\underset{c\ne m}{\sum}|c,E_{c}^{m}\rangle\right)+|1\rangle\langle1|\otimes P\left(\frac{1}{\sqrt{2}}\underset{c\ne m}{\sum}|c,F_{c}\rangle\right)\right]_{aTE}.
\end{array}\label{eq:Alice-Transmit-Eve when Bob gets 1}
\end{equation}
After Bob gets his outcomes Eqs. (\ref{eq:Alice-Transmit-Eve when Bob gets 0})
and (\ref{eq:Alice-Transmit-Eve when Bob gets 1}) may be traced out
the transit register $T$ and include Bob's classical register $b$
to keep his measurement result. Now the Eqs. (\ref{eq:Alice-Transmit-Eve when Bob gets 0})
and (\ref{eq:Alice-Transmit-Eve when Bob gets 1}) can be written
like,

\begin{equation}
\begin{array}{lcl}
\rho_{aEb}^{0} & = & \frac{1}{2}\left[|0\rangle\langle0|_{a}\otimes\left\{ \underset{c\ne m,c\ne n}{\sum}|E_{c}^{m}\rangle\langle E_{c}^{m}|+\frac{1}{2}\,|\left(E_{m}^{m}-E_{n}^{m}\right)\rangle\langle\left(E_{m}^{m}-E_{n}^{m}\right)|\right\} _{E}\otimes|0\rangle\langle0|_{b}\right.\\
 & + & \left.|1\rangle\langle1|_{a}\otimes\frac{1}{2}\left\{ \underset{c\ne m,c\ne n}{\sum}|F_{c}\rangle\langle F_{c}|+\frac{1}{2}\,|\left(F_{m}-F_{n}\right)\rangle\langle\left(F_{m}-F_{n}\right)|\right\} _{E}\otimes|0\rangle\langle0|_{b}\right],
\end{array}\label{eq:Alice-Eve-Bob-bit_0}
\end{equation}
and
\begin{equation}
\rho_{aEb}^{1}=\frac{1}{2}\left[|0\rangle\langle0|_{a}\otimes\sum_{c\ne m}|E_{c}^{m}\rangle\langle E_{c}^{m}|\otimes|1\rangle\langle1|_{b}+|1\rangle\langle1|_{a}\otimes\frac{1}{2}\sum_{c\ne m}|F_{c}\rangle\langle F_{c}|\otimes|1\rangle\langle1|_{b}\right].\label{eq:Alice-Eve-Bob-bit_1}
\end{equation}
Adding up Eqs. (\ref{eq:Alice-Eve-Bob-bit_0}) and (\ref{eq:Alice-Eve-Bob-bit_1}),
the non-normalized density operator which represents in one key-bit
generation round is,

\begin{equation}
\begin{array}{lcl}
\rho_{aEb} & = & \rho_{aEb}^{1}+\rho_{aEb}^{0}\\
 & = & \frac{1}{2}\,|0\rangle\langle0|_{a}\otimes\left[\left\{ \underset{c\ne m,c\ne n}{\sum}|E_{c}^{m}\rangle\langle E_{c}^{m}|+\frac{1}{2}\,|\left(E_{m}^{m}-E_{n}^{m}\right)\rangle\langle\left(E_{m}^{m}-E_{n}^{m}\right)|\right\} \otimes|0\rangle\langle0|_{b}+\underset{c\ne m}{\sum}|E_{c}^{m}\rangle\langle E_{c}^{m}|\otimes|1\rangle\langle1|_{b}\right]\\
 & + & \frac{1}{2}\,|1\rangle\langle1|_{a}\otimes\left[\frac{1}{2}\left\{ \underset{c\ne m,c\ne n}{\sum}|F_{c}\rangle\langle F_{c}|+\frac{1}{2}\,|\left(F_{m}-F_{n}\right)\rangle\langle\left(F_{m}-F_{n}\right)|\right\} \otimes|0\rangle\langle0|_{b}+\frac{1}{2}\underset{c\ne m}{\sum}|F_{c}\rangle\langle F_{c}|\otimes|1\rangle\langle1|_{b}\right]
\end{array}.\label{eq:Non-normalized density operator}
\end{equation}
For computing the conditional entropy $H\left(a|b\right)$, we will
show how the Eq. (\ref{eq:Non-normalized density operator}) is utilized
to get the statistics for all combinations of Alice's and Bob's sifted
key. Now, trace out Bob's register from Eq. (\ref{eq:Non-normalized density operator})
to keep the composite state of Alice's register and Eve's memory which
is important to calculate $S(a|E)$. The final expression of the required
density matrix is,

\begin{equation}
\begin{array}{lcl}
\rho_{aE} & = & \frac{1}{M}\left[|0\rangle\langle0|_{a}\otimes\left(\underset{c\ne m,c\ne n}{\sum}|E_{c}^{m}\rangle\langle E_{c}^{m}|+\frac{1}{4}\,|E_{m}^{m}\rangle\langle E_{m}^{m}|-\frac{1}{4}\,|E_{m}^{m}\rangle\langle E_{n}^{m}|-\frac{1}{4}\,|E_{n}^{m}\rangle\langle E_{m}^{m}|+\frac{3}{4}\,|E_{n}^{m}\rangle\langle E_{n}^{m}|\right)_{E}\right.\\
 & + & \left.|1\rangle\langle1|_{a}\otimes\left(\underset{c\ne m,c\ne n}{\sum}\frac{1}{2}\,|F_{c}\rangle\langle F_{c}|+\frac{1}{8}\,|F_{m}\rangle\langle F_{m}|-\frac{1}{8}\,|F_{m}\rangle\langle F_{n}|-\frac{1}{8}\,|F_{n}\rangle\langle F_{m}|+\frac{3}{8}\,|F_{n}\rangle\langle F_{n}|\right)_{E}\right],
\end{array}\label{eq:Alice-Eve required final expression}
\end{equation}
where $M$ is the normalization factor that can be calculated as,

\[
\begin{array}{lcl}
M & = & \underset{c\ne m,c\ne n}{\sum}\langle E_{c}^{m}|E_{c}^{m}\rangle+\frac{1}{2}\,\langle E_{n}^{m}|E_{n}^{m}\rangle+\frac{1}{4}\,\langle\left(E_{m}^{m}-E_{n}^{m}\right)|\left(E_{m}^{m}-E_{n}^{m}\right)\rangle\\
 & + & \frac{1}{2}\underset{c\ne m,c\ne n}{\sum}\langle F_{c}|F_{c}\rangle+\frac{1}{4}\,\langle F_{n}|F_{n}\rangle+\frac{1}{8}\,\langle\left(F_{m}-F_{n}\right)|\left(F_{m}-F_{n}\right)\rangle
\end{array}.
\]
We modify the derivation for $S(a|E)$ using $\rho_{aE}$ in comparison
with the seminal work \cite{IK21}. In our modified calculation, we
express the terms of $\rho_{aE}$ in Eve's two bases states, i.e.,
$\{E_{c}^{m}\}$ and $\{F_{c}\}$ which correspond to the bit values
(i.e., $0$ and $1$) in Alice's register\footnote{The above \emph{Theorem }allows our expression of Eq. (\ref{eq:Alice-Eve required final expression})
unlike the Eq. (5) in Ref. \cite{IK21}.}.

Applying this above \emph{theorem} we calculate the conditional von
Neumann entropy,

\begin{equation}
S(a|E)\ge\underset{c\ne m,c\ne n}{\sum}\left(\frac{K_{c}^{0}+K_{c}^{1}}{M}\right)S_{c}+\left(\frac{K_{m}^{0}+K_{m}^{1}}{M}\right)S_{m}+\left(\frac{K_{n}^{0}+K_{n}^{1}}{M}\right)S_{n},\label{eq:Value of S(A|E)}
\end{equation}
where

\[
\begin{array}{lclcclcl}
K_{c}^{0} & := & \langle E_{c}^{m}|E_{c}^{m}\rangle, &  &  & K_{c}^{1} & := & \frac{1}{2}\,\langle F_{c}|F_{c}\rangle,\forall\,c\ne m,n\\
\\
K_{m}^{0} & := & \frac{1}{4}\,\langle E_{m}^{m}|E_{m}^{m}\rangle, &  &  & K_{m}^{0} & := & \frac{1}{8}\,\langle F_{m}|F_{m}\rangle,\\
\\
K_{n}^{0} & := & \frac{3}{4}\,\langle E_{n}^{m}|E_{n}^{m}\rangle, &  &  & K_{n}^{1} & := & \frac{3}{8}\,\langle F_{n}|F_{n}\rangle.
\end{array}
\]
And
\[
\begin{array}{lcl}
S_{c} & = & h\left(\frac{K_{c}^{0}}{K_{c}^{0}\,+\,K_{c}^{1}}\right)-h\left(\frac{1}{2}+\frac{\sqrt{\left(K_{c}^{0}\,-\,K_{c}^{1}\right)^{2}+4\,{\rm Re^{2}}\langle E_{c}^{m}|\frac{1}{\sqrt{2}}F_{c}\rangle}}{2\,\left(K_{c}^{0}\,+\,K_{c}^{1}\right)}\right),\\
S_{m} & = & h\left(\frac{K_{m}^{0}}{K_{m}^{0}\,+\,K_{m}^{1}}\right)-h\left(\frac{1}{2}+\frac{\sqrt{\left(K_{m}^{0}\,-\,K_{m}^{1}\right)^{2}+4\,{\rm Re^{2}}\langle\frac{1}{2}E_{m}^{m}|\frac{1}{2\sqrt{2}}F_{m}\rangle}}{2\,\left(K_{m}^{0}\,+\,K_{m}^{1}\right)}\right),\\
S_{n} & = & h\left(\frac{K_{n}^{0}}{K_{n}^{0}\,+\,K_{n}^{1}}\right)-h\left(\frac{1}{2}+\frac{\sqrt{\left(K_{n}^{0}\,-\,K_{n}^{1}\right)^{2}+4\,{\rm Re^{2}}\frac{3}{4\sqrt{2}}\langle E_{n}^{m}|F_{n}\rangle}}{2\,\left(K_{n}^{0}\,+\,K_{n}^{1}\right)}\right).
\end{array}
\]
Here, we briefly describe the \emph{parameter estimation} for the
required statistics to get the values in the above equations. Let
$p_{\upsilon c}\,(p_{\upsilon\psi})$ be the observable parameter
when Bob's measurement outcome is $|c\rangle\,(|\psi\rangle)$ using
the $Z\,(X)$ basis when Alice sends state\footnote{Here, the generalized state is $|\upsilon\rangle\in\{|m\rangle,|n\rangle,|\psi\rangle\}$,
these statistics $\left(p_{\upsilon c(\psi)}\right)$ come from the
rounds where Alice and Bob do the same or different basis measurement
(see Table $1$ in Ref. \cite{IK21}).} $|\upsilon\rangle$. We may write in the form of the observable parameters
$K_{c}^{0}=p_{mc},$ $K_{c}^{1}=p_{\psi c},$ $K_{m}^{0}=\frac{1}{4}\,p_{mm},$
$K_{m}^{1}=\frac{1}{4}\,p_{\psi m},$ $K_{n}^{0}=\frac{3}{4}\,p_{mn},$
and $K_{n}^{1}=\frac{3}{4}\,p_{\psi n}.$

\[
\begin{array}{lcl}
{\rm Re\,}\langle E_{c}^{m}|\frac{1}{\sqrt{2}}F_{c}\rangle & = & \frac{1}{\sqrt{2}}\left(\frac{p_{mc}}{2}+p_{\psi c}-\frac{p_{nc}}{2}\right),\\
\\
{\rm Re\,}\langle\frac{1}{2}E_{m}^{m}|\frac{1}{2\sqrt{2}}F_{m}\rangle & = & \frac{1}{4\sqrt{2}}\left(\frac{p_{mm}}{2}+p_{\psi m}-\frac{p_{nm}}{2}\right),\\
\\
{\rm Re}\,\frac{3}{4\sqrt{2}}\langle E_{n}^{m}|F_{n}\rangle & = & \frac{3}{4\sqrt{2}}\left(\frac{p_{mn}}{2}+p_{\psi n}-\frac{p_{nn}}{2}\right).
\end{array}
\]

In our study, we take only the depolarizing channel to evaluate the
satellite-based effect of the HD-Ext-B92 protocol. Suppose the depolarizing
channel $\mathcal{D}_{q}(\rho)$ with parameter ${\rm q}$ acting
on a density operator $\rho$ on a Hilbert space of dimension ${\rm d}$.
$\mathcal{D_{{\rm q}}}(\rho)$ acts as follows,

\[
\mathcal{D}_{{\rm q}}(\rho)=\left(1-\frac{{\rm d}}{{\rm d}-1}{\rm q}\right)\rho+\frac{{\rm q}}{{\rm d}-1}I.
\]
In the above, we have already mentioned the required parameter to
calculate the key rate in terms of observable statistics. The observable
statistics may be written in the effect of depolarizing channel scenario,

\[
\begin{array}{lccclclcl}
 &  & p_{mm} & = & p_{nn} & = & p_{\psi\psi} & = & 1-{\rm q},\\
 &  & p_{mc} & = & p_{nc} & = & p_{\psi c} & = & \frac{{\rm q}}{{\rm d}-1},\\
p_{m\psi} & = & p_{n\psi} & = & p_{\psi m} & = & p_{\psi n} & = & \frac{1}{2}\left(1-\frac{{\rm q}\,{\rm d}}{{\rm d}-1}\right)+\frac{{\rm q}}{{\rm d}-1}.
\end{array}
\]
The above analysis is sufficient to evaluate $S(a|E)$ using Eq. (\ref{eq:Value of S(A|E)}),
and to get the key rate we need the value of $H(a|b)$ which is analyzed\footnote{Assuming $p_{ij}$ is the joint probability when Alice's and Bob's
raw bit are ``$i$'' and ``$j$'' given that not eliminating that
iteration \cite{DP22}.} in the following,

\begin{equation}
\begin{array}{lcl}
H\left(a|b\right) & = & H\left(p_{00},\,p_{01},\,p_{10},\,p_{11}\right)-h\left(p_{00}+p_{10}\right).\end{array}\label{eq:Value of H(a|b)}
\end{equation}
To compute Eq. (\ref{eq:Value of H(a|b)}), Alice and Bob use classical
sampling i.e., the values of observable probabilities under the simulated
channel. Using Eq. (\ref{eq:Non-normalized density operator}) with
normalization term $M$,

\[
\begin{array}{lcl}
p_{00} & = & \frac{1}{2M}\left(1-p_{m\psi}\right),\\
p_{01} & = & \frac{1}{2M}\left(1-p_{mm}\right),\\
p_{10} & = & \frac{1}{2M}\left(1-p_{\psi\psi}\right),\\
p_{11} & = & \frac{1}{2M}\left(1-p_{\psi i}\right).
\end{array}
\]
These are the needful analyses that we recap above for estimating
the minimum value of the key rate in Eq. (\ref{eq:Key-rate equantion}).

\section*{Appendix B\label{sec:Appendix-B}}

We may write the first and second moments of the beam parameters in
Eq. (\ref{eq:Vector of beam-parameters}) concerning the connection
detailed in Eq. (\ref{eq:Down-Link and Up-Link condition}). The angle
of orientation of the elliptical profile $\varphi$ is presumed to
have a uniform distribution within the interval $[0,\frac{\pi}{2}]$.
The mean value and the variance in the centroid position of the beam,
in the case of up-links, are consistent for both the $x$ and $y$
directions, and they are equal to\footnote{See for details Appendix C in Ref. \cite{VSV+17}.},

\[
\begin{array}{lclcl}
\left\langle x_{0}\right\rangle  & = & \left\langle y_{0}\right\rangle  & = & 0,\\
\left\langle x_{0}^{2}\right\rangle  & = & \left\langle y_{0}^{2}\right\rangle  & = & 0.419\,\sigma_{R}^{2}\mathcal{W}_{0}^{2}\Omega^{-\frac{7}{6}}{\rm \frac{h}{L}},
\end{array}
\]
in this context, the term \textquotedbl Rytov parameter\textquotedbl{}
represents the quantity $\sigma_{R}=1.23\,C_{n}^{2}k^{\frac{7}{6}}{\rm L}^{\frac{11}{6}}$,
while $\Omega=\frac{k\mathcal{W}_{0}^{2}}{2\,{\rm L}}$ stands for
the Fresnel number, $k$ is the optical wave number. The selected
reference frame is such that $\left\langle x_{0}\right\rangle =\left\langle y_{0}\right\rangle =0$.
The mean and (co) variance of $\mathcal{W}_{i}^{2}$ can be written
as,

\[
\begin{array}{lcl}
\langle\mathcal{W}_{i}^{2}\rangle & = & \frac{\mathcal{W}_{0}^{2}}{\Omega^{2}}\left(1+\frac{\pi}{8}\,{\rm L}n_{0}\mathcal{W}_{0}^{2}{\rm \frac{h}{L}}+2.6\,\sigma_{R}^{2}\Omega^{\frac{5}{6}}{\rm \frac{h}{L}}\right),\\
\langle\Delta\mathcal{W}_{i}^{2}\Delta\mathcal{W}_{j}^{2}\rangle & = & \left(2\delta_{ij}-0.8\right)\frac{\mathcal{W}_{0}^{4}}{\Omega^{\frac{19}{6}}}\left(1+\frac{\pi}{8}\,{\rm L}n_{0}\mathcal{W}_{0}^{2}{\rm \frac{h}{L}}\right)\sigma_{R}^{2}{\rm \frac{h}{L}}.
\end{array}
\]
The same type of expressions also applies to down-links when considering
the position of the beam centroid,

\[
\begin{array}{lclcl}
\left\langle x_{0}\right\rangle  & = & \left\langle y_{0}\right\rangle  & = & 0,\\
\left\langle x_{0}^{2}\right\rangle  & = & \left\langle y_{0}^{2}\right\rangle  & = & \alpha\,{\rm L},
\end{array}
\]
additionally, for the semi-major and semi-minor axes of the elliptical
beam profile,

\[
\begin{array}{lcl}
\langle\mathcal{W}_{i}^{2}\rangle & = & \frac{\mathcal{W}_{0}^{2}}{\Omega^{2}}\left(1+\frac{\pi}{24}\,{\rm L}n_{0}\mathcal{W}_{0}^{2}\left({\rm \frac{h}{L}}\right)^{3}+1.6\,\sigma_{R}^{2}\Omega^{\frac{5}{6}}\left({\rm \frac{h}{L}}\right)^{\frac{8}{3}}\right),\\
\langle\Delta\mathcal{W}_{i}^{2}\Delta\mathcal{W}_{j}^{2}\rangle & = & \left(2\delta_{ij}-0.8\right)\,\frac{3}{8}\,\frac{\mathcal{W}_{0}^{4}}{\Omega^{\frac{19}{6}}}\left(1+\frac{\pi}{24}\,{\rm L}n_{0}\mathcal{W}_{0}^{2}\left({\rm \frac{h}{L}}\right)^{3}\right)\sigma_{R}^{2}\left({\rm \frac{h}{L}}\right)^{\frac{8}{3}},
\end{array}
\]
here, $\alpha\approx2$ $\mu{\rm rad}$ refers to the angular pointing
error. Subsequently, the understanding of the probability distribution
concerning the elliptic beam parameters (Eq. (\ref{eq:Vector of beam-parameters}))
is applied to calculate the PDT utilizing Eq. (\ref{eq:PDT Equation})
and a process of random sampling which is mentioned in Section \ref{subsec:Elliptic_Beam_Model}.
\end{document}